\documentclass[preprint,authoryear,10pt]{elsarticle}
\usepackage{timet}
\usepackage{natbib}
\usepackage{graphicx}
\usepackage{subfigure}
\usepackage{amssymb}

\journal{Physics of the Earth and Planetary Interiors}

\newcommand{\Pm}{\ensuremath{Pm}}
\renewcommand{\vec}{\bf}
\newcommand{\la}{\leq}

\begin{document}

\begin{frontmatter}

\vspace{-2.cm}

   \title{Systematic parameter study of dynamo bifurcations  in geodynamo simulations}

   \author[L. Petitdemange]{Ludovic Petitdemange}
   \address{ENS, LRA, D\'epartement de
     Physique, \'Ecole Normale Sup\'erieure, 24 rue Lhomond, 75252
     Paris Cedex 05, France LERMA, CNRS UMR 8112 \\ 
     ludovic@lra.ens.fr
     }

\begin{abstract}
  We investigate the nature of the dynamo bifurcation in a configuration applicable to the Earth's liquid outer core, i.e. in a rotating spherical shell with thermally driven motions with no-slip boundaries. Unlike in previous studies on dynamo bifurcations, the control parameters have been varied significantly in order to deduce general tendencies. Numerical studies on the stability domain of dipolar magnetic fields found a dichotomy  between non-reversing dipole-dominated dynamos and the reversing non-dipole-dominated multipolar solutions.  We show that, by considering weak initial fields, the above transition disappears and is replaced by
a region of bistability for which dipolar and multipolar dynamos coexist. Such a result was also observed in models with free-slip boundaries in which 
the geostrophic zonal flow can develop and participate to the dynamo mechanism for non-dipolar fields. We show that a similar process develops in no-slip models when viscous effects are reduced sufficiently.

   The following three regimes are distinguished: (i) Close to the onset of convection ($Ra_c$) with only the most critical convective mode (wave number) being present, dynamos set in supercritically in the Ekman number regime explored here and are dipole-dominated. Larger critical magnetic Reynolds numbers indicate that they are particularly inefficient.  (ii) in the range $3<Ra/Ra_c<Ra_c$, the bifurcations are subcritical and  only dipole-dominated dynamos exist.  (iii) in the turbulent regime $(Ra/Ra_c>10)$, the relative importance of zonal flows increases with $Ra$ in non-magnetic models.  The field topology depends on the magnitude of the initial magnetic field.  The dipolar branch has a subcritical behavior whereas the multipolar branch has a supercritical behavior.   By approaching more realistic parameters, the extension of this bistable regime increases. A hysteretic behavior questions the common interpretation for geomagnetic reversals.

  Far above the dynamo threshold (by increasing the magnetic Prandtl number), Lorentz forces contribute to the first order force balance, as predicted for planetary dynamos. When $Ra$ is sufficiently high, dipolar fields affect significantly the flow speed, the flow structure and heat transfer which is reduced by the Lorentz force regardless of the field strength. This physical regime seems to be relevant for studying geomagnetic processes. 

\end{abstract}

\begin{keyword}
Earth core \sep Magnetohydrodynamics (MHD) \sep Dynamo: theory and simulations \sep numerical solution \sep planetary interiors, 
\end{keyword}

\end{frontmatter}

\section{Introduction}

  The mechanism whereby planets maintain magnetic
fields against ohmic decay is one of the longest standing
problems in science. It is commonly believed
that the magnetic field is generated by electromotive
forces driven by electrically conducting fluid motions
in celestial bodies, namely dynamo action \citep{moffatt,book_dormy}. Dynamo action is considered to be responsible for the presence of magnetic activity for a large variety of astrophysical objects including planets, stars and galaxies. Planetary magnetic fields result from
dynamo action thought to be driven by convection in electrically
conducting fluid regions. Convection in these systems is strongly influenced by the Coriolis force resulting from global planetary rotation.

   Since the time of the first fully three-dimensional numerical models \citep[e.g.][]{gr95}, there have been significant advances in our
understanding of the fluid dynamics of planetary cores. Many features of the
Earth's magnetic field have been reproduced numerically \citep{christensen98,christensen99,busse98,taka05,christensen07} even though realistic parameters differ by several orders of magnitude in direct numerical simulations. For instance, the Ekman number for the Earth's outer core is approximately $E=10^{-15}$ whereas $E\geq 10^{-6}$ can be considered in numerical models (see below for a complete definition of this dimensionless number). Possible field generation mechanisms in planetary conducting zones have been proposed by  \citet{olson99} and dynamo coefficients have been calculated in geodynamo models \citep{schrinner07,schrinner12} using the test-field method \citep{schrinner05}. Progress both in numerical methods as well as in parallel computer architecture has made it possible to explore an extensive parameters space in order to deduce the physical ingredients responsible for the dominance of the axial dipole field \citep{christensen06,king10,schrinner12,soderlundKA12}.

   From numerical data and theoretical arguments,  \citet{christensen06} have proposed scaling laws in order to predict observables as the magnitude of velocity field and magnetic field for planets (see also \cite{stelzer13}) and for rapidly rotating stars \citep{christensen09}. However, their predictive character and their relevance have been recently questioned by \citet{orubaD14} and  \citet{tilgner14}. In addition, according to several recent numerical studies \citep{king10,soderlundKA12,kingB13}, viscosity could play an important role in numerical results. Numerical simulations often explore a physical regime in which viscous effects would dominate inertia whereas the opposite situation is believed to hold in planetary interiors. According to  \citet{davidson14}, helical motions responsible for dynamo processes of dipolar morphology in simulations would result from the importance of viscosity. The relevance of numerical results to improve our knowledge of planetary dynamos is still a question of debate that we address in particular in this paper.

  Recently, extreme runs have been carried out in order to reduce the effects of viscosity \citep{soderlund15,Yadav16Earth,schaefferJNF17}. In these runs, the flow is strongly affected by the Lorentz force which appears as one of the dominant forces, i.e. Lorentz and Coriolis forces would comprise the leading-order force balance (MAC balance). \citet{aubertGF17} have reached very low values of the large-scale viscosity by using a different numerical approach (large-eddy simulations) and they argue for a continuous path connecting today's simulations with planetary interiors. \citet{aurnouK17} argue that the influence of the Lorentz force depends on the scale and global scale force balance would be geostrophic.  Convection would be influenced by the magnetic field only below a certain scale in simulations and in the Earth's outer core. In this paper, we show that the impact of the Lorentz force depends on the buoyant forcing in our dataset.

  Observations and numerical simulations indicate that rapid global rotation and thus the ordering influence of the Coriolis force is of major importance for the generation of coherent magnetic fields \citep{stellmach04,kapyla09,brown10}.  \citet{kutzner02} demonstrated the existence of a dipolar and
a multipolar dynamo regime and  \citet{christensen06}
showed that the transition between the two regimes is governed by a local
Rossby number ($Ro_\ell$), i.e. by the influence of inertia relative to the
Coriolis force. Similar results were reported by \citet{sreenivasan06}, as well. Dipolar models were found for small Rossby
numbers; they are separated by a fairly sharp regime boundary
from multipolar models, where inertia is more important. The models transition from a dipolar morphology to a multipolar state as the local Rossby number increases above a certain value ($Ro_\ell>0.1$). By considering different initial conditions for the magnetic field, we will show below that multipolar dynamos can be generated for local Rossby numbers lower than $0.1$ even if no-slip boundaries are used.

   Paleomagnetic measurements have allowed us to reconstruct the dynamics of the magnetic field. Irregularly over geologic time, the Earth's magnetic polarity  has changed sign and such reversals have occured several hundred times during the past 160 million years.  \citet{gr95} were the first to simulate such events numerically.  \citet{olson06} have inferred some of the physical causes associated with field reversals in planetary interiors from numerical studies. Reversals would result from the importance of inertia relative to the Coriolis force. If the local Rossby number is close to the transitional value ($Ro_\ell\approx 0.1$), the dynamos are dipole-dominated and exhibit sporadic polarity reversals. In the light of our results, we will  question the explanation proposed by \citet{olson06} for geomagnetic reversals.

  By considering lower Ekman numbers and different magnetic Prandtl numbers, we extend the study by \citet{morinD09}.  Decreasing the Ekman number  allows one to explore smaller magnetic Prandtl numbers as highlighted by \citet{christensen06}. It is of primary interest to understand dynamo bifurcations for very low $E$ and $\Pm$ as these numbers are known to be $10^{-15}$ and $10^{-6}$ respectively, in the Earth's outer core and of similar order of magnitude in other planetary dynamo regions or in rapidly rotating stellar interiors.

  In section 2, we present the differential equations  and input/output parameters. Section 3 is devoted to a hydrodynamic study and a kinematic study in which the dynamo threshold as a function of hydrodynamic forcing is determined for simple cases. A systematic study of the dynamo bifurcation is addressed in section 4 and we discuss these results in section 5.  In section 6, we focus on the action of the fields on the flow through the Lorentz force. We also discuss the physical regimes which can give rise to dipolar dynamos. We conclude and apply our results to planetary magnetism in section 7.

\section{Equations and dimensionless parameters}

  Our dynamo models are solutions of the MHD-equations for a conducting 
Boussinesq fluid in a rotating spherical shell. The fluid motion is driven by
convection due to an imposed temperature difference, \(\Delta T\) (where $T$ denotes the temperature), between 
the inner and the outer shell boundaries. The fundamental length scale
of our models is the shell width \(L\), we scale time by \(L^2/\nu\), with 
\(\nu\) the kinematic viscosity, and temperature is scaled by \(\Delta T\) and the magnetic field is considered in units of 
\(\sqrt{\varrho\mu\eta\Omega}\), with \(\varrho\) denoting the density, 
\(\mu\) the magnetic permeability, \(\eta\) the magnetic diffusivity and 
\(\Omega\) the rotation rate. With these units, the dimensionless momentum, 
temperature and induction equations are

\begin{eqnarray}
E\left(\frac{\partial{\vec{v}}}{\partial t}+({\vec{v}}\cdot{\bf \nabla}){\vec{v}}-\nabla^2\vec{v}\right)
+2{\vec{z}}\times{\vec{v}}+{\bf \nabla} P = \nonumber  \\
Ra\frac{{\vec{r}}}{r_o}T
+\frac{1}{\Pm}(\nabla\times\vec{B})\times\vec{B}\, ,\label{eq:dc:2}\\
\frac{\partial T}{\partial t}+{\vec{v}}\cdot{\bf \nabla} T   = 
\frac{1}{\Pr}\nabla^2 T \, ,\label{eq:dc:4}\\
\frac{\partial{\vec{B}}}{\partial t} = \nabla\times({\vec{v}}\times{\vec{B}})
+\frac{1}{\Pm}\nabla^2{\vec{B}} \label{eq:dc:6}.
\end{eqnarray} 
Here, the unit vector $\vec{z}$ indicates the direction of the rotation axis. We also note that the velocity field \(\vec{v}\) and the magnetic field \(\vec{B}\) are solenoidal.
The system of equations is governed by four dimensionless parameters, the 
Ekman number \(E=\nu/\Omega L^2\), the (modified) Rayleigh number 
\(Ra=\alpha_T g_0\Delta T L/\nu\Omega\), the Prandtl number \(Pr=\nu/\kappa\), 
and the magnetic Prandtl number \(Pm=\nu/\eta\). In these definitions,  
\(\alpha_T\) stands for the thermal expansion coefficient, \(g_o\) is 
the gravitational acceleration at the outer boundary, and \(\kappa\) is the 
thermal diffusivity. Another control parameter is the aspect ratio of the 
shell defined as the ratio of the inner to the outer shell radius,  
\(\chi=r_i/r_o\). It determines the width of the convection zone and is fixed in our study at 0.35. The mechanical boundary conditions are no-slip at both boundaries. Furthermore, the magnetic field matches a potential field outside the fluid shell and fixed temperatures are prescribed at both boundaries.   

  Some of the models investigated here exhibit bistability where the solution 
depends on the initial conditions for the magnetic field. A strong ($\Lambda\approx 10$) initial dipolar field gives rise to a dipolar solution whereas a multipolar solution is obtained when a weak seed field is considered as an initial condition (see section 4). Some calculations were started
from a numerical solution with slightly different parameters to test for hysteresis.  
In the bistable regime, models resulting from simulations with an initially 
weak magnetic field are referred to here as multipolar models and are 
distinguished from dipolar solutions initially started with a strong magnetic 
field.
  
The numerical solver used to compute solutions of equations 
(\ref{eq:dc:2})-(\ref{eq:dc:6}) is PaRoDy \citep[][and further developments]{dormy98}. The numerical method is similar to that described 
in  \citet{G84} except for the radial discretisation, which is treated with a finite difference scheme in physical space on a non-uniform grid denser close to the boundaries. Moreover, the pressure term has been eliminated by 
considering the double curl of the momentum equation.

  Our numerical dynamo-models are characterized by non-dimensional
output parameters. Dimensionless measures for the flow velocity are the 
magnetic Reynolds number, \(Rm=v_{\mathrm{rms}}\,L/\eta\), and the Rossby 
number, \(Ro=v_{\mathrm{rms}}/\Omega L\). In both definitions, 
\(v_{\mathrm{rms}}\) stands for the rms velocity of the flow. Similarly, 
\(B_{\mathrm{rms}}\) denotes the rms magnetic field strength.
We also measure the local Rossby number as introduced by  \citet{christensen06},
\(Ro_\ell=Ro\,\,\overline{\ell}/\pi\), based on the mean harmonic degree 
\(\overline{\ell}\) of the velocity field,
\begin{equation}
\overline{\ell}=\sum_\ell \ell\frac{<\vec{(v)}_\ell\cdot\vec{(v)}_\ell>}{<\vec{v}\cdot\vec{v}>}
\, .
\label{eq:dc:7}
\end{equation} 
The angle brackets in (\ref{eq:dc:7}) denote an average over time and $\vec{(v)}_\ell$ is the velocity component at degree $\ell$. Another definition relevant for stress-free mechanical boundaries is given by \citet{schrinner12} (see their Appendix). In this definition, the axisymmetric zonal contribution $v_{\phi}^{ax}$ is not taken into account in the calculation of the Rossby number which becomes a convective Rossby number based on the convective velocity $v_c=v_{\mathrm{rms}}-v_{\phi}^{ax}$ and in the calculation of the mean harmonic degree $\ell_c$ of the convective flow $v_c$. $<\vec{v}_c\cdot\vec{v}_c>$ corresponds to the non-zonal (or convective) energy  of the flow. We also define the convective Reynolds number $Re_c=v_c\, L/\nu$ and the zonal Raynolds number $Re_z=v_\phi^{ax}\, L/\nu$.
 

The magnetic field strength is measured by the dimensionless Lorentz number,
\(Lo=B_{\mathrm{rms}}/(\sqrt{\varrho\mu}\Omega L)\), and the classical Elsasser
number \(\Lambda=B_{\mathrm{rms}}^2/{\Omega\varrho\mu\eta}\). They are related 
through \(\Lambda=Lo^2\,Pm/E\). Moreover, following  \citet{christensen06}, we characterize the geometry of the magnetic field by the relative dipole field strength, \(f_\mathrm{dip}\) or dipolarity, 
which is defined as the ratio of the average field strength of the dipole 
field to the field strength in harmonic degrees \(\ell\le 12\) at the 
outer boundary.

A non-dimensional measure for the heat transport is given by the Nusselt 
number, \(Nu\), defined as the ratio of the total heat flow $Q$ and the 
conducted heat flow, \(Q_\mathrm{cond}=4\pi r_o r_i\varrho c \kappa\Delta T/D\) with the heat capacity \(c\). In our models, $Nu=Q/Q_\mathrm{cond}$ is measured at the outer boundary and averaged in time. 
 
Typical resolutions are 288 points in the radial direction (up to 384 points). The spectral decomposition is truncated at a hundred modes (up to $l_{max}=m_{max}=256$), in order to observe a drop by a factor of 100 or more for the kinetic and the magnetic energy spectra  of $l$ and $m$ from the maximum to the energy cut-off $l_{max}$ and $m_{max}$.

\section{Kinematic study}

  In this section, we study the ability of convection motions in rapidly rotating spherical shells to drive a dynamo.  With a weak initial magnetic field, the Lorentz force does not affect the flow. In this case, the flow is called kinematically unstable  when the initial field grows exponentially. This configuration is obtained if the magnetic Reynolds number is higher than a critical value $Rm_c$. Otherwise, exponential decay is observed. We determine the evolution of $Rm_c$ with the control parameters $Ra$ and $E$. In this section, the kinematic phase is numerically considered by explicitly ignoring the Lorentz force in the MHD equations.

\subsection{Convection in rapidly rotating spherical shells}

\begin{figure}
\begin{center}
\begin{tabular}{cc}
\subfigure[]{\includegraphics[width=7.cm]{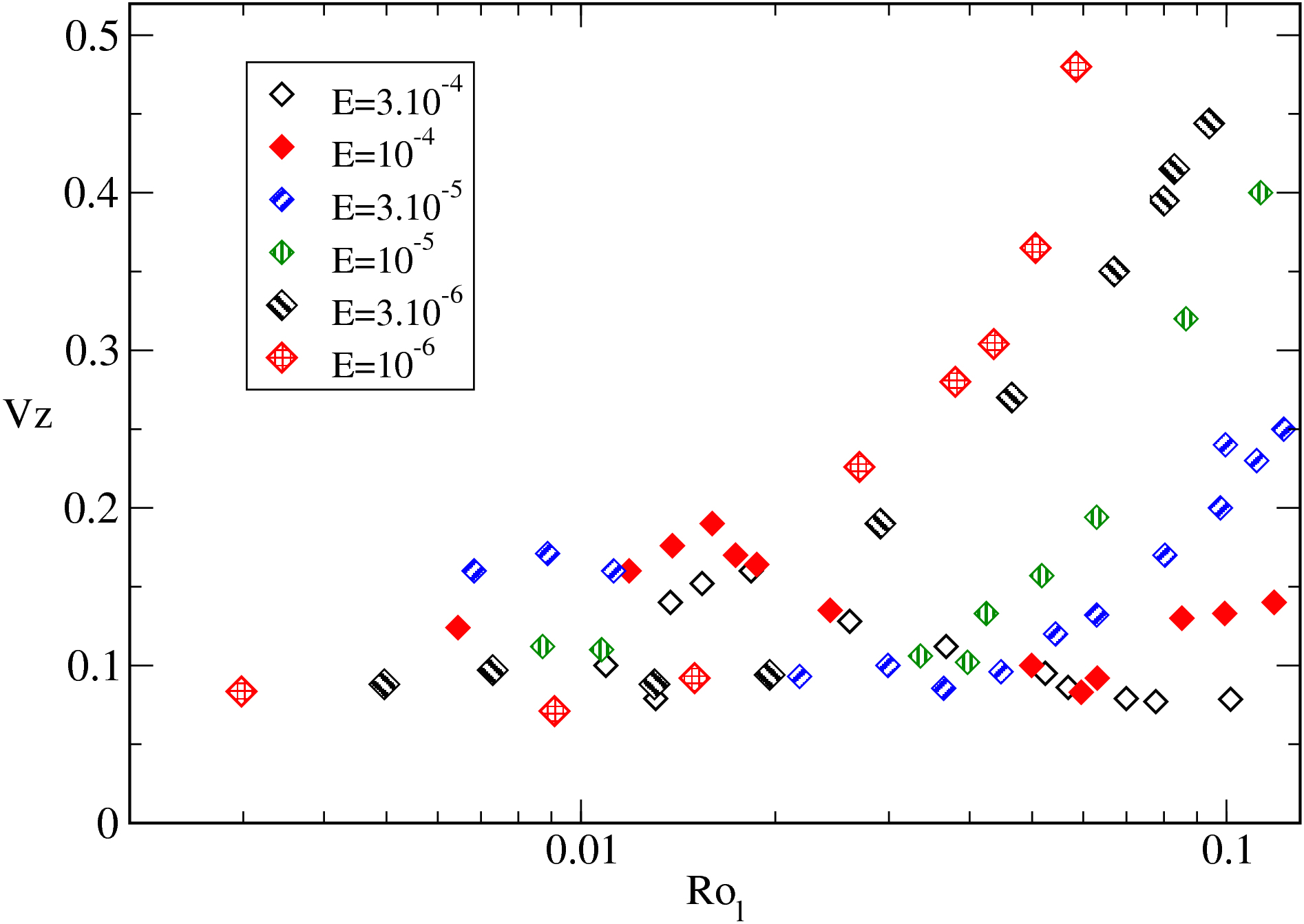}} &
\subfigure[]{\includegraphics[width=7.cm]{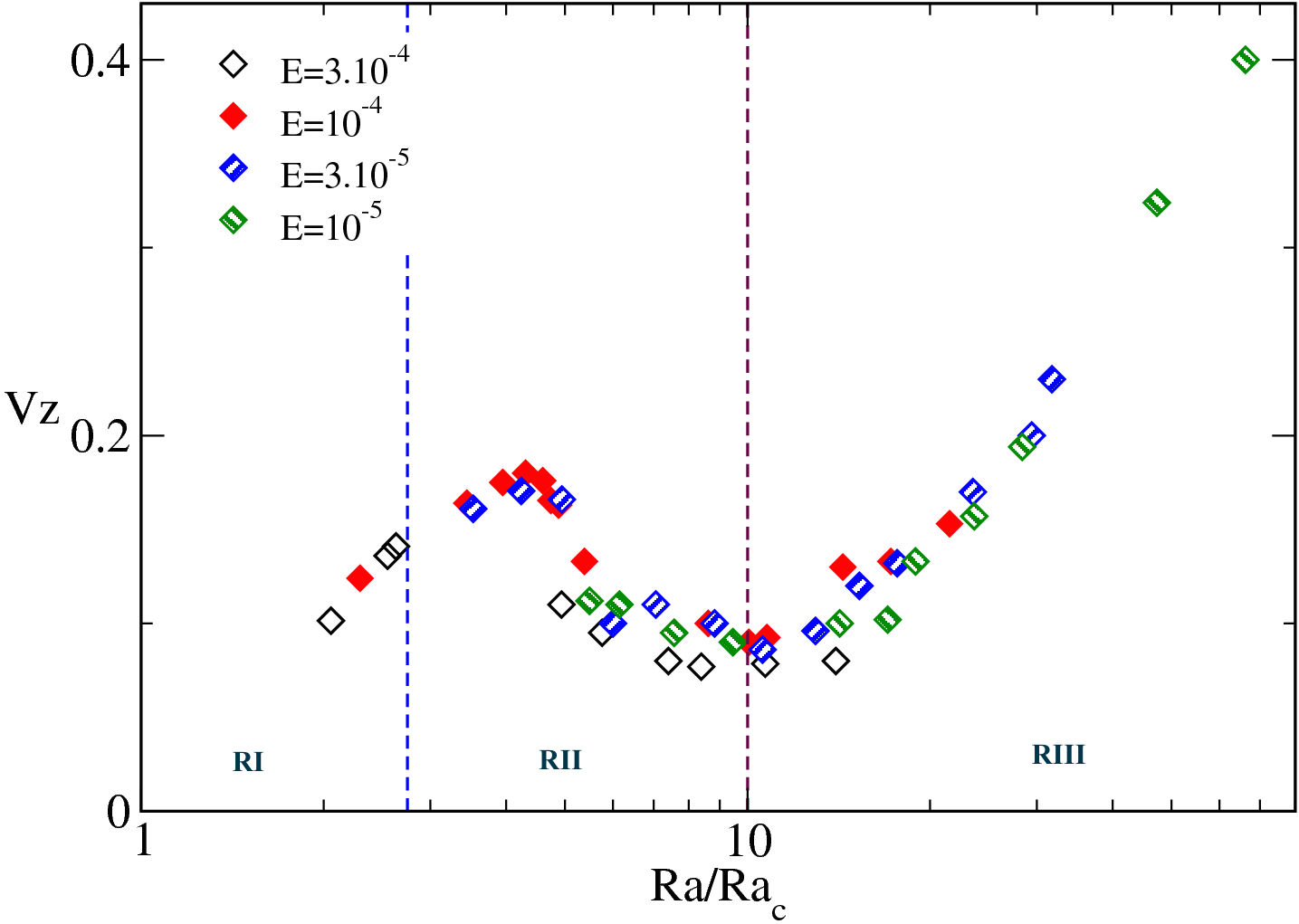}} 
\end{tabular}
\end{center}
\caption{Relative importance of zonal flows measured by $V_Z$ in hydrodynamical simulations as a function of $Ro_\ell$ (panel (a)) and $Ra$ (panel (b)).}
\label{VzNS}
\end{figure}

  We shortly review some important hydrodynamical results and we highlight in particular some properties of flows on which dipolar solutions can set in ($Ro_\ell\le 0.12$) (see Fig~\ref{VzNS}(a)). \citet{gastineWA16} present a recent study of convection in rotating spherical shells. However, the relative amplitude of zonal flows is not explicitly addressed.

  Small-scale convection transfers kinetic energy via nonlinear Reynolds stresses into the mean zonal flow, the magnitude of which is measured by $V_Z=E_\mathrm{ax}^\mathrm{tor}/E_\mathrm{KIN}$ where $E_\mathrm{ax}^\mathrm{tor}$ denotes the axisymmetric toroidal part of the mean kinetic energy $E_\mathrm{KIN}$. To obtain a Reynolds stress effect, a high degree of correlation is required between
the cylindrically radial and azimuthal velocity components of the small-scale eddies. The dynamics of  convection columns is intimately connected for Prandtl
numbers of the order unity or less with the differential rotation that is partly generated by their Reynolds stresses especially when free-slip boundaries are used \citep{grote2000,christensen02,busse06}. The use of rigid boundaries is known to dramatically affect the magnitude of zonal flows in numerical models and in experiments \citep{aubert01,aubert05,gilletJ06}. As the Ekman number decreases, the effect of the boundary layer dissipation decreases and the energy of zonal flows becomes important (see Fig~\ref{VzNS}(b)).  Asymptotic studies have shown that zonal flows could be important in the Earth's outer core.  The turbulent regime is particularly interesting for planetary applications where $E\ll 1$, $Re\gg 1$ and $Ro\ll 1$. In geodynamo simulations, the Reynolds number would never be large enough in order to ignore viscous effects in the
interior of the shell, in contrast with what is expected
in planetary interiors. However, our hydrodynamical results suggest that zonal flows can have a significant part of the kinetic energy in geodynamo simulations. We show in our MHD study that zonal flows affect the nature of the dynamo bifurcation and they allow the existence of a bistable regime.   

 In rotating spherical shells, the flow  is dominated by
columnar vortices aligned with the rotation axis distributed around the solid
inner core close to the onset of convection. Heat transfer occurs preferentially in the equatorial plane in this regime. These rolls exhibit properties of thermal Rossby waves in that they are drifting in the prograde
azimuthal direction. Slightly above the supercriticality (in RI), the Coriolis force and the pressure force dominate and they organize the flow in columns parallel to the rotation axis (Proudman-Taylor constraint). Such a flow structure is shown in Fig~\ref{VstructHD} (left panels).   The helicity $He=<\vec{v}\cdot(\vec{\nabla}\times\vec{v})>$, is predominantly positive in the southern hemisphere and
negative in the north. In this case, the brackets $<\ldots>$ denote an average over time and spatial directions. 

  Close to the onset of convection ($Ra\leq 3Ra_c$: RI), only one convection mode develops or dominates and the magnitude of $V_Z$ is limited (below $0.2$, see Fig~\ref{VzNS}(b)). The convection mode which develops for the lowest Rayleigh number when the other control parameters are fixed, is called the critical mode and this mode has the azimuthal symmetry $m_c$.  This situation is typical for convection in RI. The increase of $Ra$ allows to extend convection cells into the whole volume. 

  For higher supercriticalities ($3Ra_c\la Ra\la 10Ra_c$), $V_Z$ decreases as $Ra$ increases. The flow becomes time dependent and its equatorial symmetry is less and less pronounced. Additional convection modes develop and join the critical one. The fluid inside the tangent cylinder is still nearly stagnant. Otherwise, convection is vigorous and almost space-filling. The size of convective vortices becomes thinner in the azimuthal direction  (see Fig~\ref{VstructHD}, middle panels).  Helicity which has only one sign in each hemisphere in RI, varies along the rotation axis in RII and it changes sign with the emergence of helical motions close to the solid inner core.  

  At sufficiently supercritical Rayleigh numbers ($Ra\geq 10\, Ra_c$: RIII), $V_Z$ increases with $Ra$ (see Fig~\ref{VzNS}(b)), i.e. zonal flows play an increasingly important role as buoyant forcing increases. \citet{aubert01} have also mentioned in their experimental and theoretical study that the turbulent scaling fits their data (the flow speed and the typical length scale) when $Ra>10\, Ra_c$ for both fluids: gallium ($\Pr=0.025$) and water ($\Pr=7$) (see their figures 10 and 11).  Convection develops in RIII in the polar regions as well. Fluid motions in these regions interact with the turbulent convection outside the tangent cylinder. In this turbulent regime (RIII), convection is organized as a set of thin plume sheets rather than columnar cells. The pattern still drifts, but this is no longer the consequence
of wave propagation, but of a real zonal circulation that can be strong when compared to convective velocity (see \citet{aubert01} and references therein).  Reynolds stresses (inertial effects) and thermal wind forcing generate large-scale zonal flows even if the Coriolis force is still dominant ($Ro_\ell<0.12$).

\begin{figure}
\begin{center}
\begin{tabular}{c}
\includegraphics[width=7.cm]{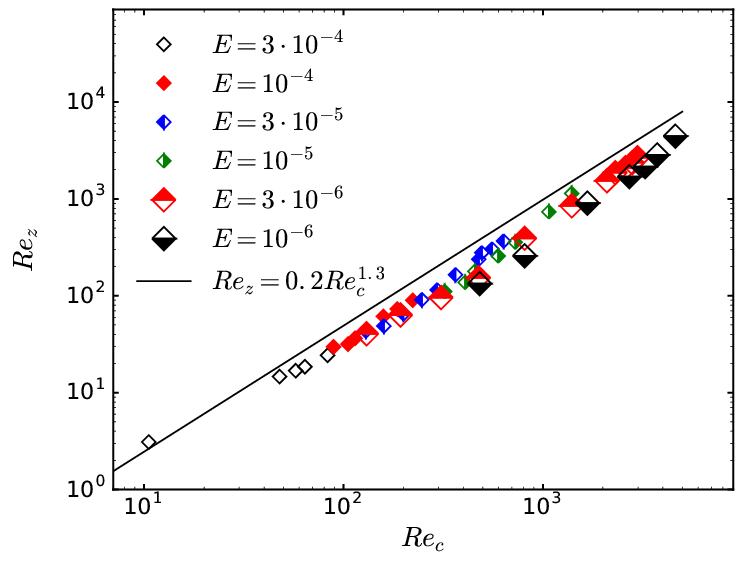} 
\end{tabular}
\end{center}
\caption{Evolution of the zonal Reynolds number $Re_z$ as a function of the convective Reynolds number $Re_c$ in hydrodynamic direct numerical simulations with $Ra>10Ra_c$, $Pr=1$ and different Ekman numbers. The line gies the $1.3$ scaling which appears to fit the low $Pr$ data according to  \citet{gilletJ06}. We note that this scaling law also describes our dataset with $Pr=1$ in the turbulent regime.   }
\label{RecRez}
\end{figure}

\begin{figure}
\begin{tabular}{ccc}
\includegraphics[width=4.5cm]{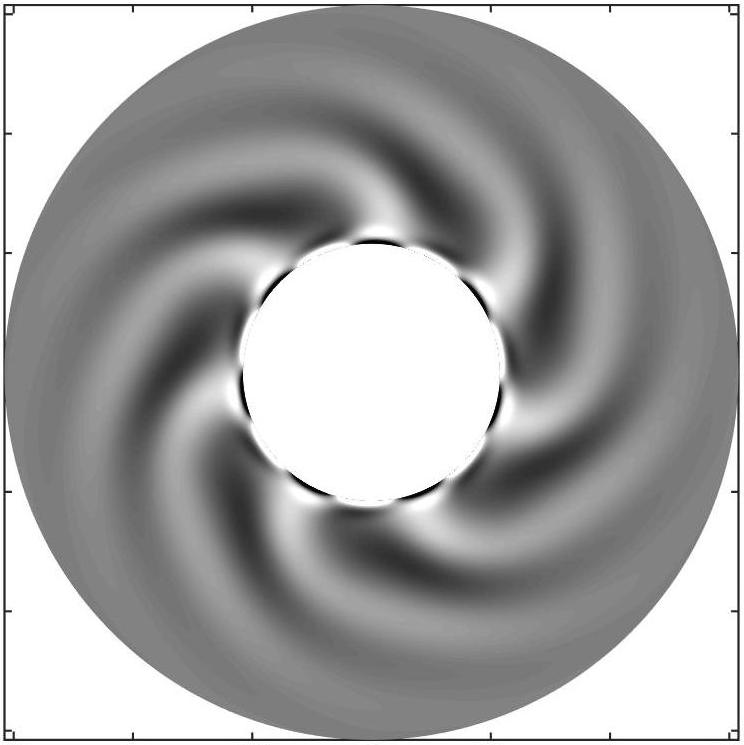} & \includegraphics[width=4.5cm]{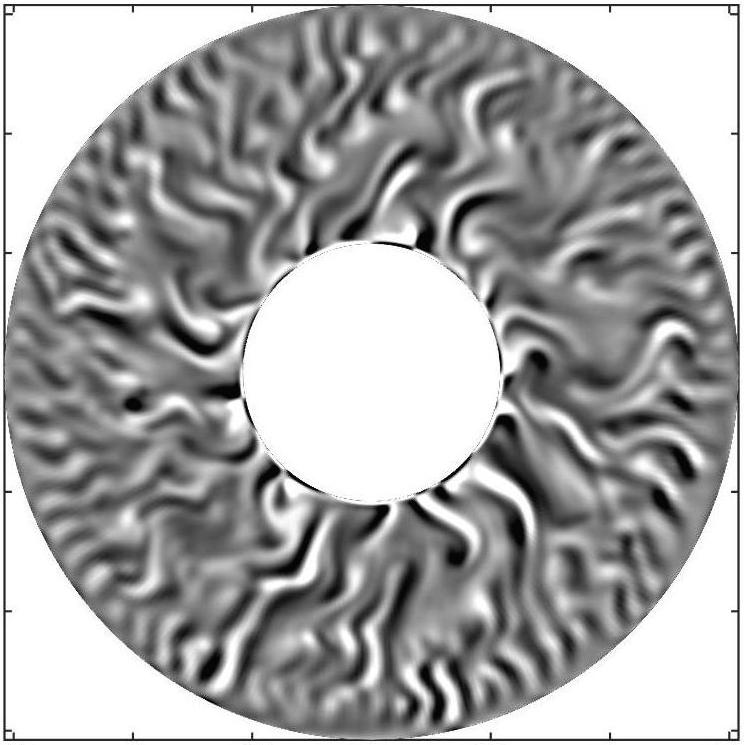} & \includegraphics[width=4.5cm]{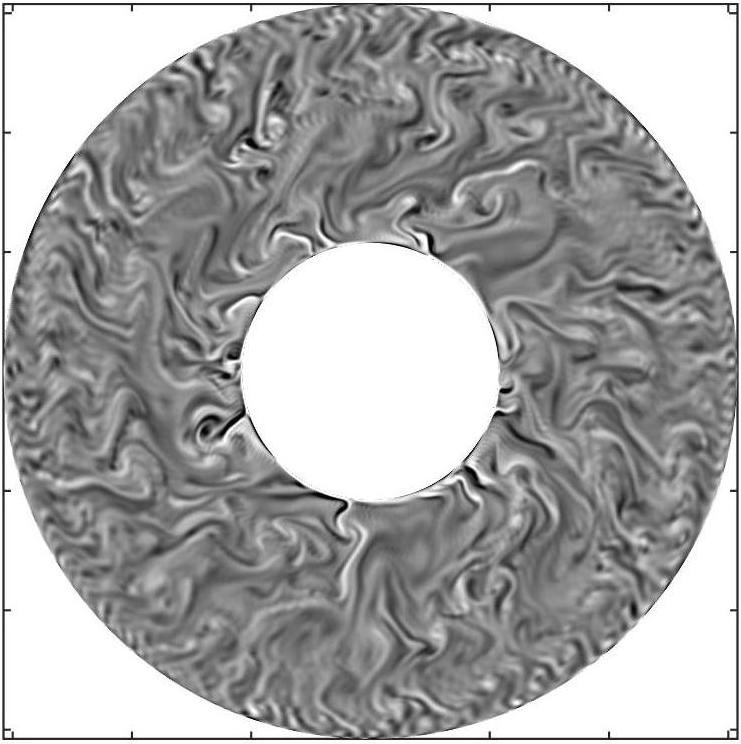} \\
\includegraphics[width=3.5cm]{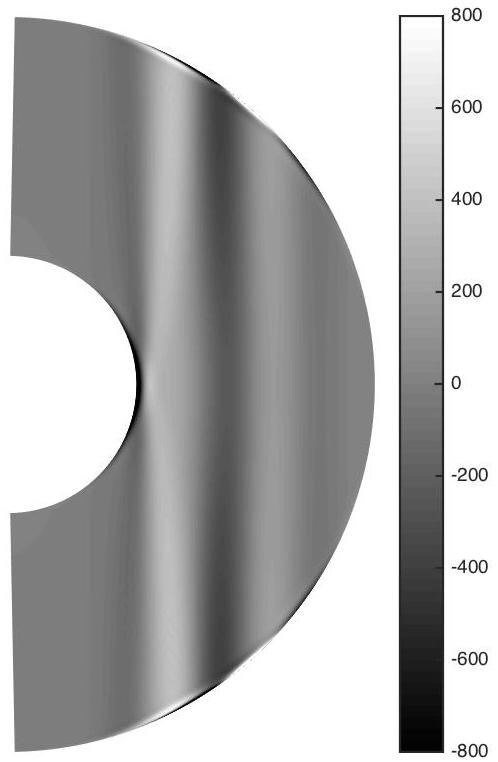} & \includegraphics[width=3.5cm]{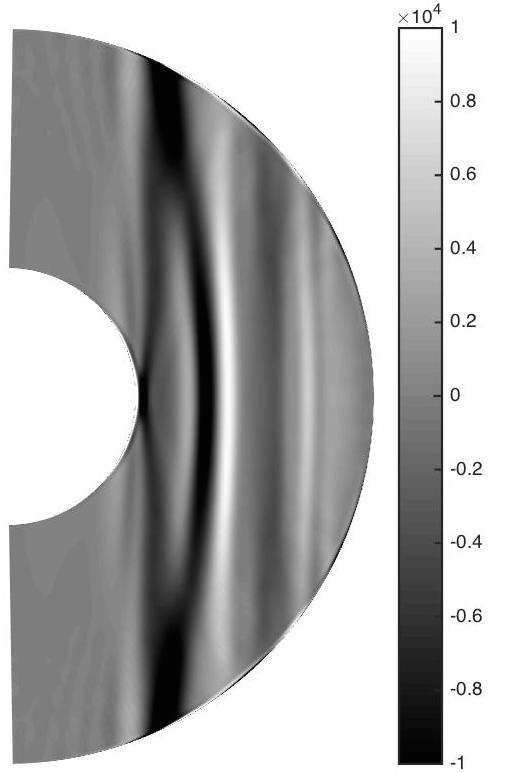} & \includegraphics[width=3.5cm]{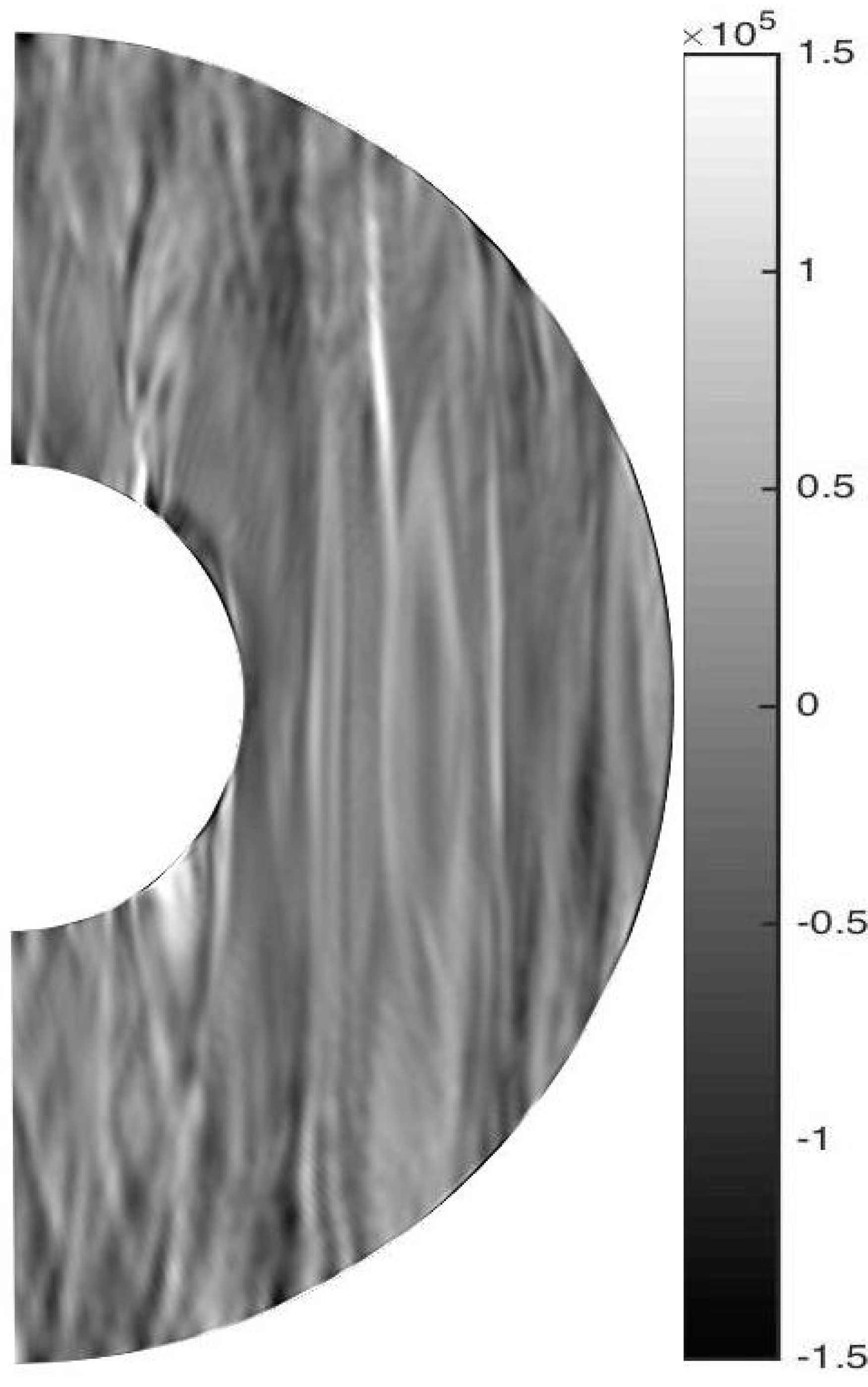} 
\end{tabular}
\caption{Axial vorticity $\omega_z=\vec{\nabla}\times\vec{v}\cdot\vec{z}$ in the equatorial plane (top panels) and azimuthal sections of $\omega_z$ (bottom panels) for hydrodynamical simulations in RI (left panels) with $E=10^{-4}$ and $Ra=160=2.297 Ra_c$ ; in RII (middle panels) with $E=3\cdot 10^{-5}$ and $Ra=900=10.47 Ra_c$ and in RIII (right panels) with $E=10^{-5}$ and $Ra=5000=47.3 Ra_c$. Such flow structures are typical for convection motions in each regime (RI, RII, RIII).}
\label{VstructHD}
\end{figure}

  Fig~\ref{VzNS}(a) shows that lowering the Ekman number promotes the development of zonal flows for lower $Ro_\ell$. The typical flow structure for convection motions in RIII is given in Fig~\ref{VstructHD} (right panels).  A prograde jet close to the equator and a retrograde jet are observed on the outer boundary \citep{christensen99,aubert05}. Such a mean zonal flow is typical for convection motions in rapidly rotating spherical shells with rigid (in RIII) or free-slip boundaries and a large aspect ratio $\chi$. Inertia perturbs the QG structure of the flow. We confirm with our dataset the scaling law $<v_\phi>\propto <v_{nz}>^{1.3}$ found by \citet{gilletJ06} in RIII (sse Fig~\ref{RecRez}).

\subsection{Kinematic dynamo action }

\begin{figure}
\begin{center}
\begin{tabular}{c}
\includegraphics[width=7.cm]{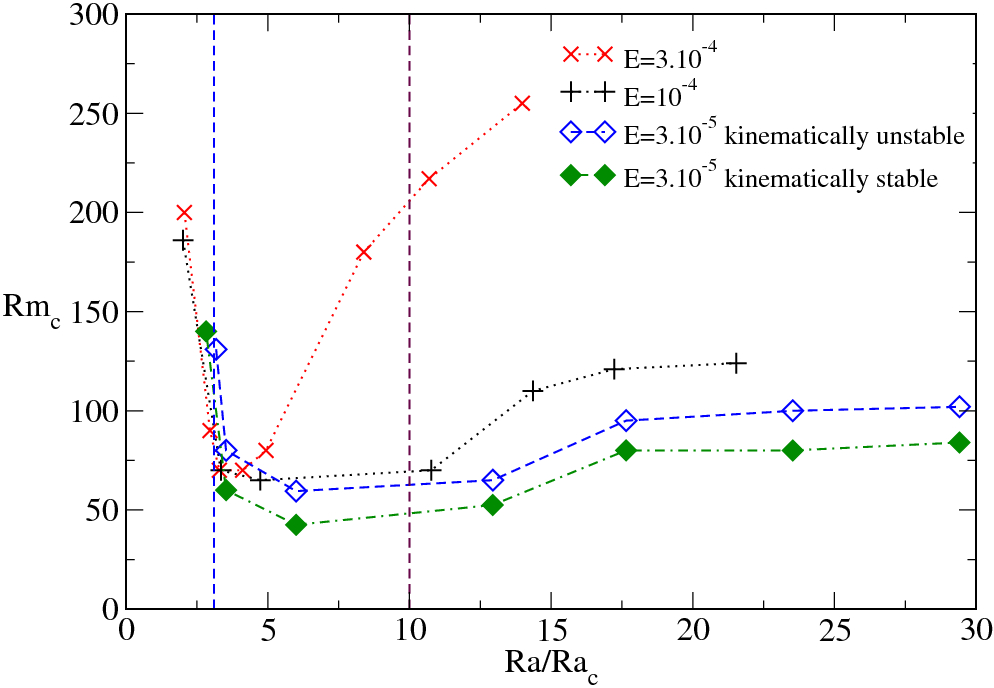} 
\end{tabular}
\end{center}
\caption{Kinematic dynamo threshold $Rm_c$ as a function of the supercriticality $Ra/Ra_c$ indicated by cross symbols ($E\geq 10^{-4}$). For $E=3\times 10^{-5}$ (diamond symbols), a weak seed field is amplified for empty symbols whereas it decreases for full symbols.  }
\label{RaRmcKIN}
\end{figure}

  At sufficiently low $Ra$ ($Ra\le 10Ra_c$), the flow remains  perfectly equatorially symmetric. Equatorially symmetric and anti-symmetric magnetic modes then decouple. In this case, we observe that the most unstable mode is anti-symmetric and it corresponds to the axial dipolar component. Close to the kinematic dynamo threshold $Rm_c$, the growth rate of the magnetic energy $\sigma$ approaches zero and evolves linearly with the distance $Rm-Rm_c$. Since $Rm=Re\, \Pm $, $Rm$ can be varied for a given flow (i.e. for a particular value of $Re$ determined by $Ra$) by changing $\Pm$. For a given buoyant forcing $Ra$ and Ekman number $E$, a series of kinematic simulations was performed by varying $Rm$ (by changing $\Pm$) and the exponential growth (or decay) rate was measured.  The threshold $Rm_c$ was determined by linearly interpolating the values of $Rm$ between the slowest growing dynamo and the slowest decaying dynamo. The series of runs was then repeated for different values of $Ra$ and $E=3\cdot 10^{-4}$ and $E=10^{-4}$ with $\Pr=1$. This method has been also recently used in cartesian models by \citet{sadekAF16}.  Since this procedure is numerically demanding, we have limited this kinematic study to $E> 10^{-5}$ (see Fig.~\ref{RaRmcKIN}). 

  In RI and part of RII, the critical magnetic Reynolds number $Rm_c$ decreases when increasing $Ra$.  It means that increasing $Ra$ promotes dynamo action sufficiently close to the onset of convection. Because of the definitions of $Ra$ and $E$, the influence of the global rotation results from the values of these numbers.  As the rotation rate increases (by decreasing $Ra$ or $E$), it suppresses vertical variations and the flow becomes almost QG. Two-dimensional flows can not give rise to dynamo action. However, a QG flow with three components can trigger dynamo action (see Roberts \& King 2013 for a recent review on geodynamo theory). Increasing $Ra$ in RI allows to obtain a more complex flow which promotes dynamo action. Slightly above $Rm_c$, the growing magnetic energy is mainly divided into two azimuthal components. One of them is the axisymmetric axial dipolar mode. The second mode has the same azimuthal symmetry as the dominant convection mode.

  In RI, convection cells extend only partly to the outer boundary in the equatorial plane. The optimal configuration for having dynamo action ($Rm_c$ reaches its minimum) corresponds to a completely developed and space-filling convection with almost no zonal flows (RII). In this regime, the most unstable mode is equatorially symmetric and it is dominated by the axial dipolar component. For $E=3\cdot 10^{-4}$, $Rm_c$ increases rapidly with $Ra$ in RII whereas $Rm_c$ is almost constant with $E\leq 10^{-4}$.

  In RIII, $Rm_c$ increases with $Ra$ (see Fig~\ref{RaRmcKIN}).  In the kinematic phase, the growing magnetic field has no preferential symmetry since the velocity field breaks the equatorial symmetry. Magnetic modes with a low axial dipole contribution and sometimes localized mainly in one hemisphere grow in time if $Rm$ exceeds $Rm_c$. Large kinetic fluctuations affect the growth rates which must be averaged over long periods of time (several magnetic diffusion times) in order to have precise values. Determining $Rm_c$ in the turbulent regime requires considerable numerical resources.  In models with high $Ra$, kinetic fluctuations and zonal flows have certainly an important role in field generation. This point is addressed in more details in the next sections by highlighting the saturation mechanism of non-dipolar dynamos. Additional equatorially anti-symmetric flow contributions set in  and couple magnetic modes of both symmetries. Since $Rm_c$ increases with $Ra$, we can note that the emergence of small-scale kinetic fluctuations and large-scale zonal flows has the effect of reducing dynamo action even though the effects of global rotation are still dominant ($Ro_\ell$ smaller than $0.1$).

\section{A systematic study of dynamo bifurcations in geodynamo models}

\subsection{Previous studies and method }

  In this section, numerical results for the full MHD equations including the back-reaction of the Lorentz force are presented. The models are integrated in time until a statistically steady state is reached. Two initial conditions for the magnetic fields are considered: a strong dipolar field (with $\Lambda\approx 10$) or a weak seed field where all spherical harmonics lower than 20 (for the degree $l$ and the number $m$) are initialized with random amplitudes corresponding to the Elsasser number of $\Lambda\approx 10^{-2}$. The initial velocity field and the temperature perturbation correspond to the solutions described in the previous section.  Some calculations were started from a saturated state of another model with slightly different parameters to test for hysteresis.

    According to our hydrodynamical study, the turbulent regime which is relevant for planetary interiors is poorly explored with the Ekman numbers in the range $10^{-3}\ge E\ge 10^{-4}$ which \citet{morinD09} utilised.  Systematic parameter studies have also been done with the same numerical setup \citep{christensen99,kutzner02,christensen06,king10,schrinner12,soderlundKA12,yadav16}. Given that viscosity plays an important role in simulations with $E\geq 10^{-3}$, we limit the range to lower values. In addition, those particular behaviors are observed in simulations when $E\geq 10^{-3}$: the bifurcation can take the form of an isola \citep{morinD09}, the magnetic field can be dominated by an equatorial dipole component \citep{aubert04} or a subcritical dynamo can be obtained close to the onset of convection \citep{christensen01}. We do not observe such behaviors for smaller Ekman numbers.

  In our dataset, the buoyant forcing $Ra$ is varied significantly up by a factor of 14 from $Ra=500$ to $Ra=7000$ for the lowest value of $E$. $\Pm$ is also varied by a factor of 24 from $\Pm=0.5$ to $\Pm=12$ for $E=10^{-4}$ and almost by a factor 67 from $\Pm=0.075$ to $\Pm=5$ for $E= 10^{-5}$. By considering such a huge parameter space, we  obtain a large amount of dynamo bifurcations in each hydrodynamical regime. Previous studies either focused on one particular hydrodynamical regime or obtained only a small amount of bifurcations with $E<10^{-3}$ \citep{morinD09}. Due to these limitations, previous studies on the nature of dynamo bifurcations with our setup were never able to precisely highlight the influence of the dimensionless parameters $\Pm$, $Ra$ and $E$. Such influences are crucial in order to understand the expected behavior with realistic parameters. The Earth's Ekman number is estimated to be $E=10^{-15}$, some eight orders of magnitude lower than what is currently achievable in numerical simulations, i.e. $E= 10^{-6}$ \citep{Yadav16Earth} or $E=10^{-7}$ \citep{schaefferJNF17}. In our MHD study, the range $10^{-5}\le E\le 3\cdot 10^{-4}$ is considered in order to provide a large number of models and vary significantly the other dimensionless numbers. This method enables us to quantitatively understand the influence of the different physical processes responsible for magnetic field generation. 

 For $E=3\cdot 10^{-4}$, \citet{morinD09} observed a supercritical dynamo bifurcation with $\Pm=6$ and a subcritical one  by lowering $\Pm$ to $3$. Decreasing the Ekman number to $E=10^{-4}$ allowed to have a supercritical bifurcation for $\Pm=3$. The behavior observed with $\Pm=6$ (supercritical bifurcation) is shifted towards lower values of $\Pm$ as $E$ is decreased.  \citet{morinD09} argued that this behavior could be extrapolated to realistic parameters because both $E$ and $\Pm$ are very low in planetary interiors. They chose the Rayleigh number $Ra$ as their control parameter, even though changing this number affects significantly the flow because it directly controls the buoyant driving.  To highlight the nature of dynamo bifurcations, we show the evolution of $\Lambda$ in $(\Pm,Ra/Ra_c)$ plane for $3\cdot 10^{-4}\ge E\ge 10^{-5}$.

  Fig.~\ref{RaPmbif} shows our MHD dataset in $(Ra/Ra_c,\Pm)$ planes for different Ekman numbers. Dipolar solutions below or on the red dotted curve are only obtained with a strong initial field. For $E=3\cdot 10^{-4}$ and $E=10^{-4}$, these curves correspond to the kinematic dynamo threshold in Fig~\ref{RaRmcKIN}. We show that the nature of dynamo bifurcations depends on the hydrodynamical regime. These regimes are separated in Fig~\ref{RaPmbif} by vertical dotted lines.

\subsection{Dynamo bifurcation in RI} 
  
\begin{figure*}
\begin{tabular}{cc}
\subfigure[$E=3\cdot 10^{-4}$]{\includegraphics[width=7.cm]{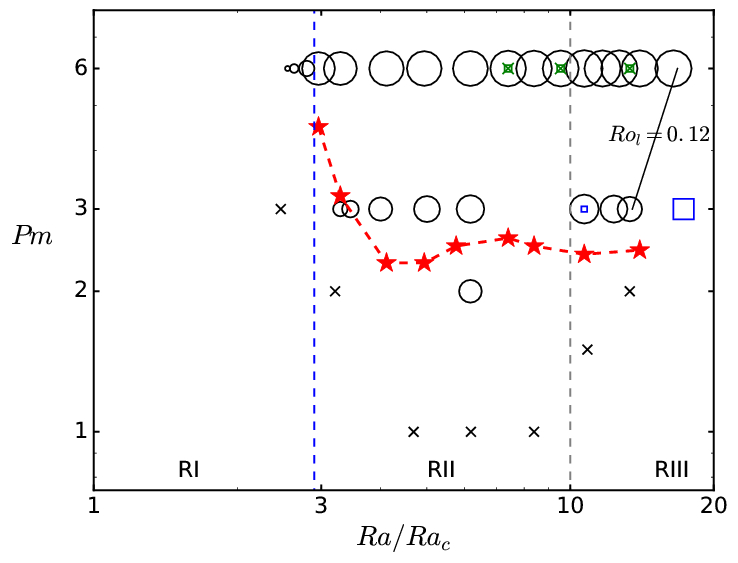}} & 
\subfigure[$E=10^{-4}$]{ \includegraphics[width=7.cm]{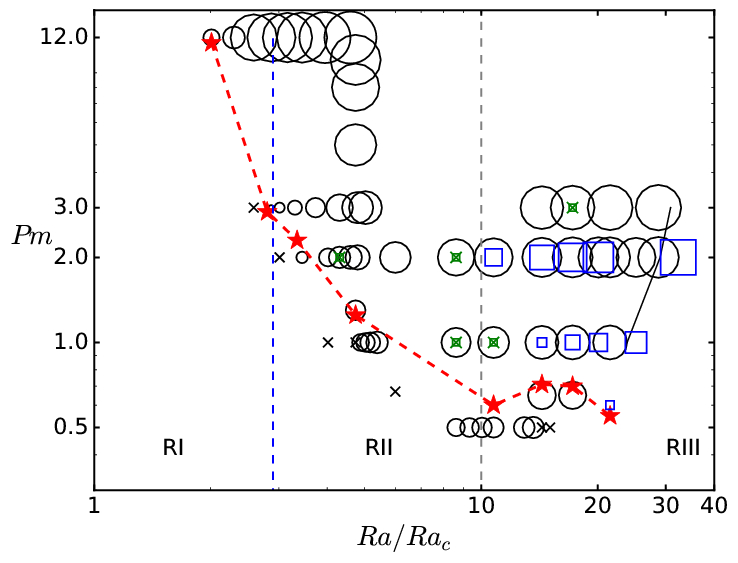}} \\
\subfigure[$E=3\cdot 10^{-5}$]{\includegraphics[width=7.cm]{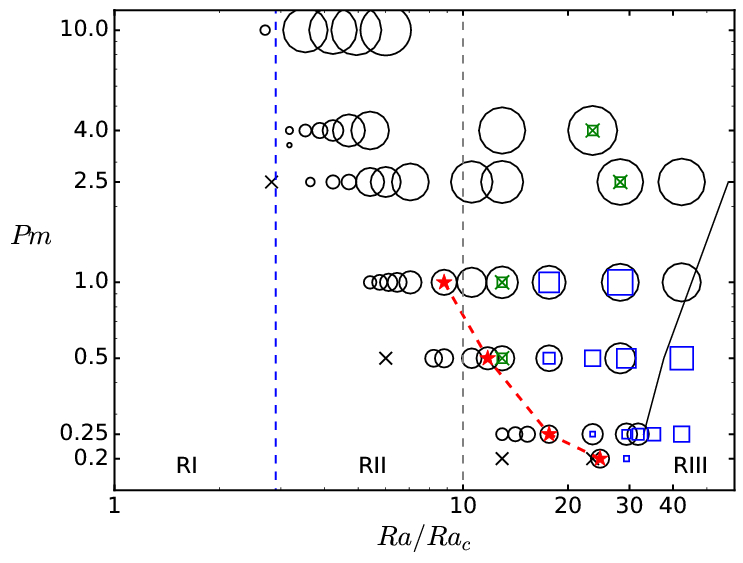}} & 
\subfigure[$E=10^{-5}$]{ \includegraphics[width=7.cm]{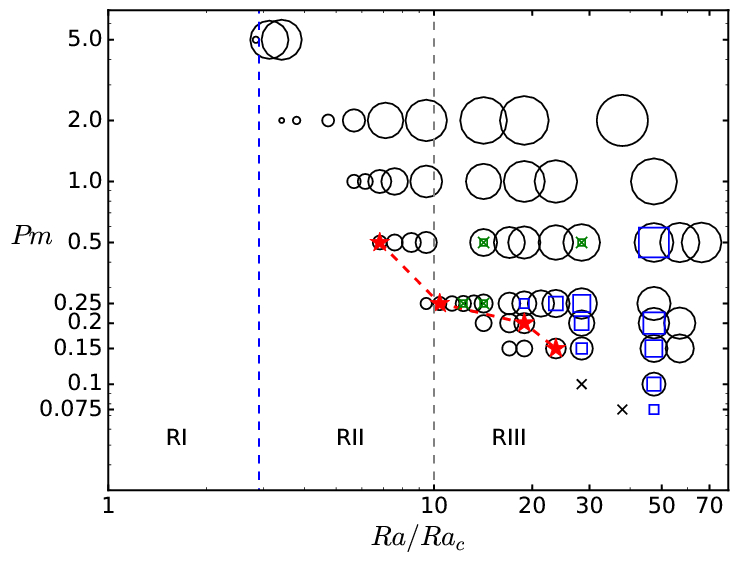}}
\end{tabular}
\caption{Regime diagrams for dynamos with rigid boundaries driven by an imposed temperature contrast at different values of the Ekman number. Circles show dipolar dynamos, squares non-dipolar dynamos and crosses failed dynamos. The crossed out squares (in green) indicate that a dipolar solution is obtained regardless of the magnitude of the initial field. For circles and squares, the size of the symbol has been chosen according to the value of the Elsasser number ($\propto log(\Lambda)$). Dynamo solutions below the red dotted curves can be only obtained by considering a strong initial field.  For $E=3\cdot 10^{-4}$ and $E=10^{-4}$, these red dotted curves correspond to the kinematic dynamo threshold in Fig~\ref{RaRmcKIN}.}
\label{RaPmbif}
\end{figure*}

  Given that $Re$ is low in RI, high values of $\Pm$ have to be considered so that $Rm=Re\, \Pm$ exceeds $Rm_c$. In addition, as seen in the kinematic study, $Rm_c$ is high in RI. We explore dynamo bifurcations in RI for $E=3\cdot 10^{-4}$, $\Pm=6$ ; $E=10^{-4}$, $\Pm=12$ ; $E=3\cdot 10^{-5}$, $\Pm=10$ and $E=10^{-5}$ and $\Pm=5$ (see Fig~\ref{RaPmbif}).

\begin{figure}
\begin{center}
\begin{tabular}{cc}
\multicolumn{2}{c}{\subfigure[Dynamo bifurcation diagram with $E=10^{-4}$ and $\Pm=12$]{\includegraphics[width=9.cm]{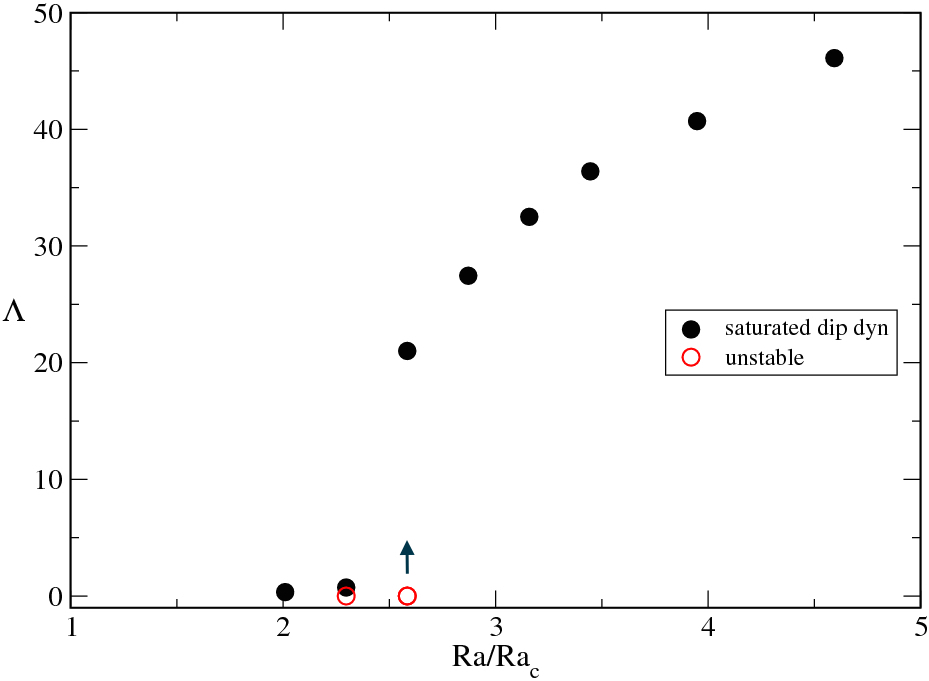}}} \\
\subfigure[]{\includegraphics[width=7.cm]{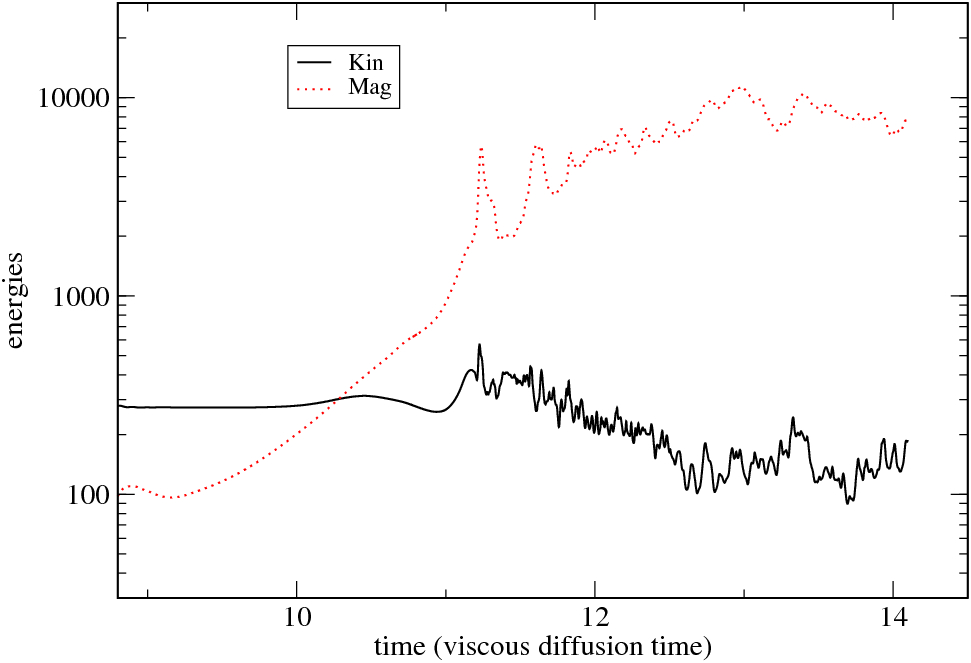}} &
\subfigure[]{\includegraphics[width=7.cm]{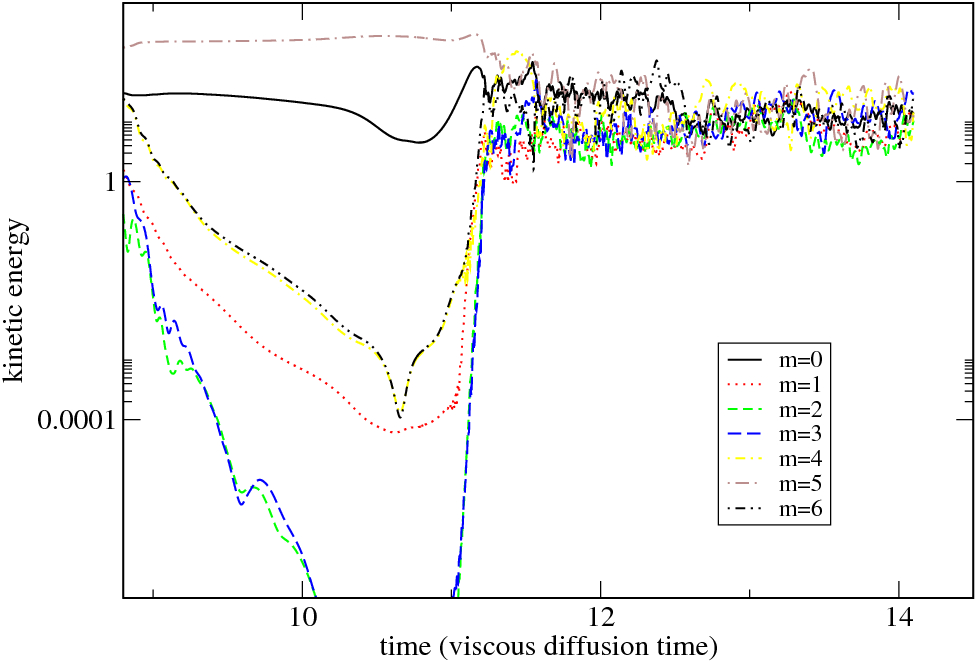}}
\end{tabular} 
     \caption{$\Lambda$ as a function of $Ra$ with the parameters $E=10^{-4}$ and $\Pm=12$ (top panel). $\Pm$ is sufficiently high to explore RI close to the dynamo onset.  In panels (b) and (c), temporal evolution of the kinetic and magnetic energies are shown for a simulation with $Ra=180=2.584\times Ra_c$ which has been initialized with the saturated solution with $Ra=160$. These panels illustrate the generation of additional azimuthal convection modes in dynamo models whereas the azimuthal symmetry of the non-magnetic velocity field is unchanged.}
     \label{biflamBact}
\end{center}
  \end{figure}

  Fig~\ref{biflamBact}(a) shows a typical dynamo bifurcation diagram for RI. The parameters are $E= 10^{-4}$, $\Pm=12$.  Clearly, a supercritical dipolar branch is obtained. \citet{morinD09} and \citet{dormy16} presented very similar diagrams with $E=3\cdot 10^{-4}$ and $\Pm\ge 6$.

  The magnitude of the field strength is very low in the vicinity of the dynamo threshold for models in RI. As a result, the flow is almost unchanged by Lorentz forces. In particular, the structure of the flow is still dominated by only one convection mode and the dynamo solution can be classified as a laminar dipolar dynamo. In Fig~\ref{biflamBact}(a), laminar dynamos have $\Lambda<1$ and  $Ra\le 160$ ($Ra\le 2.3Ra_c$).

  We observe an increase of the magnetic energy (measured by the Elsasser number $\Lambda$) by more than one order of magnitude when $Ra$ is increased from $Ra=160$ ($2.3Ra_c$) to $Ra=180$ ($2.6 Ra_c$) for $E=10^{-4}$ and $\Pm=12$ (see Fig.~\ref{biflamBact}(a)). A similar sharp increase is obtained for $E=3\cdot 10^{-4}$ and $\Pm\ge 6$. Such a jump is only observed at low $Ra$ (in RI). Then, a further increase of $Ra$ induces a moderate variation of $\Lambda$. Such a sharp and pronounced variation for the magnetic energy has recently been reported by \citet{dormy16} as the manifestation of a new strong-field dynamo branch in geodynamo simulations with high $\Pm$. This strong-field branch is believed to be relevant in the geodynamo context \citep{roberts78}. We only observe such sharp variations of $\Lambda$ in the laminar regime.

  In our kinematic study, we highlighted the inverse relationship between $Rm_c$ and $Ra$ in RI.  $Rm=Re\, \Pm$ increases with the vigour of convection. But, when keeping $\Pm$ fixed and increasing $Ra$, not only the flow amplitude, but also the efficiency of the dynamo increases as indicated by the decreasing critical magnetic Prandtl number in our kinematic simulations. The dynamo supercriticality $Rm-Rm_c$ therefore rises more significantly than the increase in flow amplitude due to the larger $Ra$ would imply. This suggests a somewhat faster increase of magnetic energy with $Ra$. However, the very drastic rise at a certain Rayleigh number indicates a bifurcation to another magnetic mode. The action of the Lorentz force must be taken into account in order to understand the abrupt variations of $\Lambda$.

  A sharp increase of $\Lambda$ with $Ra$ is typical of dynamo bifurcations in the laminar regime (see Fig~\ref{biflamBact}) and the field strength increases by one order of magnitude. Such variations can be regarded as a transition between laminar dynamos of weak dipolar fields and dynamos with stronger dipolar fields which must affect significantly the flow. However, their hydrodynamic counterparts are still in RI. In particular, without the action of magnetic field through Lorentz forces, the flow is still dominated by only one convection mode when $Ra$ is increased from $Ra=160$ to $Ra=180$ with $E=10^{-4}$. When the dynamo obtained with $Ra=160$ was utilized for the initial conditions of the model with $Ra=180$, a more turbulent dipolar dynamo sets in with a significant increase of the magnetic energy and the development of many convection modes (see Fig~\ref{biflamBact} panels (b) and (c)). The same process occurs when a transition takes place from a weak-field dynamo to a strong-field dynamo in \citet{dormy16}. In Fig~\ref{biflamBact}, for $Ra=140$ and $Ra=160$ ($Ra\le 2.3Ra_c$), weak dipolar fields are obtained ($\Lambda<1$) and only one convection mode is observed (laminar dynamos). According to the evolution of $\Lambda$ with $Ra$, a laminar dynamo with $\Lambda$ slightly greater than unity for $Ra=180$ should be obtained. This scenario seems to take place at the beginning of the simulation when $Ra$ is increased to $Ra=180$. However, when $\Lambda$ approaches unity, the Lorentz force enters in the main force balance. Kinetic and magnetic perturbations with a variety of azimuthal symmetry develop and the global manifestation of this transition is a marked increase of $\Lambda$, visible in Fig.~\ref{biflamBact}(a). We report here in Fig~\ref{RaPmbif} abrupt variations of $\Lambda$ for 4 different values of $E$. \citet{morinD09} and \citet{dormy16} observed similar variations only with $E=3\cdot 10^{-4}$. Thus, we have noticed that all these abrupt variations occur in the same way (regardless the value of $E$). Variations result from a transition between laminar dynamos with $\Lambda<1$ and strong-field dynamos. 

  This result is in agreement with magnetoconvection studies in which the development of convection is affected by strong fields. In the presence of strong imposed magnetic fields and rotation, the first order force balance is magnetostrophic, i.e. a balance between the Coriolis, pressure gradient, and Lorentz terms. Studies of linear magnetoconvection have shown that the azimuthal wavenumber of convection decreases to $m=\mathcal{O}(1)$ when a strong magnetic field ($\Lambda>1$) is imposed in the limit $E\rightarrow 0$ (\cite{bookchandra}; \cite{eltayeb70}; \cite{fearn83}; \cite{cardin95}) whereas for the non-magnetic case, linear asymptotic analyses predict that the azimuthal wavenumber of axial columns varies as $m=\mathcal{O}(E^{-1/3})$ \citep{roberts68,jones2000,dormy04}. When the field strength becomes important ($\Lambda\sim 1$), the spatial structure of convection modes are affected by the Lorentz force. As predicted by magnetoconvection, we observe that the flow structure depends on the magnitude of the magnetic field $\Lambda$.

  \subsection{Dynamo bifurcation in RII}

  In this section, we highlight the nature of the dynamo bifurcation in RII $(3Ra_c\la Ra\la 10Ra_c)$.   Considering lower values of $\Pm$ allows to study the nature of the dynamo bifurcation in RII. Regardless of the magnitude of the initial magnetic field, the axial dipolar component is finally the dominant magnetic component, i.e. a dipole-dominated dynamo is obtained. Crossed out squares (in green) indicate different magnitude for the initial field has been tested.   Contrary to laminar dynamos, the saturated field acts on the velocity field by reducing the kinetic energy and the amplitude of velocity fluctuations close to the dynamo threshold (see the figure 8 in \citet{morinD09}). The saturated field differs from the kinematically growing mode mainly by the presence of extended toroidal magnetic fields close to the equatorial plane.

  In Fig.~\ref{RaPmbif}, supercritical bifurcations are observed with the parameters $E=10^{-4}$ and $\Pm=3$ (panel (b)), $E=3\cdot 10^{-5}$ and $\Pm\ge 2.5$ (panel (c)) and $E=10^{-5}$ and $\Pm\ge 1$ (panel (d)). For these models, the field strength increases gradually with the control parameters $Ra$ or $\Pm$ and saturated dipolar dynamos only exist if $Rm$ exceeds the kinematic critical Reynolds number $Rm_c$. For lower $\Pm$, strong initial dipolar fields are maintained over time with $Rm<Rm_c$ (below the red dotted curves). Otherwise, no dynamo action can be generated with weak initial fields. For these parameters, a turning point $Rm_t$ exists below which strong initial fields cannot be maintained. Such a behavior corresponds to subcritical bifurcations. 

  When $Ra$ is slightly higher than $3Ra_c$ (close to RI) and $Rm$ slightly higher than $Rm_c$, a process called ``mode selection'' is observed. This process explains the existence of laminar dynamos in which only one convection mode is dominant in RII. In the kinematic phase, the growing magnetic energy is primarily distributed on the axisymmetric component and on another azimuthal mode with a symmetry close to that of the critical convection mode $m_c$. At this time, the velocity field is not affected by this weak growing magnetic mode and the kinetic energy is distributed on several azimuthal modes as it is typical in RII. In the saturated phase, the Lorentz force promotes the kinetic azimuthal mode with the same symmetry as the initially growing magnetic mode and it has a negative influence on the other kinetic convection modes. Consequently, a laminar dynamo is finally the stable solution if $\Lambda$ is smaller than 1 as in RI. Such a mode selection process was also reported by \citet{morinD09}. Dynamos with stronger fields ($\Lambda>1$) are generated by increasing $Ra$ or $\Pm$ in which the dominance of only one convection mode is less pronounced.

  In RII, the field strength $\Lambda$ depends on $\Pm$ and $Ra$. The smaller $E$, the faster $\Lambda$ increases with $Ra$ and $\Pm$ (see also Appendix A).

  Abrupt variations of $\Lambda$ has been obtained in RI. Such variations do not appear in RII. Weak and strong dynamo branches observed in RI seem to collapse in RII and give rise to the usual dipolar branch. In RII, $\Lambda$ (size of the circles in Fig~\ref{RaPmbif}) increases with $\Pm$ and can reach values higher than $40$. In this case, the Lorentz force affects the flow structure (see section 5 and 6). Even if only one dipolar branch exists in RII and RIII, the relative importance of the Lorentz force depends strongly on $\Pm$. In simulations with high-enough $\Pm$ values, this force is one of the dominant forces.

  \subsection{Dynamo bifurcation in RIII}

\begin{figure}
 \begin{tabular}{c}
 \subfigure[Visualisation of the bistable regime]{\includegraphics[width=9.cm]{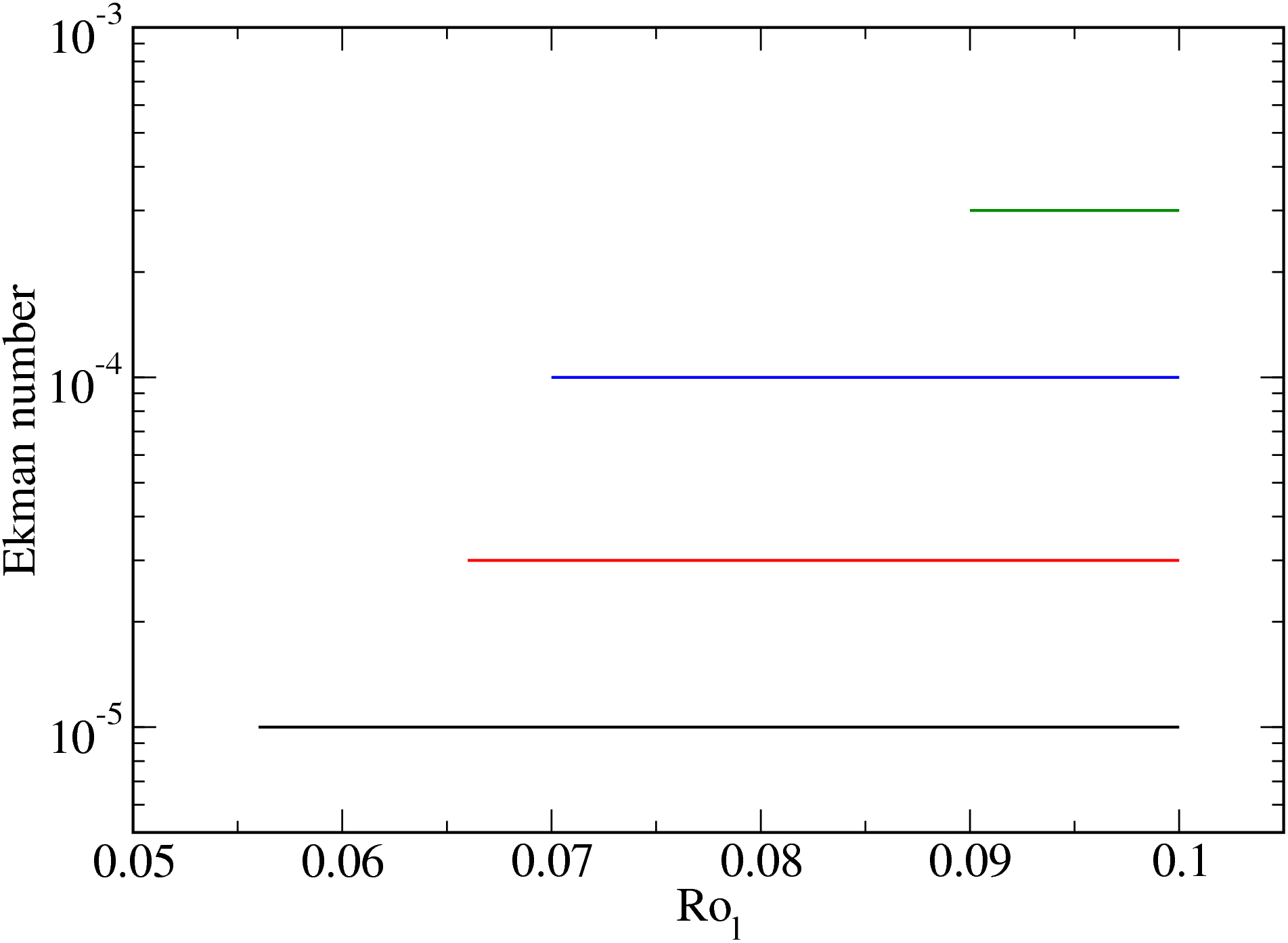}}
 \end{tabular} 
      \caption{Evolution of the bistable regime in a plane ($Ro_\ell,E$). By considering weak initial fields, multipolar solutions can be obtained into the stability domain of dipolar fields which gives rise to a bistable regime. The solution depends on the magnitude of the initial field and on $\Pm$.  }
      \label{bistability}
   \end{figure}

  In RIII, the magnetic field topology depends on the magnitude of the initial field at low values of $\Pm$ (see Fig.~\ref{RaPmbif}).  With a weak initial field, multipolar  solutions are obtained with $Rm$ close to $Rm_c$ and $E$ sufficiently low whereas for $\Pm$ or $E$ sufficiently high, dipole-dominated dynamos are finally the stable solutions. For $E=3\cdot 10^{-4}$, only one bistable example has been obtained (see Fig~\ref{RaPmbif}(a) and Fig~\ref{bistability}). But, the size of the bistable area in $(Ra/Ra_c,\Pm)$-plane increases as $E$ decreases (see also Fig.~\ref{bistability}). The stability domain of dipolar solutions is limited on the right by the criterion $Ro_\ell<0.12$ \citep{christensen06}.

  Multipolar dynamos are only generated from a flow in the turbulent regime (RIII).  The magnitude of the magnetic energy is proportional to the difference $Rm-Rm_c$ for the multipolar branch and only sets in from a small perturbation when $Rm>Rm_c$ (see Appendix). This bifurcation is thus a supercritical one.  The use of strong initial fields inhibited multipolar dynamos close to $Rm_c$ in previous systematic studies. For instance, \citet{christensen06} claimed that the dynamo threshold for multipolar dynamos corresponds to $Rm>1000$. By considering a weak initial field, we have clearly determined the nature of the  dynamo bifurcation for the multipolar branch and the existence of multipolar dynamos with $Ro_\ell$ much lower than $0.1$ with $Rm$ close to $Rm_c$ (see Fig.~\ref{RaRmcKIN}). 

  When $\Pm$ is sufficiently high, regardless of the initial magnetic field, the dynamo is dominated by the axial dipolar component. Such solutions obtained by considering a weak initial field are shown in Fig.~\ref{RaPmbif} by green crossed out squares. Such solutions provide the upper bound of the stability domain of multipolar dynamos in $(Ra/Ra_c,\Pm)$-plane. Crossed out squares appear for lower $\Pm$ as the Ekman number decreases. However, this stability domain is shifted down by lowering $E$ and its size increases by reaching much lower values of $\Pm$.

  In Fig.~\ref{bistability} we report the extension of the bistable regime for different Ekman numbers $E$. As $E$ decreases, multipolar solution with lower $Ro_\ell$ can be obtained by considering a weak initial field. This result suggests that multipolar solutions characterized by reversing fields could be relevant solutions for realistic parameters: $E=10^{-15}$ and $Ro=10^{-6}$. However, geomagnetic reversals are rare events and the Earth's magnetic field is dominated by the dipolar component.

  The existence of a bistable regime in geodynamo models was not so clearly reported by previous systematic parameter studies using the same numerical setup. \citet{christensen06} only mentionned that the dipolarity depends on the initial magnetic conditions for one set of parameters. The existence of a bistable regime induces a hysteretic behavior.   For $E=10^{-4}$ and $\Pm=2$, $\Pm=1$ and $E=3\cdot 10^{-5}$ and $\Pm=0.25$ (see Fig~\ref{fighyst}), we have tested the existence of a hysteretic behavior as $Ra$ is varied. Dipolar fields collapse if $Ra$ is increased such as $Ro_\ell$ exceeds the transitional value $0.12$ and the multipolar field configuration appears to be the only stable solution. Since the multipolar branch can extend into the stability domain of dipolar dynamos, hysteretic behavior is observed if $Ra$ is decreased from this state, i.e. multipolar dynamos are maintained in the stability domain of dipolar dynamos. As stated by \citet{busse09} and \citet{schrinner12} for stress-free models, the emergence of a bistable regime results from the action of zonal flows. As described in our hydrodynamical study, zonal flows gain in strength when the Ekman number is decreased in RIII even if $Ro_\ell$ is much lower than 0.1 in models with no-slip boundaries.

  Strong dipolar fields can be maintained over time whereas weak fields are not amplified (dipolar models below the red dotted curves), i.e. the dipolar branch is subcritical in RIII.

\section{Discussion on dynamo bifurcations}

\subsection{Subcriticality in RII}

\begin{figure}
\begin{tabular}{ccc}
\subfigure[In RI]{\includegraphics[width=4.cm]{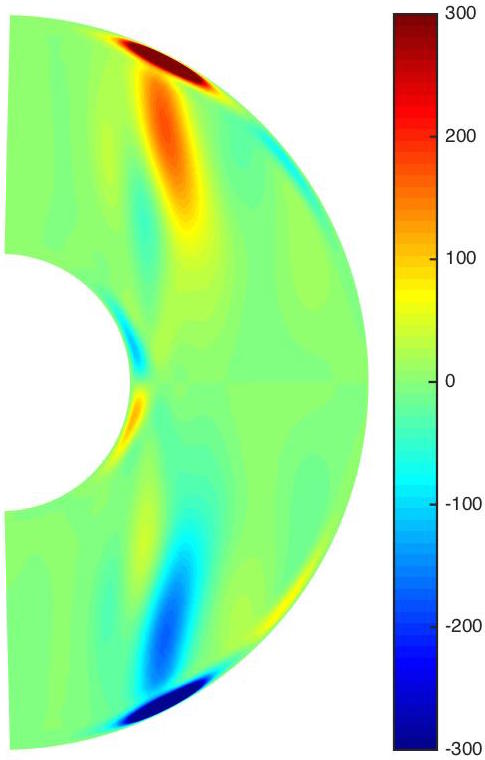}} &
\subfigure[In RII]{\includegraphics[width=4.cm]{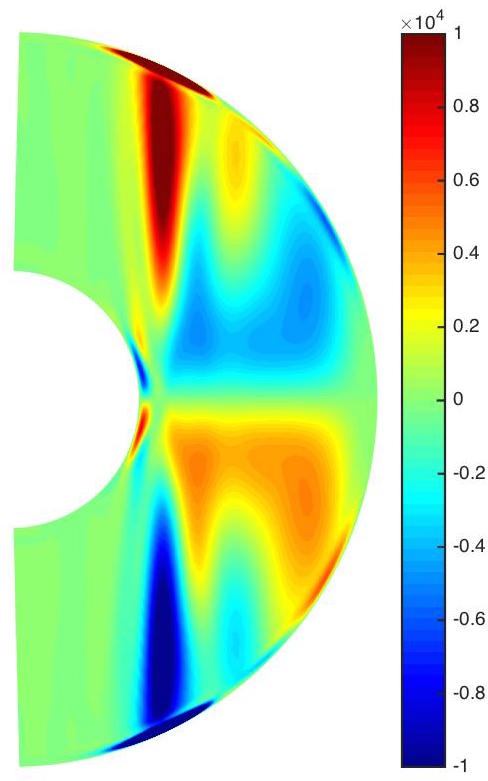}} &
\subfigure[in RIII]{\includegraphics[width=4.cm]{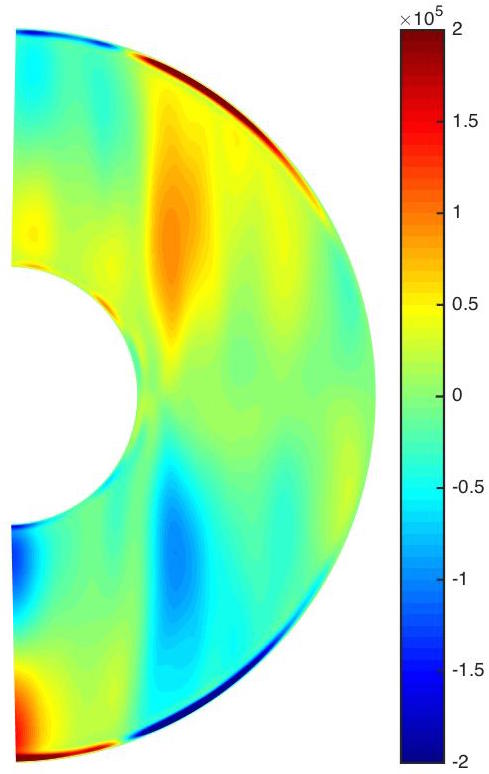}}
\end{tabular}
\caption{Action of the magnetic field on the helicity distribution. Helicity from the associated non-magnetic runs have been subtracted from the helicity calculated in MHD simulations dominated by a dopilar field. The difference is then show in each regime. The parameters are $E=10^{-4}$, $Ra=340=4.88Ra_c$ and $\Pm=1$ (on the middle), (on the right) $Ra=750=10.77Ra_c$ and $\Pm=0.5$ and (on the left) $Ra=155=2.55Ra_c$ with $E=3\cdot 10^{-4}$ and $\Pm=6$. The action of Lorentz forces associated with dipolar solutions is highligthed close to the dynamo threshold in the hydrodynamical regimes: RI, RII and RIII (from left to right).}
\label{HemMHD-HNS}
\end{figure}

\begin{figure*}
\begin{tabular}{cc}
\subfigure[$E=3\cdot 10^{-4}$]{\includegraphics[width=7.cm]{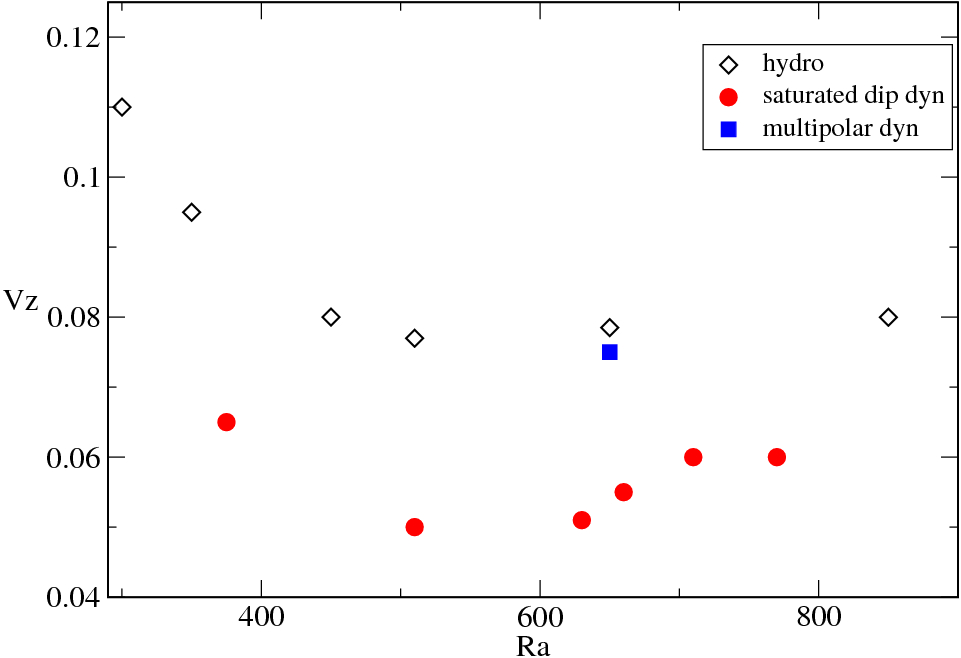}} &
\subfigure[$E=10^{-4}$]{\includegraphics[width=7.cm]{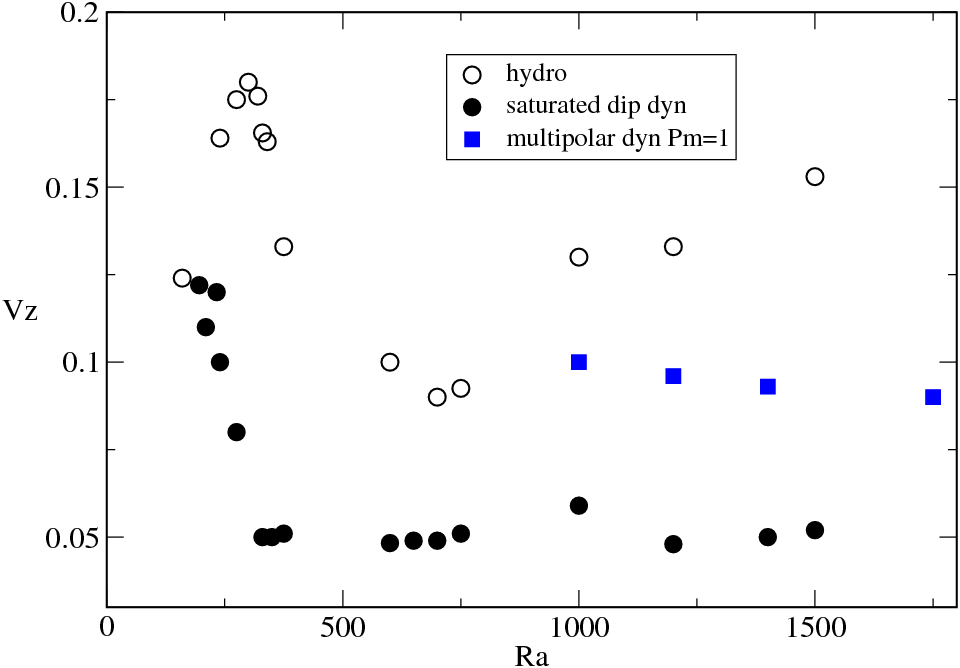}} \\
\subfigure[$E=3\cdot 10^{-5}$]{\includegraphics[width=7.cm]{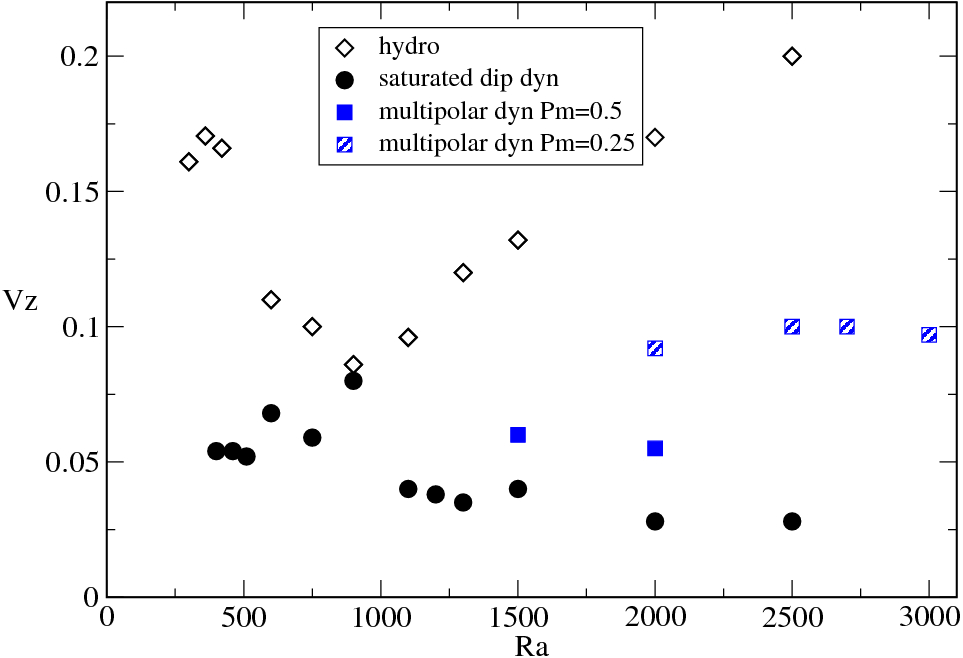}} &
\subfigure[$E=10^{-5}$]{\includegraphics[width=7.cm]{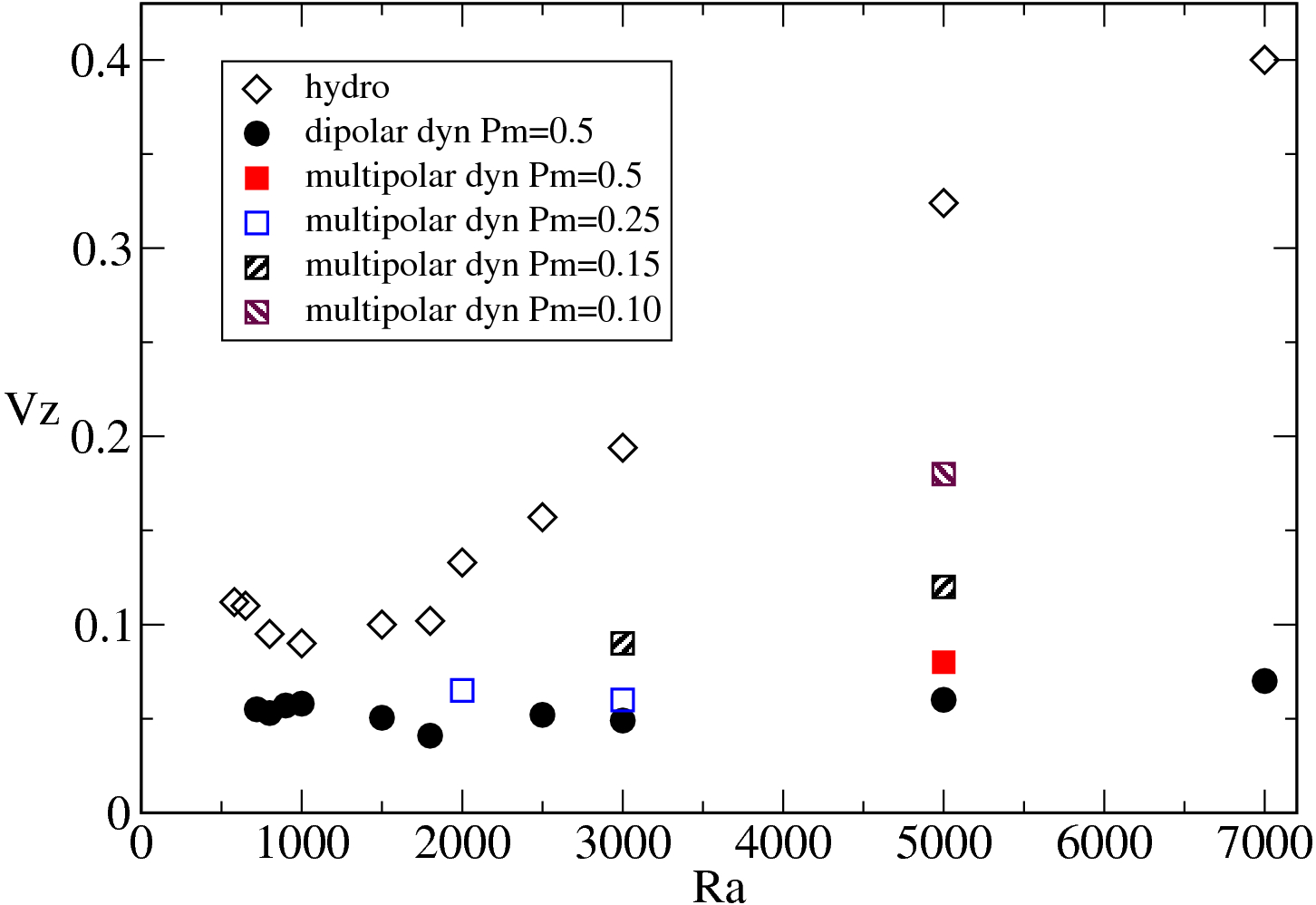}}
\end{tabular}
\caption{Comparison of the relative importance of zonal flows in hydrodynamical models and in dipolar and multipolar MHD simulations for different Ekman numbers and for $Ro_l<0.12$.}
\label{VzMHDH}
\end{figure*}

  For dynamo models with $Rm<Rm_c$ (below the dotted red curves), the Lorentz force associated with strong initial dipolar fields modifies the flow in such a way that it promotes dynamo action which in turn maintains magnetic activity.   \citet{schrinner07} and \citet{schrinner12}  pointed out the importance of the $\gamma$-effect in the generation of dipolar fields. This effect is crucial to advect the mean azimuthal magnetic component latitudinally and radially  (see \cite{schrinner12} for details). The $\gamma$-effect corresponds mathematically to the anti-symmetric part of the $\alpha$ tensor which is related to the spatial variations of the velocity field \citep{moffatt}.  The $\gamma$-effect operates only if the field strength is not negligible, as it results from $\overline{\bf{u'}\times\bf{b'}}$ where $\bf{u'}$ and $\bf{b'}$ denote the fluctuating parts of the velocity and magnetic fields respectively. As noted by \citet{sreenivasan11}, it can induce subcriticality when:  (i) the initial field is strong enough and (ii) the flow has a non-zero fluctuating part. Convection in rapidly rotating shells obviously satifies the latter condition.

  An additional condition must also be mentioned for the $\gamma$-effect; if convection cells only partly reach the outer sphere, as it is the case in  models in RI or close to RI, no $\gamma$-effect can exist. By considering a strong initial  dipolar magnetic field, convection cells extend on the whole volume, and in particular, up to the outer sphere close to the equatorial plane. Fig.~\ref{HemMHD-HNS} enables us to better understand the impact of the magnetic field on the helicity distribution and it was created by subtracting helicity obtained in non-magnetic simulations from helicity of corresponding MHD runs for a subcritical dynamo ($Rm_t<Rm<Rm_c$). In RII, the Lorentz force associated with a dipolar field in such a model acts to transport helicity towards the outer sphere at low latitudes, while it is approximately zero in hydrodynamical runs. By contrast, in RI,  the action of the field on helicity is shown in Fig~\ref{HemMHD-HNS}(a) when $Rm$ is slightly above $Rm_c$. This result is typical for laminar dynamos ($\Lambda<1$). Helicity is mainly modified in boundary layers. In RIII, helicity is mainly globally decreased by the magnetic tension of dipolar fields (see Fig.~\ref{HemMHD-HNS}(c)). 

  Here, helicity is simply used as a proxy for the flow structure which does not exist close to the outer sphere without dipolar fields in RII. A strong initial dipolar field extend radially convection cells and induces the presence of $\gamma$-effect which is a scenario for subcriticality in RII.

  \citet{sreenivasan11} argued that dipolar magnetic fields enhance the kinetic helicity and are therefore easier to maintain than fields with a more complicated field topology. However, as noted by these authors, the relation between kinetic helicity and induction mechanisms is not straightforward. Moreover, \citet{schrinner07} showed that the kinetic helicity is indeed a bad proxy measure for induction effects ($\alpha$-effect) in these models.   As noted in our kinematic study, the growing mode has a preferred dipolar symmetry in  RI and RII if $Rm$ exceeds $Rm_c$. A mode selection mechanism as proposed by \citet{sreenivasan11} to explain the dominance of dipolar fields in nonlinear models is not necessary for non-turbulent flows ($Ra<10Ra_c$).

\subsection{Subcriticality in RIII and multipolar dynamos}

  In order to understand the existence of multipolar dynamos in the stability domain of dipole-dominated dynamos ($Ro_\ell<0.12$), we study the saturation mechanism of multipolar dynamos by approaching the dynamo threshold. Since the multipolar branch is supercritical, dynamo models with $Rm$ very close to $Rm_c$ can be obtained. By comparing the flow of such models with their hydrodynamical counterparts, the action of the Lorentz force can be highlighted. At sufficiently low Ekman numbers, important zonal flows develop in RIII even if $Ro_\ell$ is lower than 0.1 (see Fig.~\ref{VzNS}a). If $Rm$ is slightly above $Rm_c$, a weak field is exponentially amplified with time and saturates by reducing the energy of zonal flows which initially caused its amplification (see Figs.~\ref{VzNS} and \ref{VzMHDH}).    The convective energy distribution is not significantly affected by the presence of multipolar fields. The mean energy of zonal flows in multipolar dynamos (squares in Fig~\ref{VzMHDH}) is reduced, but, it is still higher than that observed in associated dipolar dynamos (full circles). At levels sufficiently above the dynamo threshold $Rm_c$ (by increasing $\Pm$), zonal flows are substantially quenched by Lorentz forces and the magnitude of zonal flows in these multipolar models becomes comparable to that of dipolar models (Fig~\ref{VzMHDH}). Once zonal flows have reached a comparable level for the dipolar and multipolar branch, the system seems to prefer the former. Consequently, a transition from a multipolar dynamo to a dipolar dynamo is observed if $\Pm$ is high enough. Such transitions are reported in Fig~\ref{RaPmbif} by crossed out squares. Such a behavior was also obtained in models with stress-free boundary conditions \citep{schrinner12} where the multipolar branch corresponds to $\alpha\omega$ oscillatory dynamos. Such transitions represent the upper boundary in $(Ra/Ra_c,\Pm)$-plane for the stability domain of the multipolar branch when $Ro_\ell<0.1$. Close to this upper boundary, a transient multipolar solution is first obtained for a period of time which can be of the same order as one magnetic diffusion time. This period is shorter for higher $\Pm$.

    Even if zonal flows play a constructive role in the generation of multipolar dynamos, we did not find good correlations between the azimuthal magnetic component from the DNS with that generated by the $\omega$-effect resulting from the stretching of poloidal fields by zonal flows. This lack of correlation suggests that multipolar dynamos in geodynamo models result from an $\alpha^2\omega$ mechanism where small-scale convection cells and large-scale zonal flows contribute constructively to the generation of azimuthal fields which are then partially converted into poloidal components by an $\alpha$-effect. An example of such a mechanism in global spherical shells was highlighted by \citet{schrinner11a} where the dynamo coefficients were determined with the test-field method \citep{schrinner05}.

  Relative to multipolar dynamos with $Ro_\ell>0.1$, we note the emergence of large scale magnetic components as an equatorial dipole mode if $Rm$ is close to 100 with $Ro_\ell<0.1$ and $E\ge 3\cdot 10^{-5}$. Dynamos influenced by an equatorial dipole mode can have a dipole field strength $f_{dip}$ higher than 0.5.  In order to distinguish the different dynamo branches, the non-axisymmetric contribution must be filtered out in the definition of $f_{dip}$. Zonal flows are known to be more important in models with higher $Ro_\ell$. Non-axisymmetric magnetic modes are damped by the differential rotation in models with $Ro_\ell>0.1$ (see Moffatt, 1978 Chap. 3.11), but, it can be built by columnar convection less affected by zonal flows having lower $Ro_\ell$. For $\Pm$ and $E$ sufficiently low, multipolar dynamos with important zonal flows ($Vz>0.25$) can be obtained even if $Ro_\ell$ is much lower than 0.1. In this case, we note that the relative importance of axisymmetric magnetic components increases and periodic reversals of the axial dipole mode are observed (see Appendix B). For instance, such a behavior appears for the multipolar dynamo with $E=10^{-5}$ and $\Pm\la 0.15$. \citet{sheykoFJ16} reported a similar result for a slightly lower Ekman number. A study of reversals in geodynamo models with $Ro_\ell$ much lower than 0.1 is postponed to a future paper.

\section{Action of magnetic fields in geodynamo models}

 In this section, we contrast dynamo models with non-magnetic, but
otherwise identical, rotating convection models in order to quantify the
influence of Lorentz forces on convective dynamics. While
comparisons between dynamo and purely hydrodynamical simulations
have been conducted (e.g., \cite{christensen99}; \cite{aubert05}), these studies are typically limited to
convection less than 40 times critical and dipolar magnetic field
geometries. \citet{soderlundKA12} have considered more turbulent flows, but they  used hyperdiffusivities in modelling the most turbulent flows. In addition, they focused mainly on the origin of the transition between dipolar and multipolar fields. Here, the action of multipolar and dipolar fields on the flow is determined for the different hydrodynamical regimes close to and far above the dynamo threshold. This approach enables one to understand the nature of dynamo bifurcations presented in the previous section and the force balance in dynamo models.

\subsection{Close to the dynamo threshold }

  Close to the dynamo threshold $Rm_c$ in RI or close to the turning point $Rm_t$ in the non-laminar regimes, the Lorentz force must play a minor role. However, this force modifies sufficiently the flow in RI which allows the saturation of the exponential magnetic field growth and it generates the necessary conditions for the maintenance of dipolar fields in RII and RIII.

\begin{figure}
\begin{tabular}{cc}
\subfigure[Supercritical $(Ra=360)$]{\includegraphics[width=7.cm]{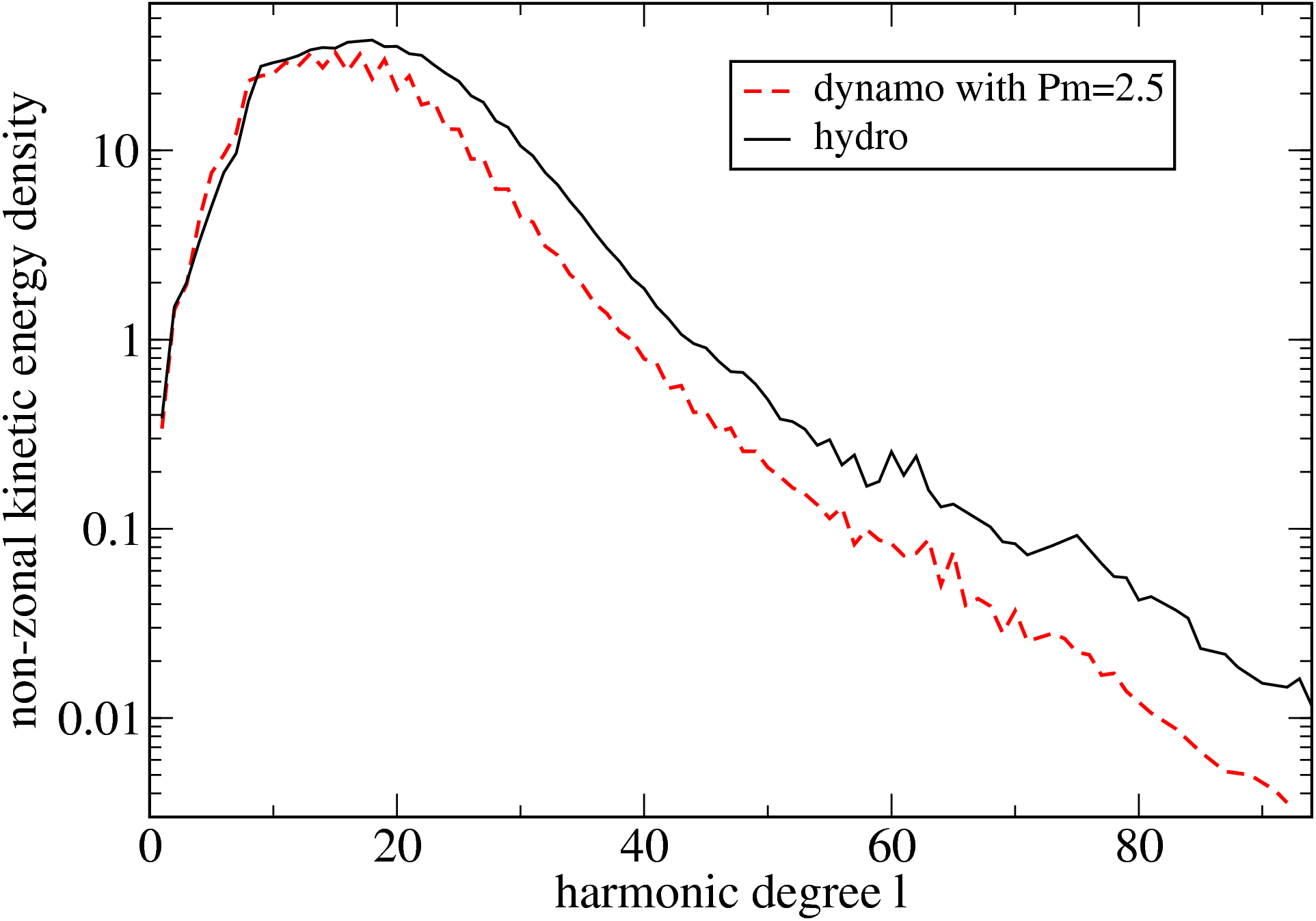}} &
\subfigure[Subcritical $(Ra=2000)$]{\includegraphics[width=7.cm]{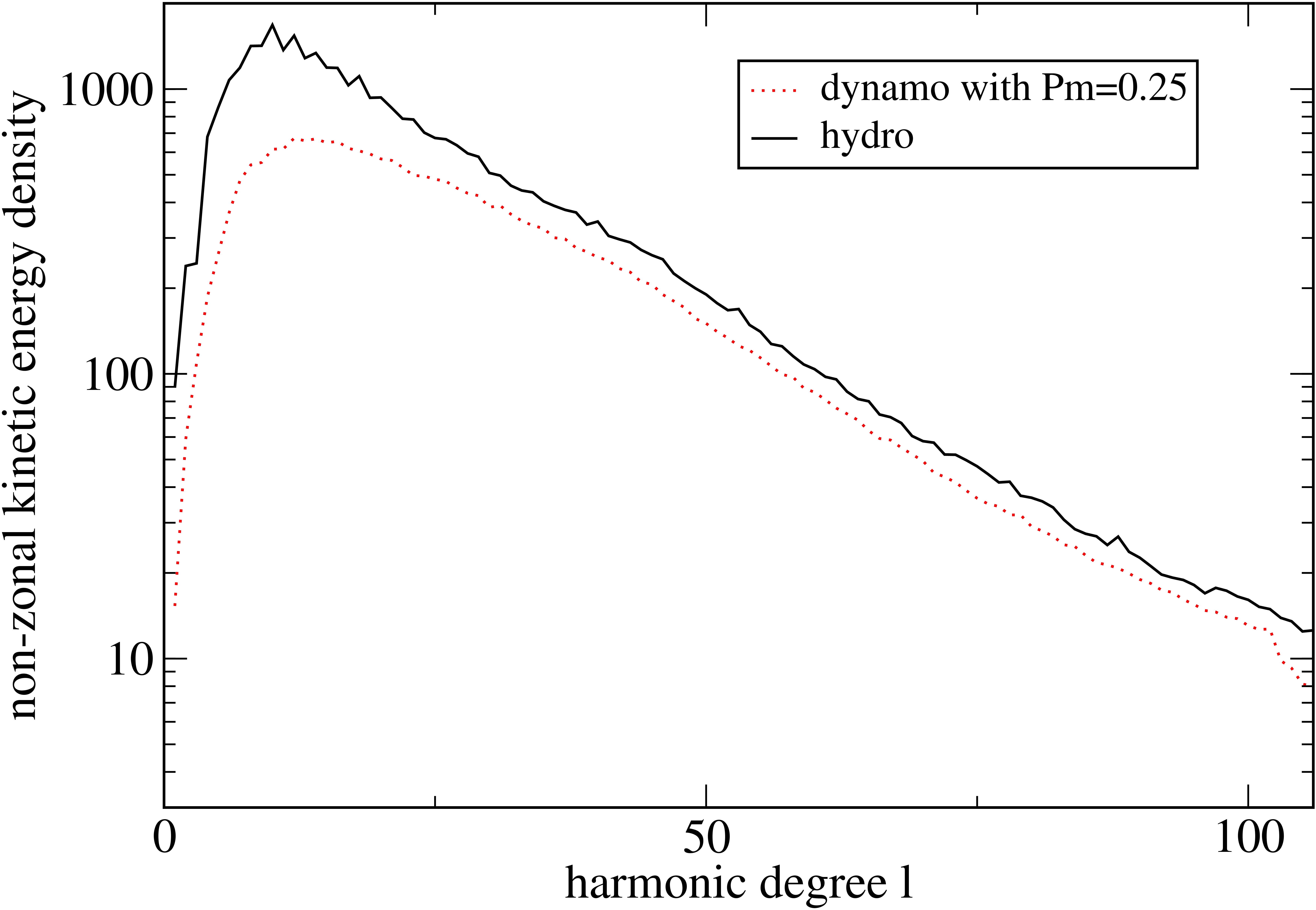}}
\end{tabular}
\caption{Kinetic energy distribution of the convective flow (non-zonal) as a function of the harmonic degree $l$ with $E=3\cdot 10^{-5}$. Dynamo results for dipolar solutions are compared to their associated hydrodynamic models for magnetic Reynolds numbers close to their critical values.  }
\label{SpecUcMHDH}
\end{figure}

  \citet{aubert05} noticed that the Lorentz force associated with dipolar solutions affects zonal flows significantly. We note here that the non-zonal velocity field $v_c$ is also affected. In Fig.~\ref{SpecUcMHDH}, the kinetic energy density distribution of $v_c$ is plotted for dynamo runs and non-magnetic runs. Close to $Rm_c$ in RII (see Fig.\ref{SpecUcMHDH}a), the presence of the dipolar magnetic field does not change the convective flow at large-scales (corresponding to small harmonic degree $l$). By contrast, dynamo action converts a fraction of the kinetic energy at small-scales into magnetic energy. Very close to the dynamo threshold, the Lorentz force affects slightly the main force balance in which the Coriolis force is mainly balanced by pressure gradients. For non-magnetic convection in rapidly rotating shells, the VAC balance is obtained with $Ra<10Ra_c$ \citep{aubert01}. It is important to note that the Lorentz force enters in this force blance by reducing the kinetic energy of convection at small-scales. Only one example for each hydrodynamical regime is given in this manuscript, but, the robustness of this result has been tested for different Ekman numbers $E$ and buoyant forcing $Ra$ and it persists if $Rm_c>Rm>Rm_t$ in this regime.

  In RIII ($Ra>10Ra_c$), inertia acts to generate zonal flows which in turn allow multipolar solutions and  a bistable regime to exist (see Figs.~\ref{VzNS} and \ref{VzMHDH}b). The force balance in non-magnetic models corresponds to the CIA balance \citep{aubert01}. In this regime, the dipolar branch is subcritical and the influence of the Lorentz force on the flow is minimum close to the turning point $Rm_t$.  Large-scale dipolar magnetic fields maintained by dynamo action do not allow zonal flows to dominate in the kinetic energy spectrum because they quench the axisymmetric toroidal kinetic part. First order differences result in zonal flows between dipolar dynamos and corresponding non-magnetic models.   The Lorentz force of subcritical dipolar models does not change the kinetic energy distribution of $v_c$ at small-scales (high harmonic degree $l$). A significant amount of convective kinetic energy is converted into magnetic energy at large-scales (see Fig.~\ref{SpecUcMHDH}). The strong initial dipolar field reduces the importance of inertia in this hydrodynamical regime. Fig~\ref{SpecUcMHDH} illustrates the modification of the flow structure by the Lorentz force of dipolar dynamos close to $Rm_t$ in RIII.

\subsection{Far above the dynamo threshold}

  So far, we have focused on dynamos close to their threshold. By comparing the MHD flow and the non-magnetic flow, we highlighted the action of dipolar fields in each dynamical regime. However, natural dynamos are expected to maintain magnetic activity far above their dynamo threshold where the Lorentz force is one of the dominant forces. We highlight the influence of dipolar fields on  flow structure and heat transfer. We focus on dipolar dynamos because of their geophysical applications. 

\begin{figure}
\begin{tabular}{cc}
\subfigure[Supercritical $(Ra=330)$]{\includegraphics[width=7.cm]{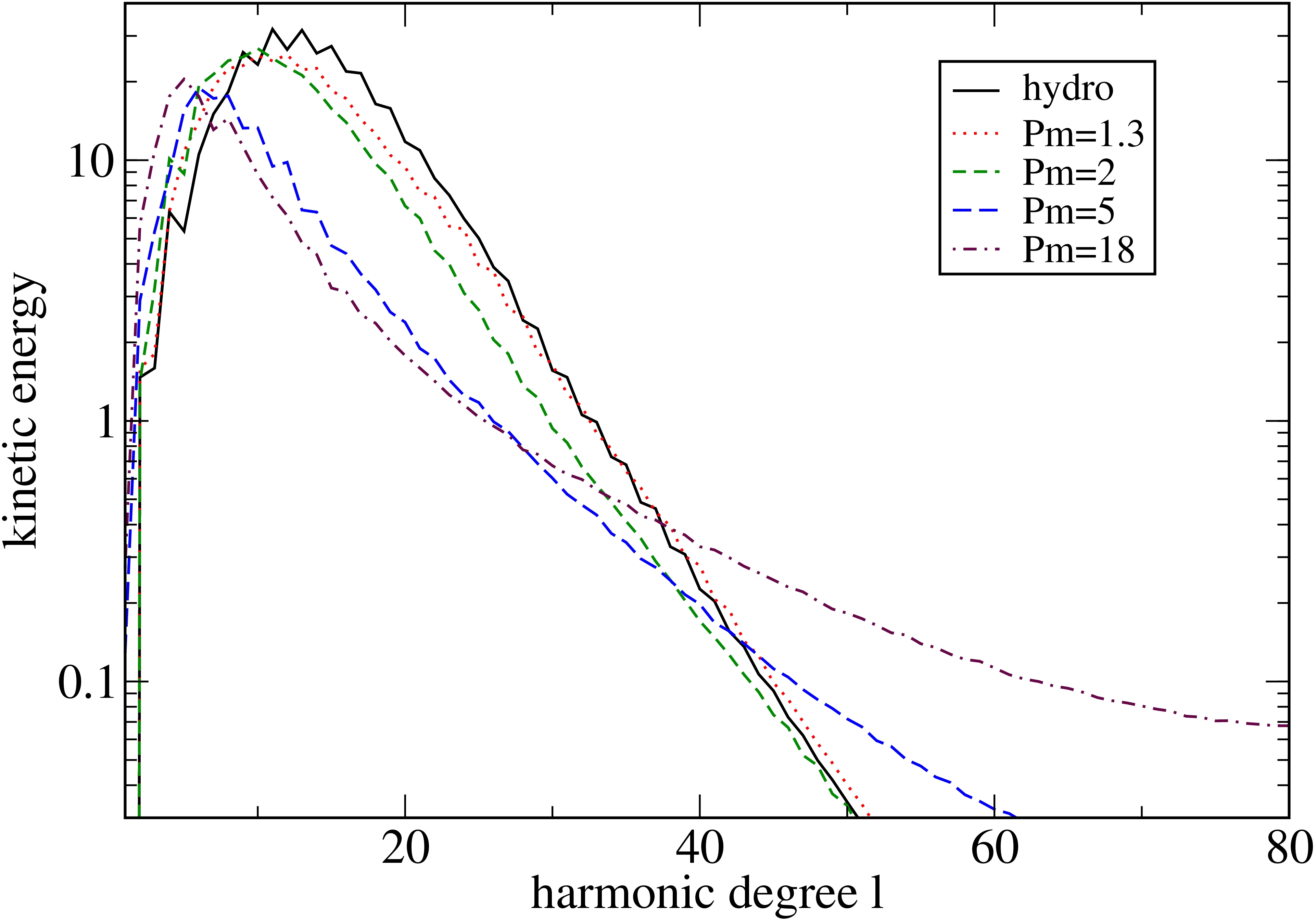}} &
\subfigure[Subcritical $(Ra=1200)$]{\includegraphics[width=7.cm]{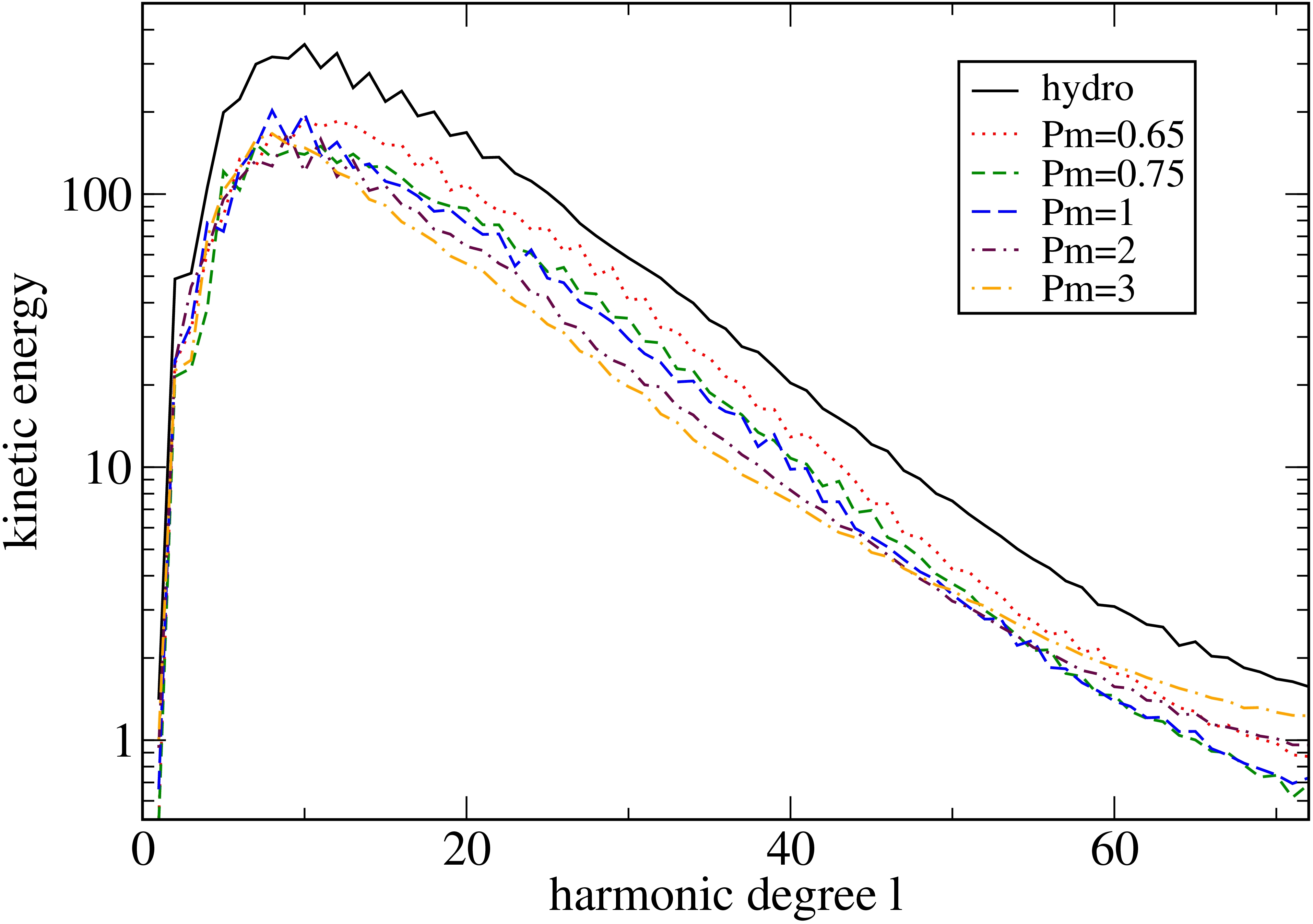} }
\end{tabular}
\caption{Kinetic energy distribution of the convective flow (non-zonal) as a function of the harmonic degree $l$ with $E=10^{-4}$ close to the dynamo threshold for a typical supercritical model (left panel) and a typical subcritical one (right panel). For $E=10^{-4}$ and $\Pr=1$, $Ra_c=69.65$.}
\label{SpecUce4}
\end{figure}

\begin{figure}
\begin{tabular}{ccc}
\subfigure[$Ra=510=6\,Ra_c$]{\includegraphics[width=4.7cm]{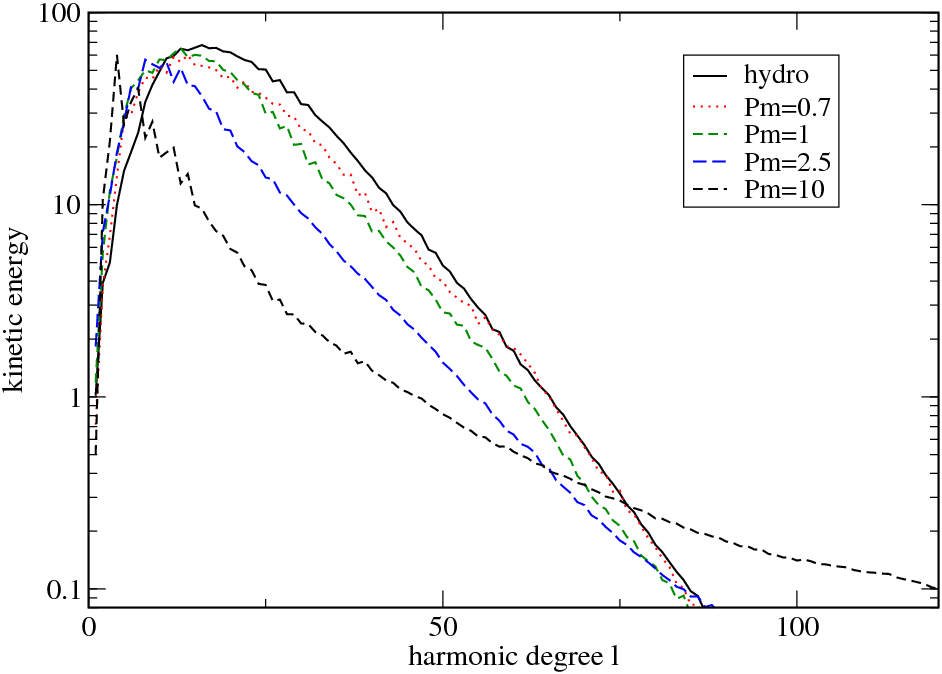}} &
\subfigure[$Ra=1100\approx 12.9\, Ra_c$]{\includegraphics[width=4.7cm]{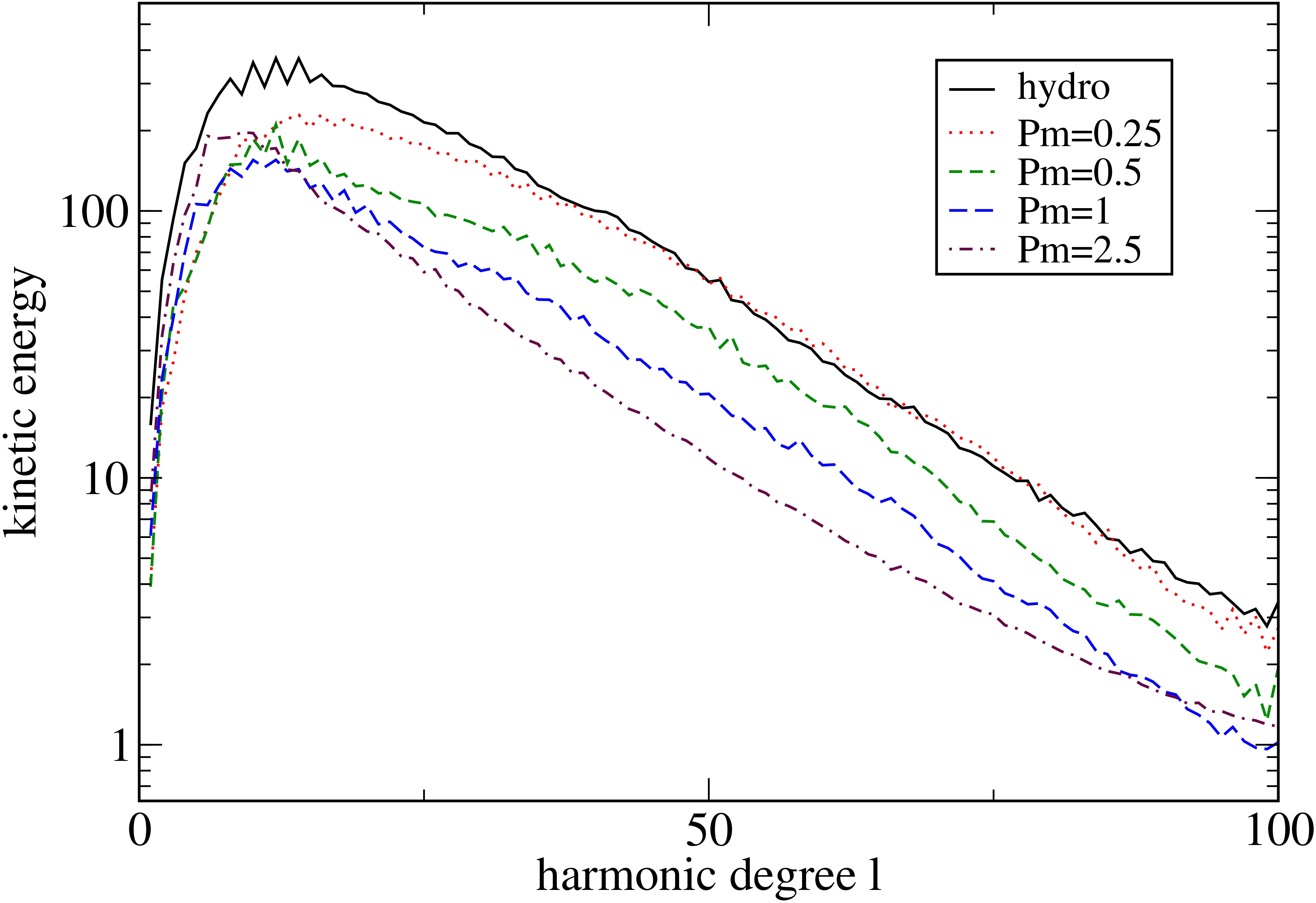}} &
\subfigure[$Ra=2000=23.5\, Ra_c$]{\includegraphics[width=4.7cm]{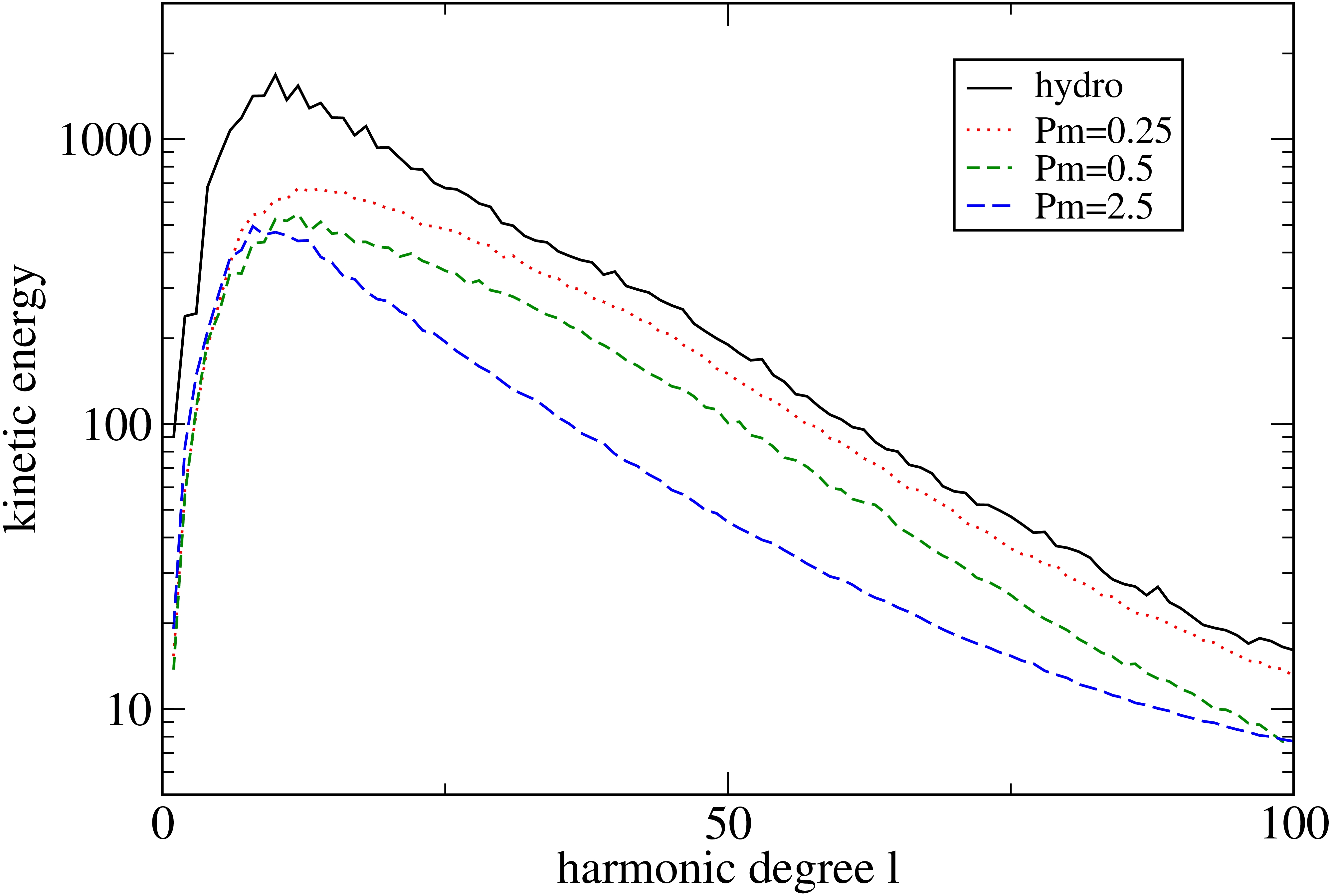}}
\end{tabular}
\caption{Kinetic energy distribution of the convective flow (non-zonal) as a function of the harmonic degree $l$ for the parameters $E=3\cdot 10^{-5}$ and different $Ra$.}
\label{SpecUc3e5}
\end{figure}

  \citet{kingB13} suggest that the typical length scale $L_u=\pi/\bar{l}$ in numerical models is controlled mainly by the influence of rotation, viscosity and buoyancy. From this result, they argued that the magnetic field does not  significantly affect the flow structure. However, this measure is a volume-averaged quantity of the influence of forces which act on the flow at different length scales. Figs.~\ref{SpecUce4} and \ref{SpecUc3e5} enable us to understand the increasing effect of the Lorentz force on the convective flow distribution as $Rm-Rm_c$ increases by controlling $\Pm$. The magnetic energy (measured by $\Lambda$) increases as  $Rm-Rm_c$ becomes greater, In the non-turbulent regime with $E=10^{-4}$, the typical length scale $L_u$ does not significantly depend on $Rm$ whereas the kinetic energy distribution is  substantially modified at any scale by dipolar fields if $Rm$ exceeds sufficiently $Rm_c$. In RIII at $E=10^{-4}$, the Lorentz force similarly decreases the kinetic energy at all scales. Viscous effects become less important as the Ekman number is decreased to $E=3\cdot 10^{-5}$ (see Fig.~\ref{SpecUc3e5}). However, we notice that the left panels in Figs.~\ref{SpecUce4} and \ref{SpecUc3e5} are very similar. In RII, the kinetic energy spectrum as more and more concentrated at large scales as the Lorentz force increases ($\Pm$ increases). At the same time, the energy of small convection cells increase with $Rm$ (at large harmonic degree $l$). The result is that the average quantity $L_u$ is almost unaffected by dipolar fields even if the flow structure depends on $\Lambda$. In RIII (right panels in Figs.~\ref{SpecUce4} and \ref{SpecUc3e5}), The kinetic energy continues to be distributed on many length scales as the magnetic influence increases. Close to the transitional value $Ro_\ell\approx 0.1$ when inertia plays a role, we observed that the flow speed was significantly reduced  by the Lorentz force. This effect becomes more noticeable as the Ekman number decreases in the turbulent regime. The growing influence of magnetic fields as $E$ decreases was also noted by \citet{soderlundKA12}.
 
  When we consider high $\Pm$ simulations, the Lorentz force constrains the convection to develop at large length scales. It is typical for dynamos in the MAC balance in which the Lorentz force is one of the dominant forces \citep{starchenkoJ02}. Such dynamos, also called strong field dynamos, can be obtained without abrupt transitions by considering dynamos largely above their dynamo threshold (by increasing $\Pm$). A discussion on this particular point can be found in \citet{dormy16} and in \citet{aubertGF17}. 

  We also observe that dipolar fields affect the azimuthal distribution (the typical harmonic order $m$) of the kinetic energy (not shown). As predicted by magnetoconvection studies, strong dipolar fields increase the size of convection cells in the azimuthal direction. However, the critical azimuthal wavenumbers of non-magnetic convection \citep{christensen06} for $E\geq 10^{-5}$ and $\Pr=1$ are smaller than $15$. This number is already low and it does not allow to observe a significant transfer of kinetic energy into larger scales.

\subsection{Influence of dipolar fields on heat transfer}

\begin{figure}
\begin{tabular}{cc}
\subfigure[]{\includegraphics[width=7.cm]{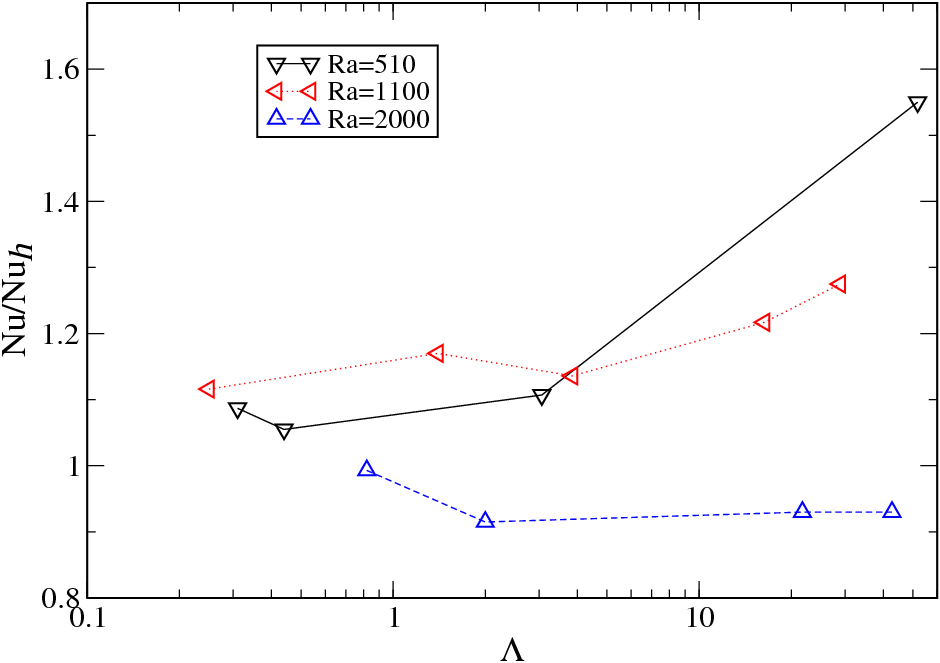}} &
\subfigure[]{\includegraphics[width=7.cm]{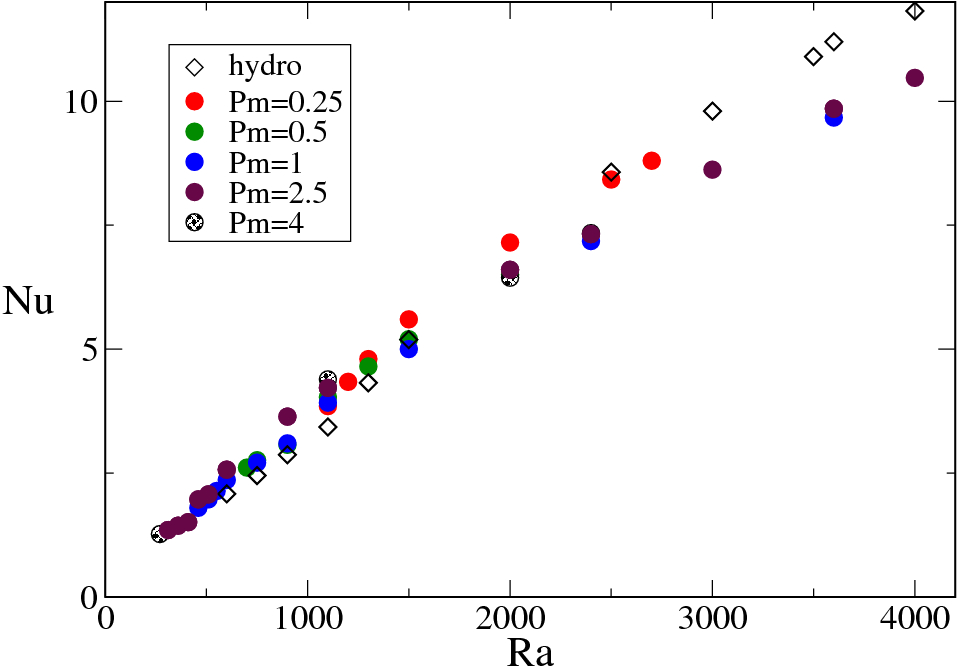}}
\end{tabular}
\caption{Typical influence of dipolar fields on heat transfer efficiency with $E=3\cdot 10^{-5}$. Variation of the ratio of $Nu$ for dynamo and  hydrodynamic cases $Nu_h$ as a function of $\Lambda$ for three values of $Ra$ (panel (a)): $Ra=510=6\, Ra_c$, $Ra=1100=12.9\, Ra_c$ and $Ra=2000=23.5\, Ra_c$.  $Nu$ as a function of $Ra$ is shown for different $\Pm$ (panel (b)).  Since $Rm$ increases with $\Pm$, the more we consider high $\Pm$, the more the Lorentz force influences convection. However, this influence depends on the hydrodynamical regime.  }
\label{ElsNu}
\end{figure}

  From hydrodynamical studies, we know that for the range of parameters $E\geq 3\cdot 10^{-6}$ and $Ro_l\le 0.11$, heat transfer scales as $Nu\sim Ra^{6/5}$ \citep{king10}. By considering lower Ekman number ($10^{-6}$ and $3\cdot 10^{-7}$), \citet{gastineWA16} have found a steeper scaling law $Nu=0.15\,\tilde{R}a^{3/2}E^2$ where $\tilde{R}a$ is here the usual Rayleigh number $\tilde{R}a=Ra\, \Pr/E$. The latter scaling holds in the rapidly rotating regime, i.e. $\tilde{R}a\, E^{8/5}<0.4$. Due to this criterion, we note that the rotation-dominated regime as described by \citet{gastineWA16} corresponds to RII with $E>10^{-6}$. Consequently, the important zonal flows which develop in low-$E$ models, do not affect the heat transfer efficiency for this range of parameters. However, their free-slip counterparts show that dominant zonal flows can reduce the heat transfer \citep{yadav16}. The increase of $Vz$ by lowering $E$ could also reduce the heat transfer with $E\le 10^{-6}$ in no-slip models.

  Fig~\ref{ElsNu} enables us to understand the influence of dipolar fields on the efficiency of heat transfer in geodynamo simulations. By considering different $\Pm$, the field strength of self-sustained dipolar dynamos has been varied by more than one order of magnitude. Different values for $Ra$ were considered in order to explore different dynamical regimes (RII and RIII). In the non-turbulent regimes ($Ra<10Ra_c$, RII), the magnetic field enhances heat transfer since the ratio $Nu/Nu_h$ is higher than $1$ for $Ra=510=6Ra_c$. $Nu$ increases with $\Lambda$ even if the field strength becomes large in this range of $Ra$. In this regime, the increase of the magnetic influence allows the quasi-geostrophic structure of the flow which limits heat transfer to break. The influence of rotation becomes less pronounced as $\Lambda$ increases and the result is an enhancement of heat transfer efficiency in MHD runs. By contrast, when $Ra$ is sufficiently high, dipolar fields reduce heat transfer even if the field strength is low ($\Lambda<1$). Close to the transitional value $Ro_\ell\approx 0.1$, heat transfer in hydrodynamic runs is more efficient than in MHD simulations (see right panel in Fig~\ref{ElsNu}). Dipolar fields in turbulent models reduce the heat flux associated with the non-zonal flow (see Fig~\ref{SpecUc3e5}). The results show that the influence of dipolar fields on heat transfer depends on the buoyant forcing $Ra$. These results hold in the range  $10^{-4}\geq E\geq 10^{-5}$ and are in agreement with the dataset presented by \citet{christensen06}.

  Results for heat transfer efficiency must be compared with those obtained by \citet{Yadav16Earth} (see their figure~4). The authors performed simulations with $\Pm=1$ and interpreted their numerical results differently since they did not notice the existence of different dynamical regimes in geodynamo simultions which are explored by varying $Ra$. The results presented in Fig~\ref{ElsNu} are not changed by taking the definition for $\Lambda$ introduced by \citet{soderlundKA12}. In their plot of $Nu/Nu_h$ as a function of $\Lambda$, the field strength has been varied by changing $Ra$ in \citet{Yadav16Earth}. As $\Lambda$ increases with $Ra$, it is not possible to distinguished whether the observed decrease of the heat transfer efficiency is due to  the increase of $\Lambda$ or that of $Ra$ in their study. By doing a similar analysis with a larger parameter space (different $\Pm$), we can claim that the magnetic influence of dipolar fields depends on the buoyant forcing. The field strength has a minor impact.

\section{Summary and applications to planetary interiors}

  In our numerical survey, dimensionless parameters were varied significantly in order to determine their influence in geodynamo models. This method enables us to argue on the validity of numerical results, as realistic parameters cannot be reached by direct numerical models. In particular, we highlighted the existence of different dynamical regimes in which dipolar dynamos can be generated. The influence of strong initial fields on the flow depends on the hydrodynamical regime. It is necessary that this influence promotes dynamo action in order to obtain a subcritical bifurcation. Otherwise, bifurcations are supercritical. Surprisingly, our study seems to show that the Ekman number does not directly affect the nature of the dynamo bifurcation and the magnitude of $Ra/Ra_c$ is in fact the key parameter. However, lowering the Ekman number allows to maintain dipole-dominated dynamos for higher and higher buoyant forcing $Ra/Ra_c$ (in RIII) where multipolar dynamos are also stable solutions.    This behavior enables us to extrapolate our results to realistic parameters. However, it is important to note that caution should be exercised in interpreting the results obtained with Ekman numbers which differ considerably from realistic values. Previous studies have noted an influence of the magnetic Prandtl number on the nature of the bifurcation. In fact, this number allows to select the hydrodynamical regime as the magnetic Reynolds number ($Rm=Re\, Pm$) has to exceed a critical value for dynamo action.

 In the laminar regime $(Ra<3Ra_c)$, the power spectrum of the non-magnetic velocity field for the longitudinal Fourier mode consists only of the critical convection mode and its harmonics. Slightly above the kinematic dynamo threshold $Rm_c$, the exponentially growing mode is the axial dipole mode. Then, the Lorentz force slightly modifies helical motions and the field strength is low (close to $Rm_c$). The bifurcation is supercritical. Dipolar fields play a major role when $\Lambda$ exceeds $1$ and the convective flow structure is thus changed by the emergence of additional convection modes. The manifestation of this change is a sharp increase of the magnetic energy interpreted as a strong-field dynamo branch by \citet{dormy16} where the force balance would be magnetostrophic.  High $\Pm$ has to be considered in order to have dynamo action close to the onset of convection. In this dynamical regime, the inertia term has a minor role, which means that the $Ro\ll 1$ condition is satisfied. The Reynolds number in such models, however, is also very low, whereas quasi-geostrophic turbulence develops in planetary interiors. Abrupt variations of the magnetic field strength (by more than one order of magnitude) as $Ra$ increases, are only obtained close to the onset of convection where fields with $\Lambda$ higher than $1$ induce an important change of the flow strucuture. 

  Numerical dynamos with $E$ higher than $10^{-5}$ in spherical shells develop in RII where convection is completely developed and zonal flows are minimal. This parameter space was favored by previous systematic parameter studies because smaller Ekman numbers require important numerical resources.   In this dynamical regime (RII), we have shown that the dynamo bifurcation can be supercritical for $Ra$ close to values corresponding to the laminar regime, or subcritical. The latter situation is obtained with low $\Pm$. We argue that subcriticality is probably induced in RII by the existence of the $\gamma$-effect which is one of the ingredients required for dipolar field generation \citep{schrinner12}. Convective motions and a non-negligible magnetic field are necessary conditions for the development of the $\gamma$-effect. This effect was utilized by \citet{sreenivasan11} to explain the dominance of dipolar fields in geodynamo simulations. On the other hand, we have shown that the axial dipole mode is the growing mode  in the kinematic phase and the other modes either decrease in time or grow more slowly. In fact, the mode selection as proposed by \citet{sreenivasan11} which is a nonlinear process, is not necessary in this dynamical regime (RII).

 In non-turbulent regimes, heat transfer efficiency improves as magnetic field strength increases. Dipolar fields reduce the geostrophic constraint which limits heat transfer in this regime. By considering models close to the dynamo threshold, we have shown that dipolar fields modify the flow at small length scales whereas the large-scale convective flow is almost unaffected. 

 In the turbulent/inertial regime $(Ra>10Ra_c)$ where the increase of the buoyant driving promotes zonal flows, a supercritical multipolar branch exists for low $Pm$ with $Rm>Rm_c$. Zonal flows contribute to  field generation of these dynamos and the Lorentz force mainly acts to saturate magnetic activity growth by quenching zonal flows. Previous geodynamo simulation studies involving strong initial dipolar fields of high Ekman numbers have reported multipolar dynamos only with $Ro_\ell>0.1$. Such conditions prevent multipolar dynamos by limiting zonal flows. 

  The dynamo bifurcation of the dipolar branch in the turbulent regime is subcritical. The kinematic study shows that $Rm_c$ increases with $Ra$ and means that turbulent fluctuations and zonal flows have a negative impact on dynamo action. Strong dipolar fields reduce such effects and allow subcriticality. The reduction of inertia effects by strong dipolar fields  limits the development of the convective flow at large-scales resulting in a decrease of heat transfer efficiency even when the field strength (measured by $\Lambda$) does not exceed unity.

  The magnetic field strength of saturated dipolar dynamos grows rapidly with $Ra$ in RII. In comparison, $\Lambda$ remains almost constant when $Ra$ is increased in RIII but $\Pm$ held fixed. Having our kinematic study in mind helps us to understand this evolution. The key parameter is the distance from the threshold $Rm-Rm_c$ and it increases rapidly with $Ra$ in RII $(3Ra_c<Ra<10Ra_c)$. In contrast, because of the increase of $Rm_c$, $Rm-Rm_c$ remains almost constant in RIII. As a result, the field strength of dipolar dynamos does not depend significantly on $Ra$ in RIII (see Appendix A).   However, the magnetic Reynolds number $Rm$ is of limited use for characterizing dynamos. This output parameter can be modified by varying the buoyant forcing $Ra$ or the magnetic Prandtl number $\Pm$ when $E$ and $\Pr$ are held fixed. We have clearly shown that the control parameters $Ra$ and $\Pm$ can have different influences on the magnetic field strength, the typical length scale or the heat transfer efficiency.

    Interestingly, the dependency of $Rm_c$ on $Ra$ shows that most of numerical dipolar dynamos with $Pm\leq 1$ and $E\geq 10^{-5}$ obtained in previous studies are dynamos with $Rm$ in the vicinity of $Rm_c$. The Lorentz force associated with such fields has a minor role (see \cite{king10}; \cite{kingB13}; \cite{soderlundKA12}), however in models with $Rm$ far above $Rm_c$, the role of the Lorentz force becomes dominant.  This situation is obtained by considering high magnetic Prandtl numbers if $E\geq 10^{-5}$ or by lowering the Ekman number with $Rm$ high enough (around 1000) (see \cite{soderlund15,Yadav16Earth,aubertGF17,schaefferJNF17}). The magnetostrophic regime which is relevant in planetary interiors corresponds to the numbers: $Ro\ll 1$, $Re\gg 1$ and $\Lambda\approx 1$. Current computational models cannot reach these extreme parameters, especially for the Reynolds number. However, turbulent effects such as generation of zonal flows  can develop in low Ekman simulations even when rigid boundaries are used. Although inertia and viscous effects play a role in simulations, the Coriolis force and the Lorentz force are dominant in models with high $\Pm$, low $E$ and high $Ra/Ra_c$. \citet{wichtC10} and \citet{teedJT15} have observed magnetostrophic events as torsional oscillations in direct numerical simulations which explore this regime ($\Pm$ sufficiently high). 

  By increasing the magnetic Prandtl number in numerical models, the importance of the Lorentz force is increased as described above. While abrupt transitions are obtained close to the onset of convection, we show that the flow structure (see section 5) and the field strength (see Fig~\ref{RaPmbif}) evolves gradually with $\Pm$ when $Ra>3Ra_c$ (in RII and RIII). We show that when $\Pm$ is sufficiently high, the field strength reaches the magnitude of the strong field dynamo branch identified with a lower bouyant forcing (see Fig~\ref{RaPmbif} with $E=10^{-4}$). In addition, the length scale of the flow is gradually increased when the Lorentz force becomes more and more dominant (see section 5). As observed in RI, we also show that the flow is organized on large scales in models with higher $Ra$ if $\Pm$ is sufficiently high as expected in the MAC regime \citep{starchenkoJ02}. Geodynamo simulations with $Rm$ largely above their threshold can improve our understanding on the Earth's dynamo (see \citet{dormy16} and \citet{aubertGF17} and references therein for a discussion).

  Paleomagnetic observations indicate that the field has reversed its polarity hundreds of times in Earth's history \citep{amit10}. Reversals and excursions are rare events, as their duration is much shorter than the period of the stable polarity chrons separating them. Such events are also very irregular. Their frequency varies significantly and can include very long periods of stable polarity. Reversals can be investigated using various tools, including numerical models, observations, laboratory magnetohydrodynamics experiments \citep{berhanu10} and theory.  Numerical dynamo models operate in a parameter regime far from what would be appropriate for modeling Earth's core, so their application to the geodynamo is questionable and efforts have to be made in order to extrapolate such numerical results. Systematic studies of numerical dynamos provide vital information about the dependence of dynamo properties on dimensionless parameters. Previous studies highlighted a dichotomy between non-reversing dipole-dominated dynamos and reversing non-dipole-dominated multipolar solutions. The initial strong dipolar field collapses in simulations if the ratio of inertia to the Coriolis term exceeds the critical value $Ro_\ell>0.1$. Dipole dominated dynamos that rarely reverse can, in some cases, be found in the boundary between both regimes. It was argued that this scenario would explain observed reversals \citep{olson06} and appears to be the only possibility for reversals in our simple numerical model (see the review by \cite{robertsK13}). Dipolar dynamos close to the multipolar regime would explore this regime episodically due to important kinetic fluctuations which temporarily do not respect the criterion for dipolar dynamos. After the rapid dipole collapse caused by the increase of inertia, the reduction of $Ro_\ell$ below the critical value allows the restoring of the dominant dipolar field with the same direction (excursion) or with the opposite direction (reversal). This scenario was observed in numerical studies with Ekman numbers greater than $10^{-4}$.

  In our study, we show the emergence of a bistable regime due to zonal flows which develop in non-magnetic runs with $Ro_\ell$ below $0.1$ and $E$ sufficiently low. In this regime, the saturated dynamo solution depends on the initial conditions for the magnetic field and we report hysteretic behavior observed for different Ekman numbers and magnetic Prandtl numbers (see Figs.~\ref{bif_e4NS} and \ref{bif_3e5NS}). The increase of $Ra$ induces the dipole collapse if $Ro_\ell$ exceeds  $0.1$ and only multipolar solutions are obtained. However, an increase of the relative importance of zonal flows was also observed as the dipole collapsed in low Ekman and low $\Pm$ models. A reduction of the buoyant forcing does not imply a transition to a dipole-dominated dynamo even if $Ro_\ell<0.1$ as the multipolar branch can extend into the dipolar regime if $E$ and $\Pm$ are sufficiently low. 

  Such a bistable regime has been also found in models with stress-free boundaries \citep{busse06,schrinner12}. As viscous boundary layers do not limit the development of zonal flows in such models, they contain a huge part of the kinetic energy \citep{christensen02} even if the Ekman number is not very low. \citet{schrinner12} have clearly shown that the differential rotation by converting poloidal magnetic field components into toroidal ones participates to the dynamo mechanism of non-dipolar dynamos which appear as oscillatory dynamos with periodic reversals of all magnetic components. Comparable situations can also appear in models with no-slip boundaries only if the Ekman number is sufficiently low (see Appendix B and \cite{sheykoFJ16}).

  From our hydrodynamic study, we deduce that the relative importance of zonal flows incrases as the Ekman number varies towards realistic values in no-slip models. As a result, the multipolar branch would persist for local Rossby numbers much lower than $0.1$. This extension of the multipolar branch into the dipolar regime reaches lower and lower values of $Ro_\ell$ by lowering the Ekman number. A decrease of $Ro_\ell$ below $0.1$ would not induce the generation of a dipole-dominated dynamo. Excursions and reversals can be explained by Olson \& Christensen's scenario  for geodynamo simulations performed with high Ekman numbers in which zonal flows are limited and the turbulent regime is not explored. This scenario also seems to be relevant if high $\Pm$ are considered. In this case, Lorentz forces prevent the emergence of zonal flows and multipolar dynamos are not present when $Ro_\ell<0.1$. However, $\Pm$ is known to be very low in liquid metals and especially in the Earth's outer core. In addition, we observe that the transition between multipolar to dipolar dynamos with high $\Pm$ can take a long period of time as this state can be a meta-stable one  whereas the duration of reversals is observed to be very short for the Earth's magnetic field.

 Although, $\Pm$ is very low in planetary interiors, magnetic effects could still prevent zonal flows when the dipole field collapses, as some other magnetic components continue to act on the flow. Such a mechanism is suggested by observations but has not yet been numerically observed. 

  \citet{olson06} estimated $Ro_\ell\approx 0.1$ for the Earth's outer core by using the results of geodynamo simulations \citep{christensen06}. \citet{kingB13} using the same dataset noticed that the length scale of the flow follows the viscous scaling law (VAC: $L_u\propto E^{1/3}D$). In other words, the presence of a dynamo-generated field does not affect significantly the size of the convection cells in these models. However, strong dipolar fields decrease the relative importance of inertia by extending the convection cells on a global scale. An estimate of $Ro_\ell$ including the action of magnetic fields is smaller than $0.1$ by several orders of magnitude. In addition, the physical conditions which affect $Ro_\ell$ have evolved in  past geophysical periods. In particular, the inner core freezes and its size increases. As a result, convection is then constrained to develop in a thinner shell in the outer core. As shown by \citet{schrinner12}, this induces an increase of $Ro_\ell$. But, \citet{olson06} scenario requires the geodynamo to reside in a narrow range of $Ro_\ell$. This requirement seems to be in contradiction with the physical conditions in the Earth's outer core, since $Ro_\ell \ll 0.1$ and with its evolution as $Ro_\ell$ has to evolve in time. Here, we do not propose a new explanation for geomagnetic reversals (see for instance \citet{petrelis09}).

  The subcriticality of the dipolar branch in the turbulent regime has implications for the long-term evolution of the geodynamo. Inner cores in terrestrial planets freeze over time and heat and light elements are released at the base of liquid outer cores. Such effects are responsible for convection motions which induce dynamo action. The growth of inner cores constrains the convection cells to develop in thinner outer cores. As shown by \citet{schrinner12}, inertia effects increase in this case. In addition, the time evolution of the geometrical constraints affects also the magnitude of the magnetic Reynolds number which depends linearly on the shell width. According to our results, two scenarios can describe the dramatic evolution of terrestrial magnetism. In both, dynamo action is ultimately lost since $Rm$ becomes lower than the turning point, $Rm_t$, and the magnetic activity fails rapidly. Either this time evolution takes place with a dominant axial dipole field as the condition $Ro_\ell<0.1$ is still satisfied, or a transition towards a multipolar dynamo occurs before the loss of dynamo action. In the latter scenario, zonal flows participate in field generation in the final period. Direct numerical simulations with important zonal flows have clearly identified hemispherical dynamos \citep{grote2000,busse06,schrinner12}. The latter scenario could be relevant in order to understand the evolution of Mars' magnetism, as \citet{stanley08} have suggested that hemispherical fields were generated in the past in the deep martian interior. Subcritical behavior in the early Mars' core is also proposed by \citet{horiW13} in order to explain the loss of magnetic activity.

\section*{Acknowledgments}

I thank Julien Aubert, Thomas Gastine and Rapha\"el Raynaud for their comments on the manuscript and useful discussions. I also thank the referee Johanes Wicht for numerous useful comments and suggestions which improved the paper. 
This work was granted access to the HPC resources of MesoPSL financed
by the R\'egion \^{I}le-de-France and the project Equip@Meso (reference
ANR-10-EQPX-29-01) of the programme Investissements d'Avenir
supervised by the Agence Nationale pour la Recherche. Numerical
simulations were also carried out at CEMAG and TGCC computing centres
(GENCI project x2014046698). L.~P. acknowledges financial support from
``Programme National de Physique Stellaire'' (PNPS) of CNRS/INSU, France.

\section*{References}

 \bibliographystyle{abbrvnat}
\setcitestyle{authoryear,open={(},close={)}}
 \bibliography{ref}

\appendix

\section{Evolution of $\Lambda$ with $Ra$ and dynamo bifurcations}

  Previous systematic dynamo numerical studies provide a large table with the values of output parameters. We prefer here to provide additional figures in order to illustrate our results. These additional figures allow us to provide some details of our numerical experiments. The values of $\Lambda$ can be observed in dynamo bifurcations diagrams. As Morin \& Dormy (2009), the control parameter is $Ra$. 

\begin{figure}
\begin{center}
\begin{tabular}{cc}
\subfigure[$E=3\cdot10^{-4}$ and $\Pm=6$]{\includegraphics[width=7.cm]{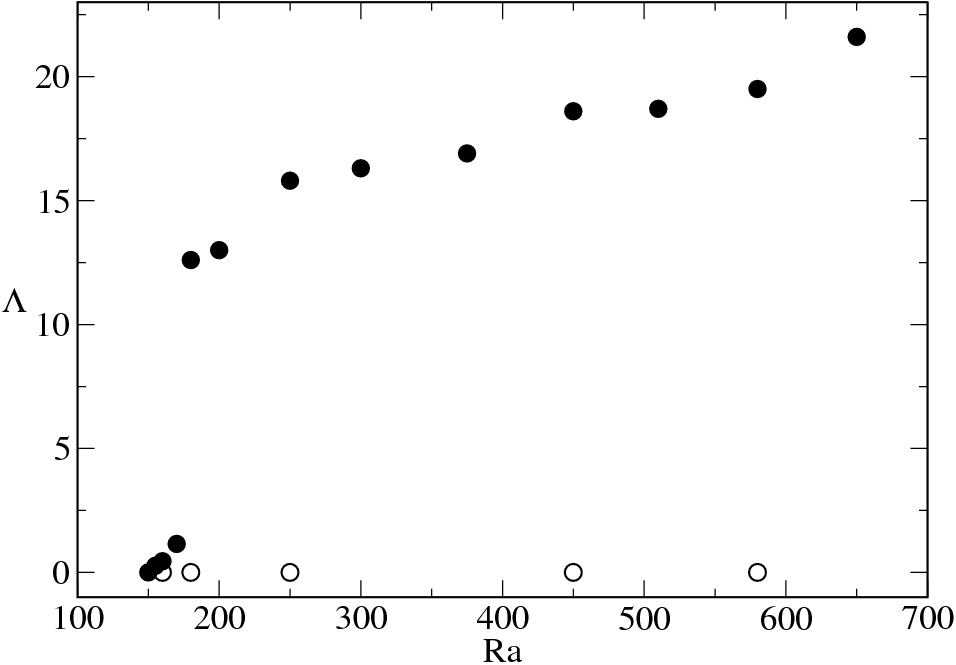}} &
\subfigure[$E=10^{-4}$ and $\Pm=12$]{\includegraphics[width=7.cm]{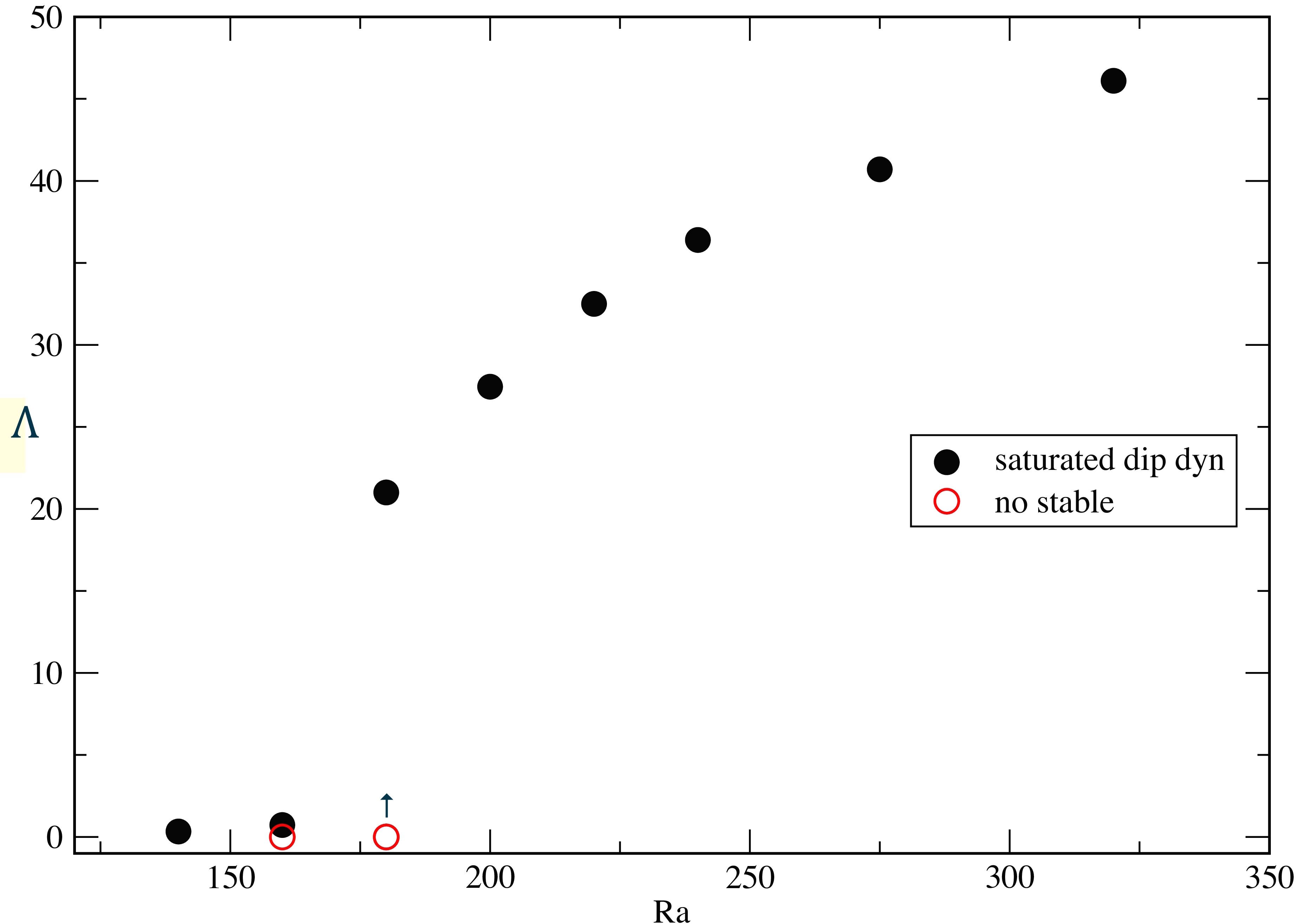}} 
\end{tabular} 
     \caption{Dynamo bifurcation diagrams are shown where empty circles correspond to initial conditions which are not stable. For $E=3\cdot 10^{-4}$ and $E=10^{-4}$, non-magnetic flows are classified as laminar flows with $Ra<220$ ($Ra<3.62Ra_c$) and $Ra<220$ ($Ra<3.25Ra_c$) respectively. }
     \label{biflam}
\end{center}
  \end{figure}

  By considering high values of $\Pm$, dynamo action can develop close to the onset of convection (in RI). Fig~\ref{biflam} contains dynamo bifurcation diagrams for the parameters $E=3\cdot 10^{-4}$, $\Pm=6$ (on panel (a)) and $E=10^{-4}$, $\Pm=12$ (on panel (b)). Regardless of the initial conditions for the magnetic field, a supercritical dipolar branch is obtained for these parameters. Close to the onset of dynamo action, the field strength saturates with a low magnitude and simulations with $Ra<180$ can be classified as laminar dynamos.  We observe an increase of the magnetic energy (measured by the Elsasser number $\Lambda$) by more than one order of magnitude when $Ra$ is increased from $Ra=160$ ($2.5Ra_c$) to $Ra=180$ ($3Ra_c$) for $E=10^{-4}$. A similar sharp increase is obtained for $E=3\cdot 10^{-4}$. Such a jump is only observed at low Rayleigh numbers. Then, a further increase of $Ra$ induces a moderate increase of $\Lambda$.  We only observe such sharp variations of $\Lambda$ if the laminar regime is explored.

\begin{figure*}
\begin{tabular}{cc}
\subfigure[$\Pm=3$]{\includegraphics[width=7.cm]{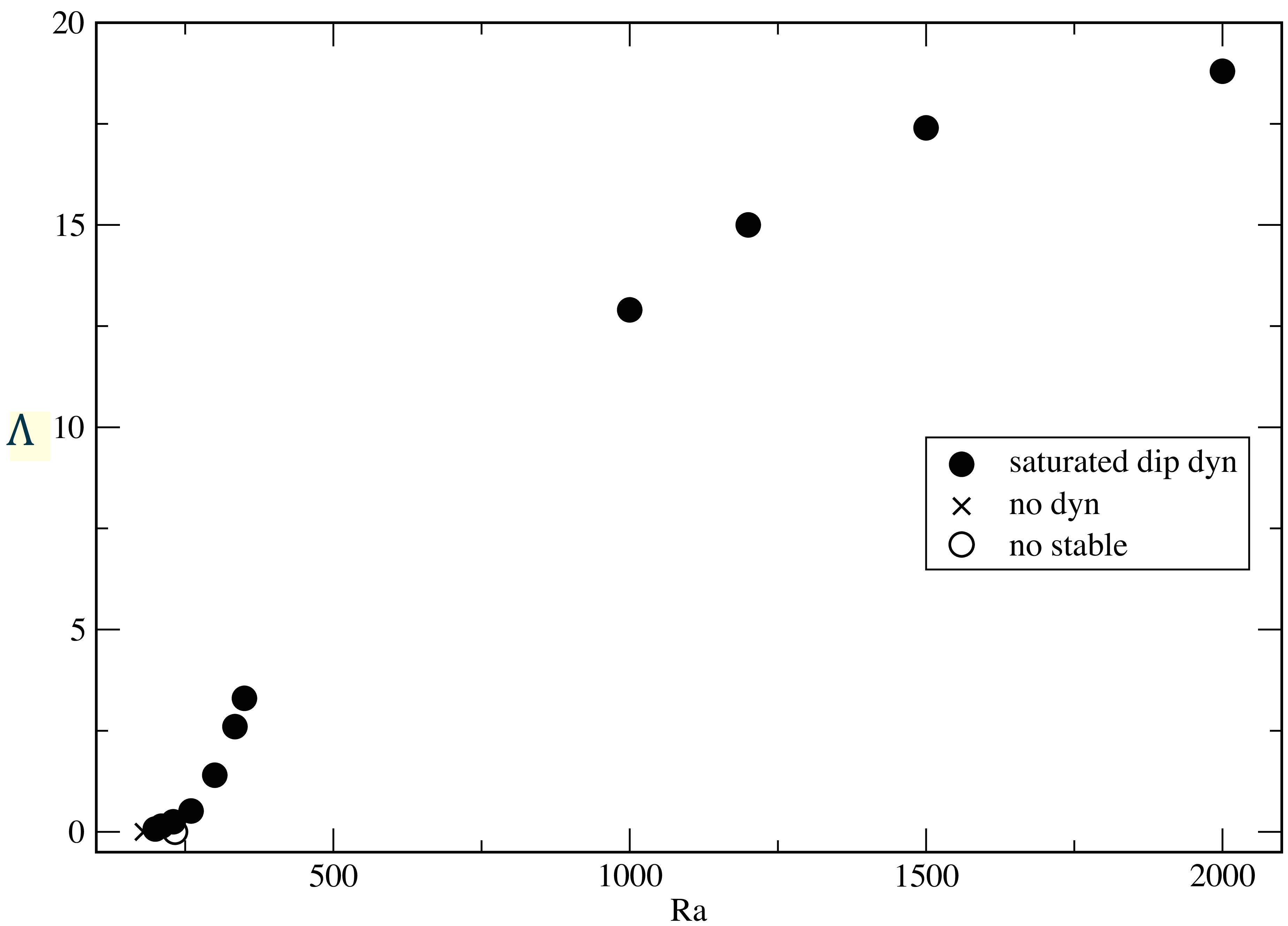}} &
\subfigure[$\Pm=2$]{\includegraphics[width=7.cm]{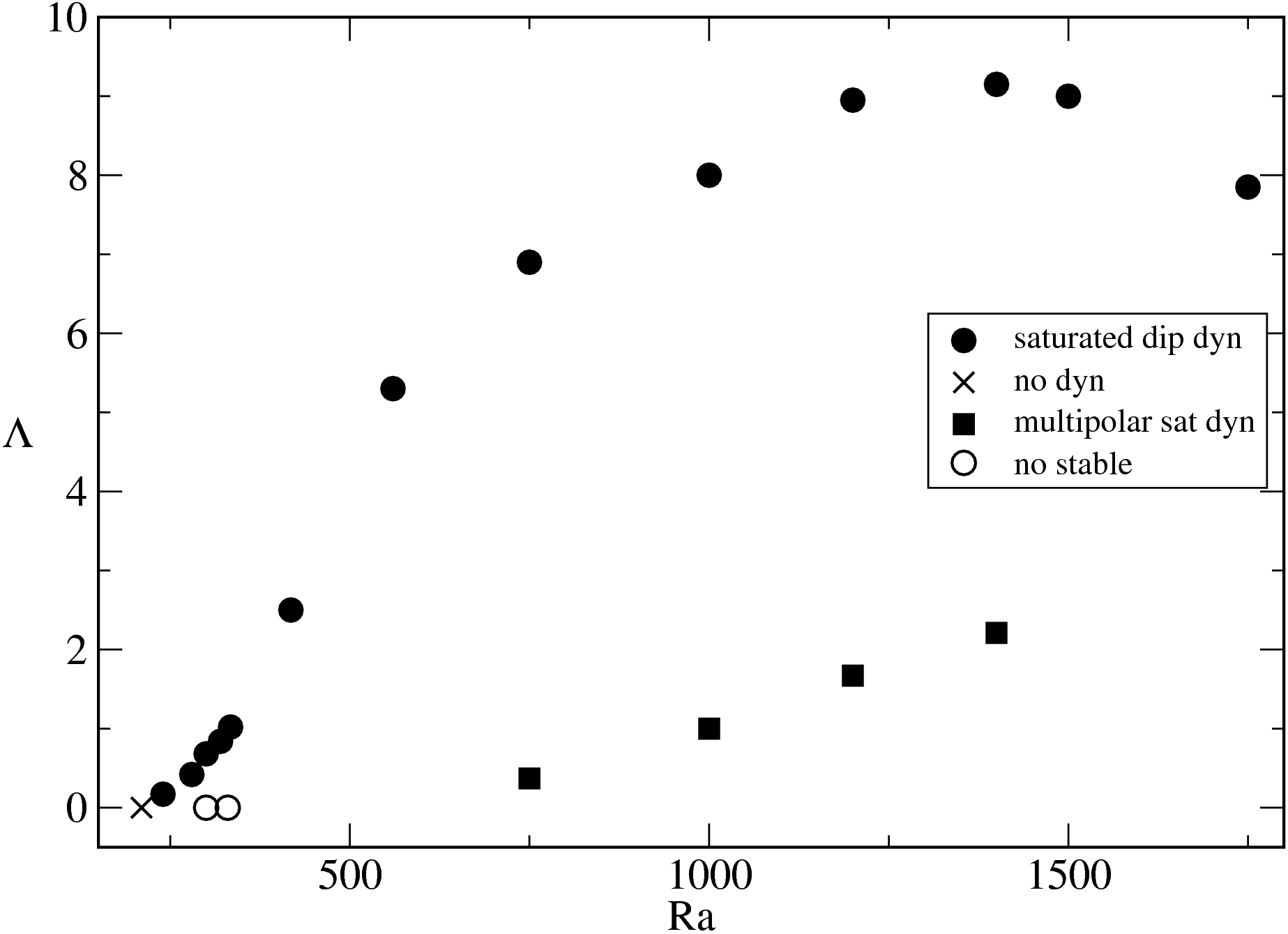}} \\
\subfigure[$\Pm=1$]{\includegraphics[width=7.cm]{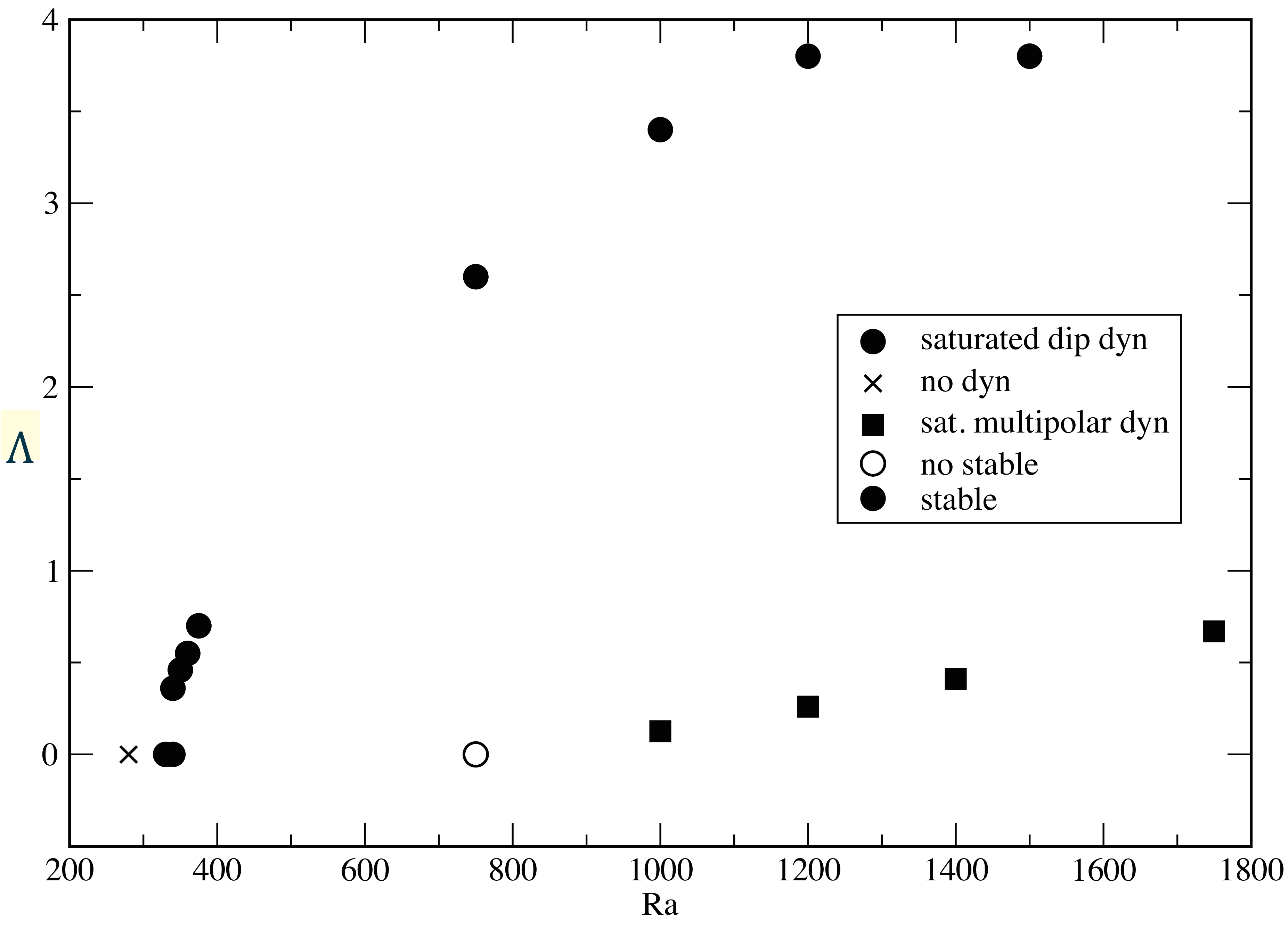}} &
\subfigure[$\Pm=0.5$]{\includegraphics[width=7.cm]{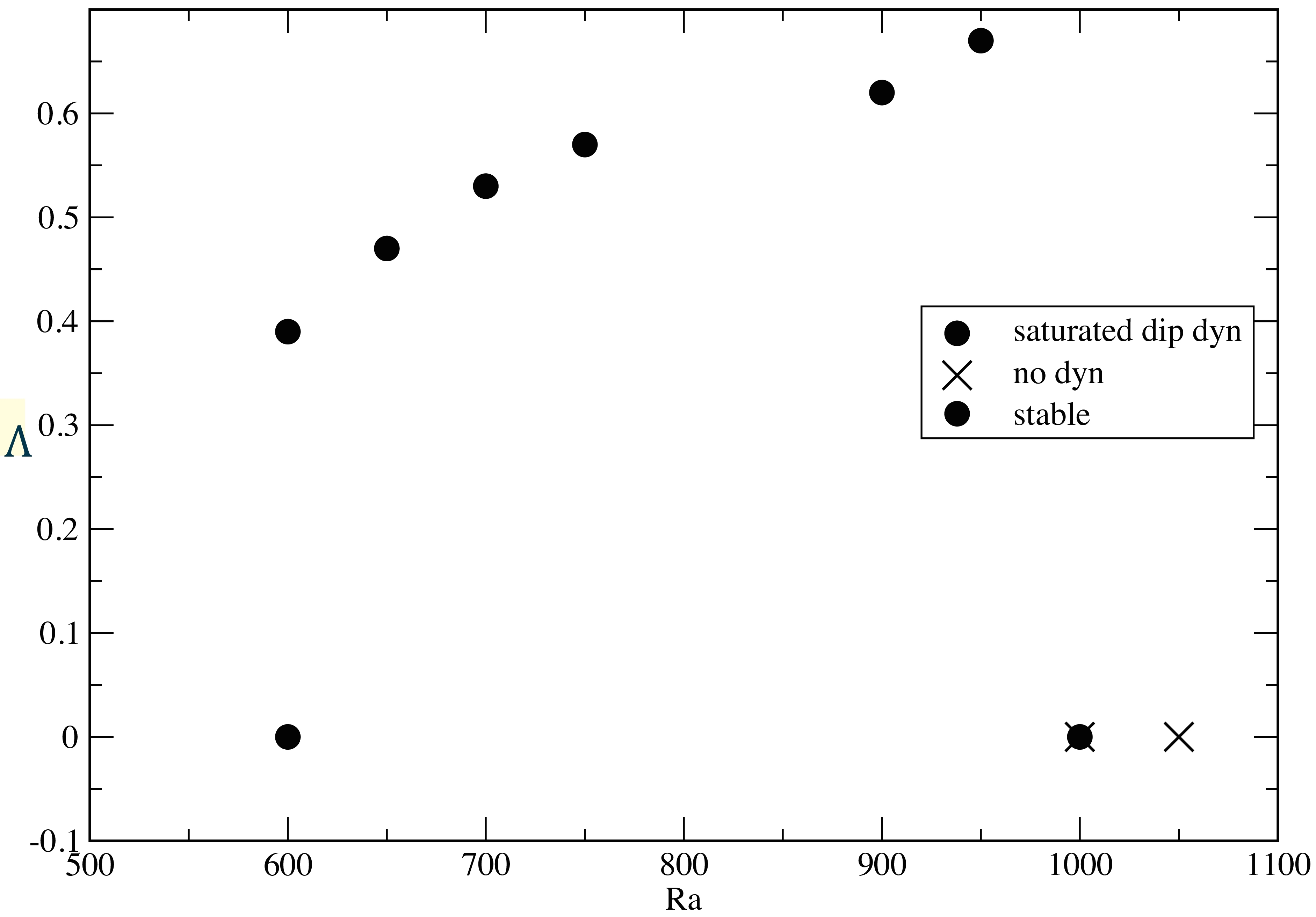}}
\end{tabular} 
     \caption{Bifurcation dynamo diagrams for $E=10^{-4}$. Full symbols correspond to stable solutions whereas empty symbols refer to initial conditions. Circles are employed for dynamos dominated by the axial dipole component at the surface and squares are employed otherwise.} 
     \label{bif_e4NS}
  \end{figure*}

\begin{figure*}
\begin{tabular}{cc}
\subfigure[$\Pm=2.5$]{\includegraphics[width=7.cm]{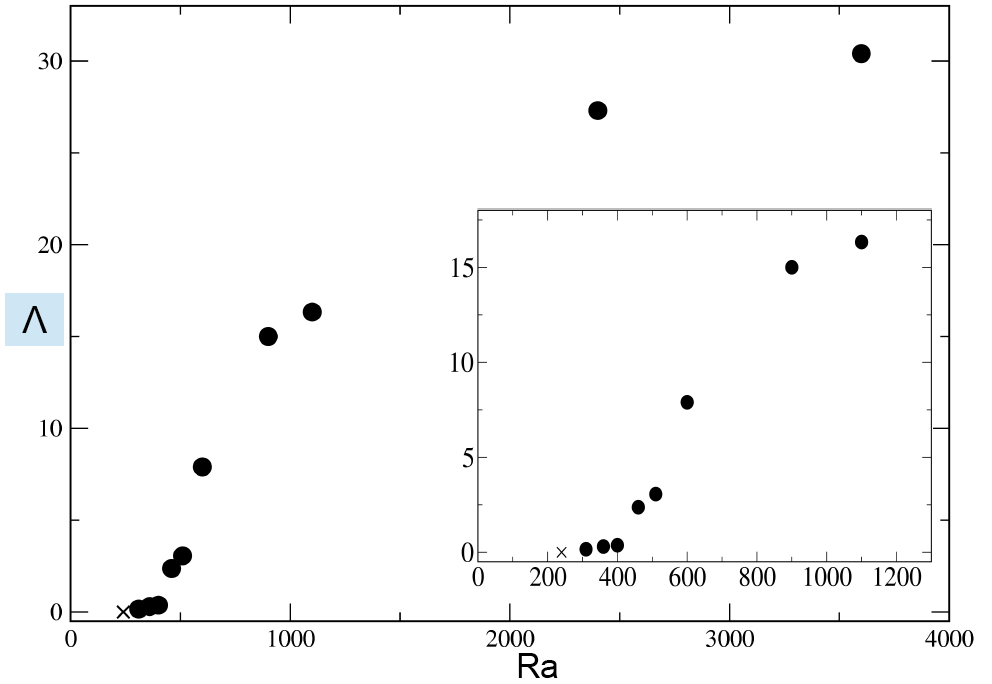}} &
\subfigure[$\Pm=1$]{\includegraphics[width=7.cm]{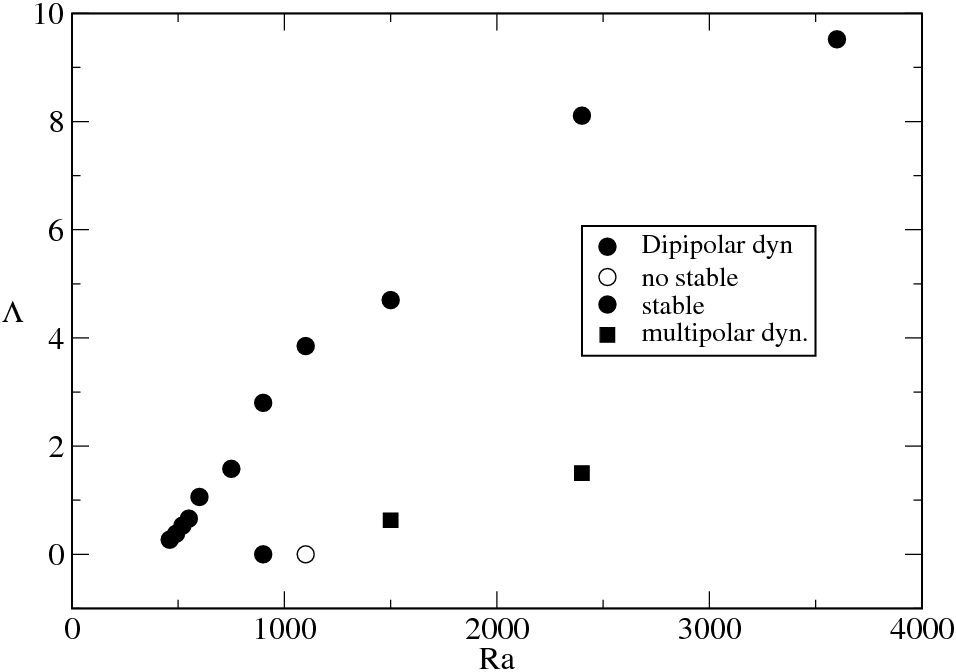}} \\
\subfigure[$\Pm=0.5$]{ \includegraphics[width=7.cm]{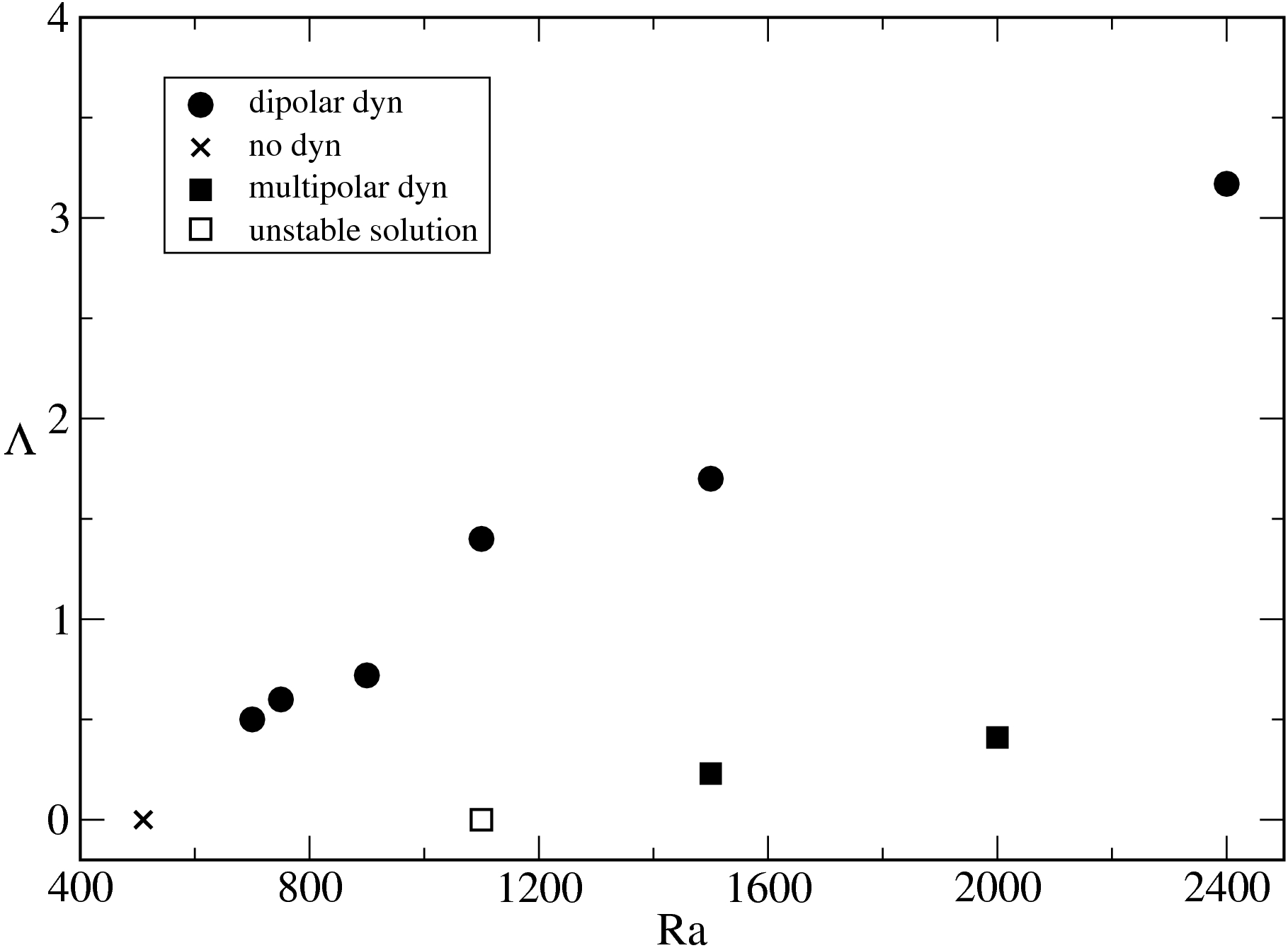}} &
\subfigure[$\Pm=0.25$]{\includegraphics[width=7.cm]{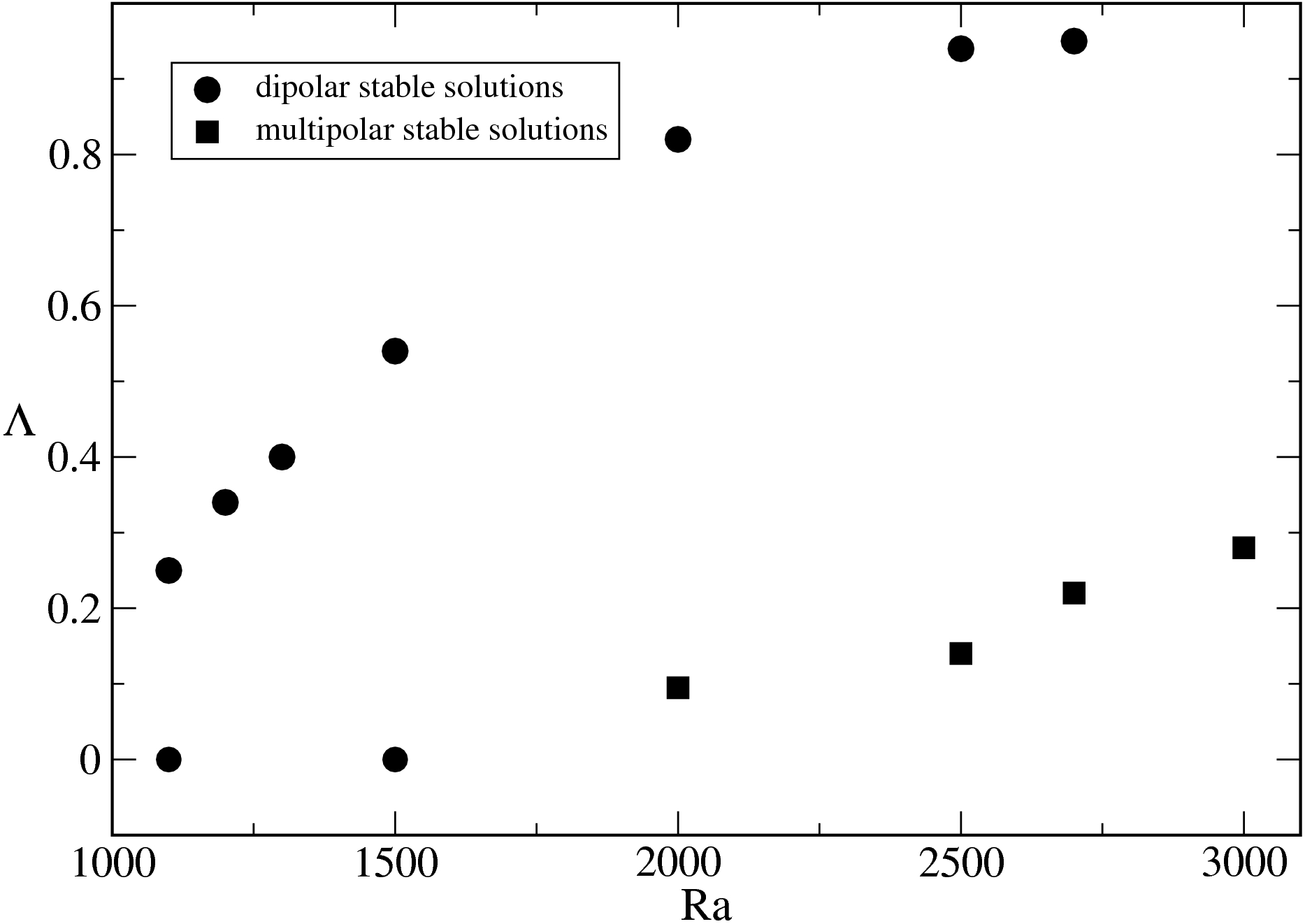}}
\end{tabular} 
     \caption{Dynamo bifurcation diagrams for $E=3\cdot 10^{-5}$. Symbols are defined in the legend of Fig.~\ref{bif_e4NS}.}
\label{bif_3e5NS}
\end{figure*}

\begin{figure}
\begin{tabular}{cc}
\subfigure[$\Pm=2$]{\includegraphics[width=7.cm]{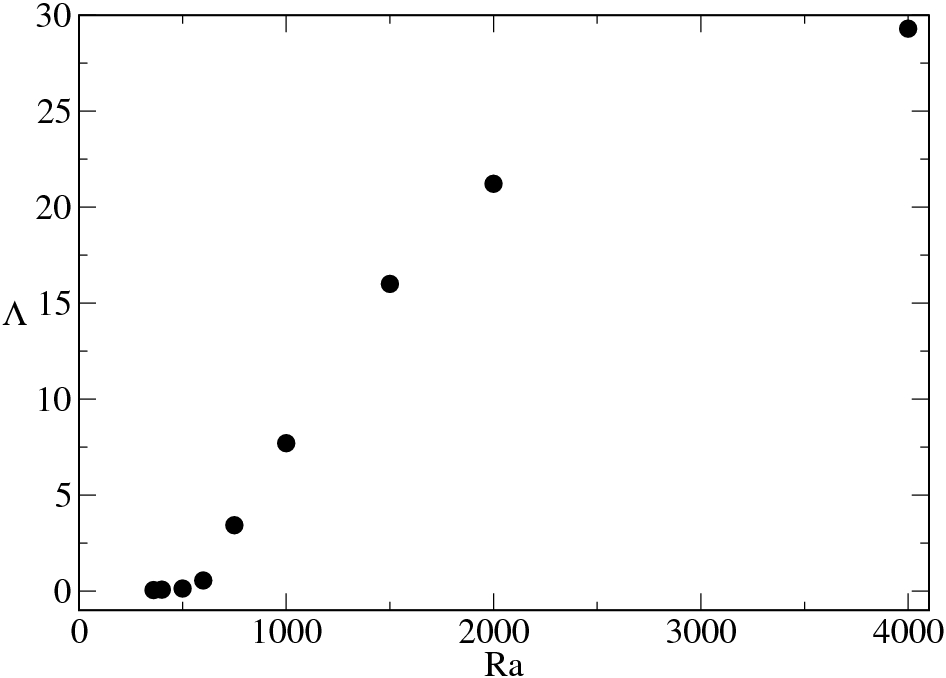}} &
\subfigure[$\Pm=1$]{ \includegraphics[width=7.cm]{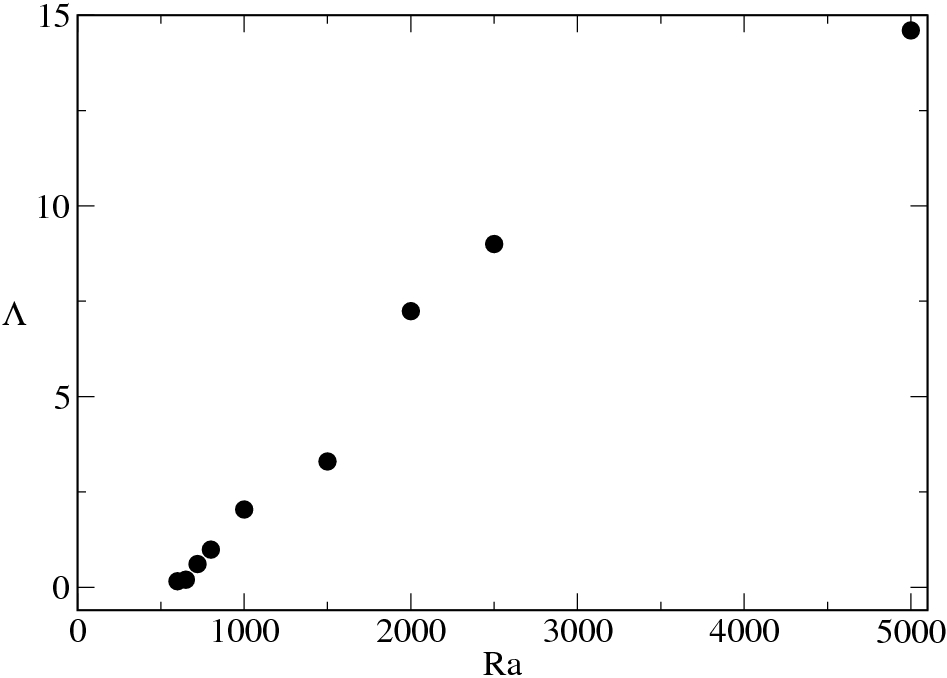}}
\end{tabular}
\caption{Magnetic field strength measured by $\Lambda$ as a function of the Rayleigh number $Ra$ for dipolar dynamos with $E=10^{-5}$. Supercritical bifurcations are observed. }
\label{bif_e5NSsup}
\end{figure}

\begin{figure}
\begin{tabular}{cc}
\multicolumn{2}{c}{\subfigure[$\Pm=0.5$]{\includegraphics[width=7.cm]{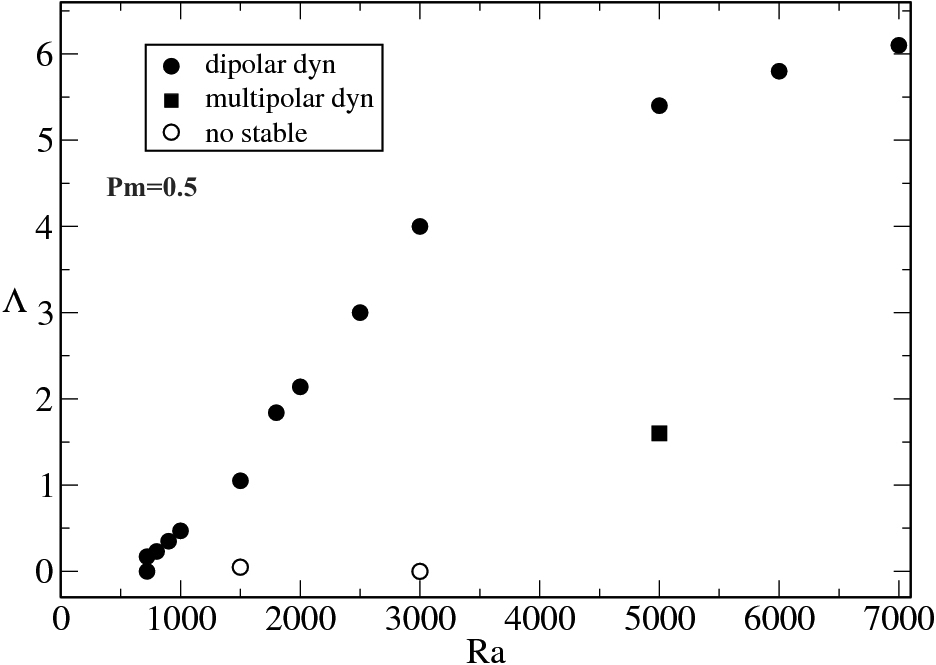}}} \\
\subfigure[$\Pm=0.25$]{\includegraphics[width=7.cm]{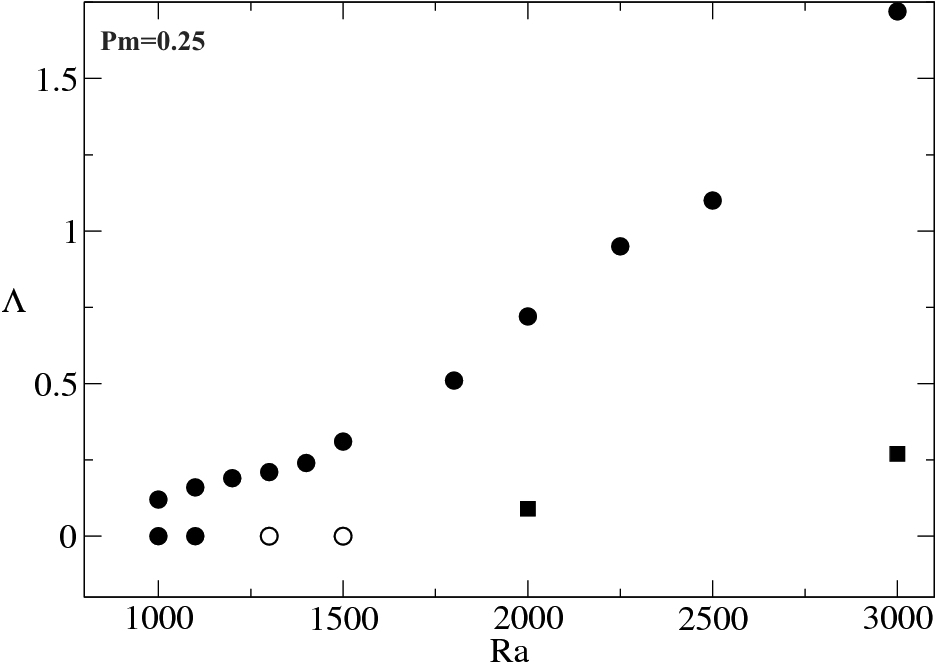}} &
\subfigure[$\Pm=0.20$]{\includegraphics[width=7.cm]{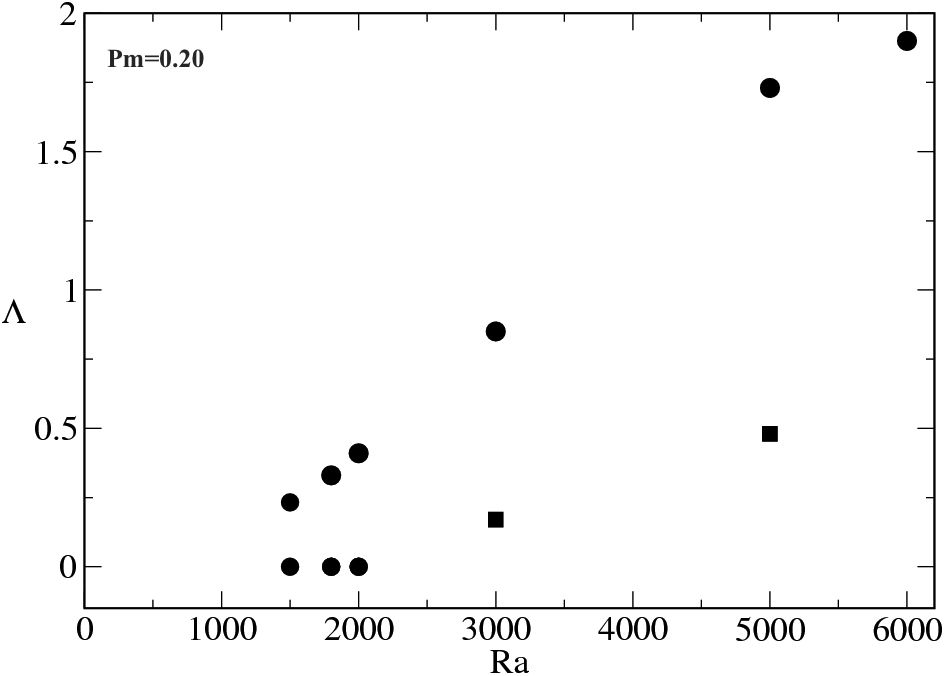}}
\end{tabular}
\caption{Magnetic field strength measured by $\Lambda$ as a function of the Rayleigh number $Ra$  with $E=10^{-5}$. The observed behaviour corresponds to subcritical bifurcations. }
\label{bif_e5NSsub}
\end{figure}

    Considering lower values of $\Pm$ allows to study the nature of the dynamo bifurcation in RII. Such diagrams are shown in Figs.~\ref{bif_e4NS},~\ref{bif_3e5NS},~\ref{bif_e5NSsup} and ~\ref{bif_e5NSsub}. Supercritical bifurcations (Fig.~\ref{bif_e5NSsup} and panels (a) for Figs.~\ref{bif_e4NS} and ~\ref{bif_3e5NS}) and subcritical bifurcations (Fig.~\ref{bif_e5NSsub} and panels (b),(c),(d) in Figs.~\ref{bif_e4NS} and \ref{bif_3e5NS}) for the dipolar branch are observed. In the latter bifurcation diagrams, multipolar solutions are also reported. Such dynamos are obtained only if the buoyant forcing is high enough (i.e. in RIII).

    Fig.~\ref{bif_e4NS} (panel (a)) shows the supercritical bifurcation as obtained by \citet{morinD09} (additional runs with higher $Ra$ have been also reported). If the Elsasser number $\Lambda$ grows rapidly with $Ra$ close to the onset of convection, it increases more slowly for more turbulent flows (with $Ra>10Ra_c$). A possible explanation for this behaviour could be that the Lorentz force could play an important role  far above the dynamo threshold and limits the growth of $\Lambda$. From another point of view, the flow properties could be changed significantly by increasing $Ra$ as shown by our kinematic study. The dynamo action could convert kinetic energy into magnetic energy differently in the laminar regime and in the more turbulent regimes. 

  In Fig.~\ref{bif_e4NS} (see panel $(b)$), a dynamo bifurcation diagram for $\Pm=2$ is given. For low $Ra$, supercritical behaviour seems to be obtained as the magnetic energy is almost proportional to the distance from the dynamo threshold. Surprisingly, $\Lambda$ as a function of $Ra$ is not monotonous and it decreases for sufficiently high $Ra$ values. This behaviour has also been reported by \citet{christensen06} for $E=3\cdot 10^{-4}$ and $\Pm=3$. If the magnitude of the field is initially low (white circles) at low $Ra$, the magnetic energy grows exponentially with a preferentially dipolar symmetry in the kinematic phase. Saturation process occurs by changing the velocity field (see section 5). From a weak initial field, nonlinear dipolar solutions are generated with $Ra<500$. Whereas, for higher $Ra$ (Regime III), the saturation mechanism does not support the dominance of the axial dipole field and a multipolar dynamo branch exists even if $Ro_l<0.1$ (multipolar runs have $0.062<Ro_l<0.095$). Since increasing $Ra$ changes the force balance by promoting inertia, this term must have an important role in the dynamo mechanism for this multipolar branch. This bistable regime can be compared to that highlighted by \citet{schrinner12} in which multipolar dynamos were identified as $\alpha\omega$ dynamos where the large-scale differential rotation plays a constructive role through the $\omega$-effect. In the present study, no-slip boundaries are used and the large-scale velocity field is mainly controlled by Ekman layers. The evolution of the magnetic energy with $Ra$ for the multipolar branch suggests supercritical behaviour for the multipolar branch.

  For $\Pm=1$ (see panel $(c)$ in Fig.~\ref{bif_e4NS}), we note that the dipolar dynamo with $Ra=340$ is only obtained if the initial field is sufficiently strong. Otherwise, no dynamo can be generated from a weak initial field. This behaviour corresponds to a subcritical bifurcation for the dipolar branch even if for $Ra=750$, the saturated dipolar dynamo can be obtained from a weak field. Then, $\Lambda$ increases slowly with $Ra$ while $Ro_l<0.1$. For $Ra=1750$, the local Rossby number does not satisfy this criteria and only a multipolar saturated dynamo is stable regardless of the initial field. The dipolar solution is not restored if $Ra$ is reduced to $1500$ from the multipolar solution with $Ra=1750$ and hysteretic behaviour is noted. This multipolar branch exists for $0.60<Ro_l<0.12$. As a result, depending on the intial magnetic conditions, either multipolar dynamos or dipolar dynamos are finally relevant solutions in the bistable regime corresponding to $0.60<Ro_l<0.10$ for $\Pm=1$. We again note that $\Lambda$ increases linearly with $Ra$ for the multipolar branch whereas for the dipolar branch, $\Lambda$ saturates in the bistable regime in which inertia affects the dynamics. 

  For $\Pm=0.5$ and $E=10^{-4}$ (panel $(d)$), the bifurcation is clearly subcritical and dynamo solutions only exist in a wedge of $Ra$. Regardless of the initial conditions for the field, at $Ra=1000$ no dynamo solutions exist, i.e. an initial strong dipolar field can not be maintained in time (cross symbol) and a weak initial field is not amplified (full circle symbol at $\Lambda=0$). This point has been reported by \citet{christensen06} in their fig.~1.  At $Ra=1000$, the local Rossby number is approximately two times lower than the critical value $0.1$ which corresponds to the collapse of the dipolar branch by inertia. Inertia has a negative influence on dynamo action by increasing $Rm_c$ in RIII (see kinematic study).

  In Fig.~\ref{bif_3e5NS}, bifurcation diagrams for $E=3\cdot 10^{-5}$ are presented.  For $\Pm=2.5$ (see panel(a)), a supercritical bifurcation for the dipolar branch with $\Lambda$ which has a non-simple dependency when $Ra<1000$. Otherwise, the non-magnetic flow is in RIII and the field strength increases slowly with $Ra$. Very close to the dynamo threshold, three dipolar dynamos are obtained with $\Lambda<0.5$ and $Ra=310$, $Ra=360$ and $Ra=400$. Then, stronger variations of $\Lambda$ are observed with $\Lambda$ higher than unity. 

  For lower values of $\Pm$ and $E=3\cdot 10^{-5}$, subcritical bifurcations are clearly obtained for the dipolar branch while a supercritical multipolar branch appears when $Ra$ is sufficiently high (see Fig.~\ref{bif_3e5NS} panel (b),(c) and (d)). For $\Pm=1$ and the range $460\leq Ra\leq 900$ or $\Pm=0.5$ and the range $700\leq Ra\leq 900$, strong initial dipolar fields are necessary for dynamo action as weak fields are not amplified. For $Ra=1100$, weak fields are amplified by a flow in RIII. Even if a field with a multipolar morphology grows in the kinematic phase, the nonlinear solution is finally dominated by the axial dipolar component. Multipolar solutions appear for higher $Ra$. For $\Pm=0.25$ and $Ra=3000$, no dipolar dynamo can be maintained as $Ro_l$ exceeds the critical value $0.12$. Since a large range of $Ra$ is studied up to the dipole collapse by inertia, we observe saturation of $\Lambda$ in RIII.

  For $E=10^{-5}$ and $\Pm=1$ or $\Pm=2$, the field strength increases rapidly when $Ra$ is close to the dynamo threshold (see Fig.~\ref{bif_e5NSsup}). For these values, the bifurcations are supercritical and the dynamo onset can be approached. RIII is explored for $E=10^{-5}$ with $Ra\geq 2000$. In this case, the increase of $\Lambda$ depends slightly on $Ra$ for $\Pm=1$ and $\Pm=2$. At lower values of $\Pm$, the bifurcation for the dipolar branch is subcritical (see Fig.~\ref{bif_e5NSsub})  and multipolar solutions are also obtained in the turbulent regime. The growth of $\Lambda$ is still less important for the dipolar branch as the non-magnetic flow is in the turbulent regime. 

 For $\Pm=0.5$ and $E=10^{-5}$, only one solution with $\Lambda=0$ (kinematically stable) is reported in Fig.~\ref{bif_e5NSsub} for $Ra=720$. However, for $Ra=800$ and $Ra=1000$, dynamo simulations with weak initial fields were tested.  The magnetic energies increase with growth rates close to zero and attaining saturation would require a very long time integration.  At $Ra=1500$ and $Ra=3000$, transitions from multipolar to dipolar solutions are observed, as is the transition shown in Fig.~\ref{run3e4Ra450Pm6}.

\section{Temporal evolution of geodynamo simulations}


\begin{figure}
\begin{tabular}{cc}
\subfigure[]{\includegraphics[width=6.5cm]{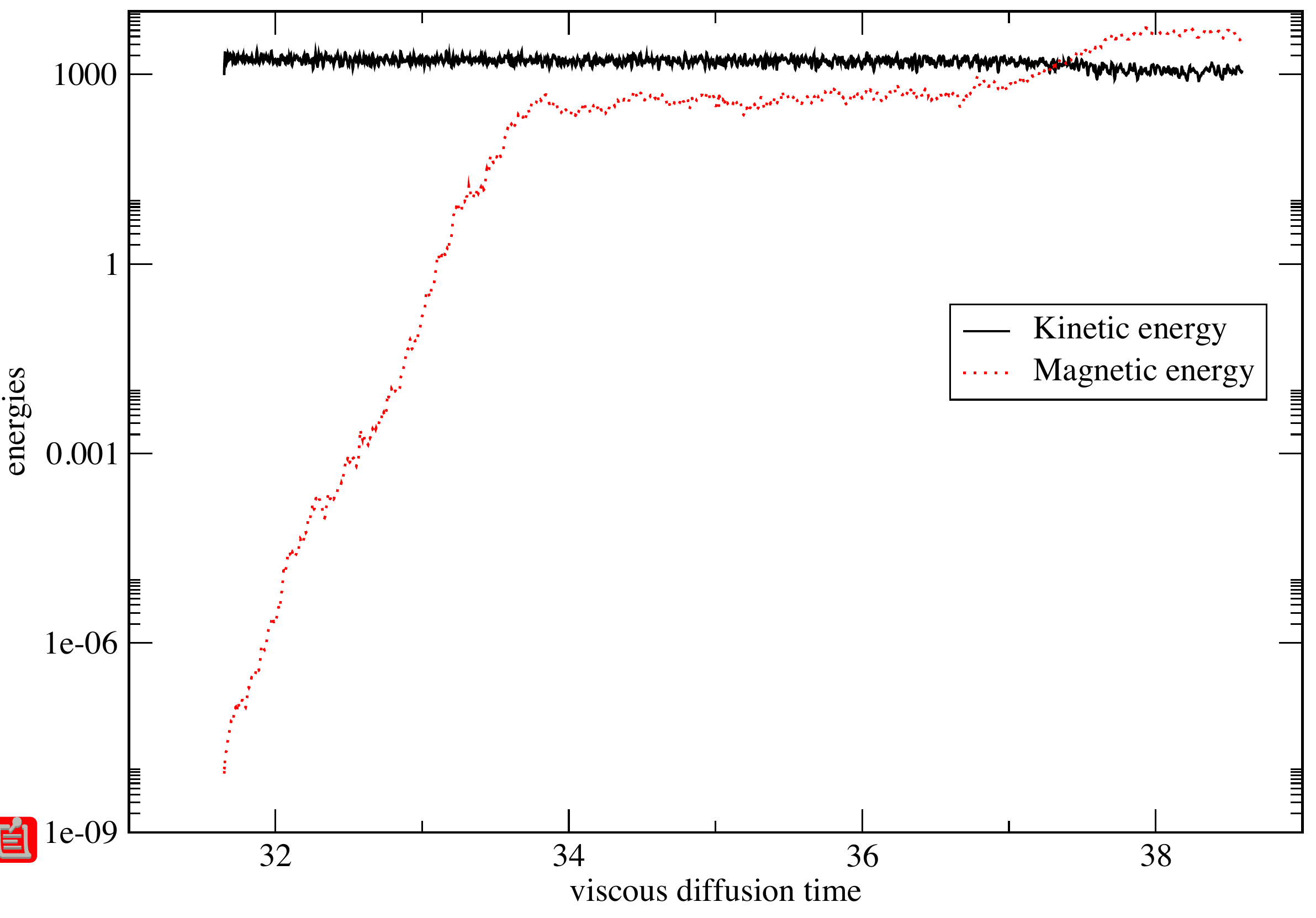}} & 
\subfigure[]{ \includegraphics[width=6.5cm]{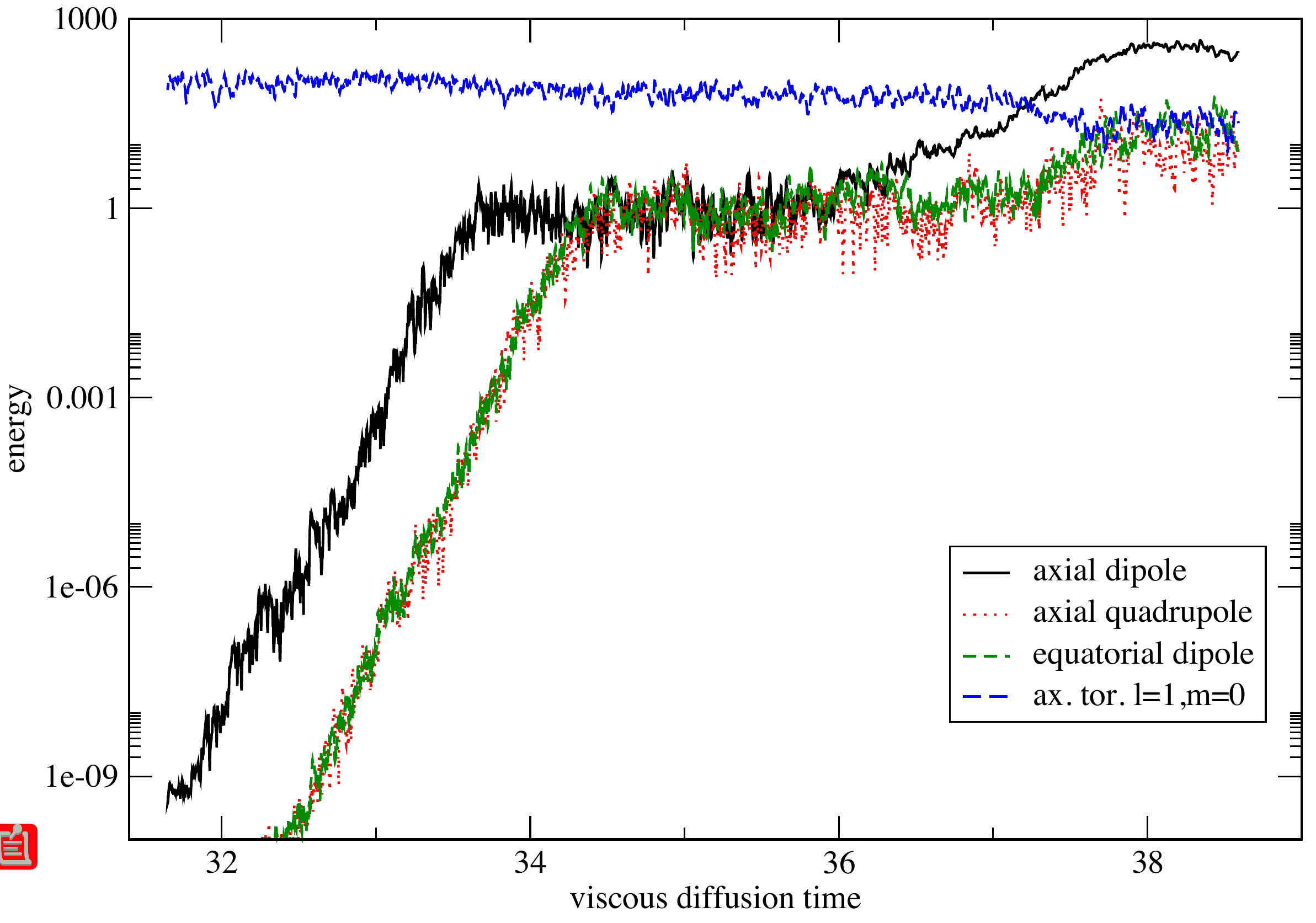}}
\end{tabular}
\caption{Typical behaviour for a dipolar dynamo obtained with a weak initial field. (a) Magnetic (red) and kinetic (black) energies as a function of time. (b)  the label $ax.tor.l=1,m=0$ corresponds to the kinetic toroidal mode with the spherical decomposition $l=1$ and $m=0$. For this simulation, the parameters are $E=3\cdot 10^{-4}$, $\Pm=6$ and $Ra=450=7.4Ra_c$.}
\label{run3e4Ra450Pm6}
\end{figure}
  
  In Fig.~\ref{run3e4Ra450Pm6}, the time evolution of kinetic and magnetic energies is shown for the parameters $E=3\cdot 10^{-4}$, $Ra=450=7.4Ra_c$ and $\Pm=6$. A dipolar solution finally sets in even when a weak initial field was considered. Such a behaviour correspond to crossed out squares in Fig~\ref{RaPmbif}. In the kinematic phase, different magnetic modes are amplified. The kinematic phase ends when the magnetic energy reaches a plateau at $t=34$. This state takes place for a period of two in units of viscous diffusion time. In this period, the axial dipole field does not dominate and the dynamo could be identified as a multipolar dynamo. But, the axial dipole mode finally grows giving rise to a dipolar solution. The transition described here corresponds to a multipolar/dipolar transition where the multipolar phase is a transient state. Only one example of such a transition is reported in Fig.~\ref{run3e4Ra450Pm6} but several were obtained and operate similarly. 

\begin{figure}
\begin{center}
\begin{tabular}{cc}
\subfigure[]{\includegraphics[width=6.5cm]{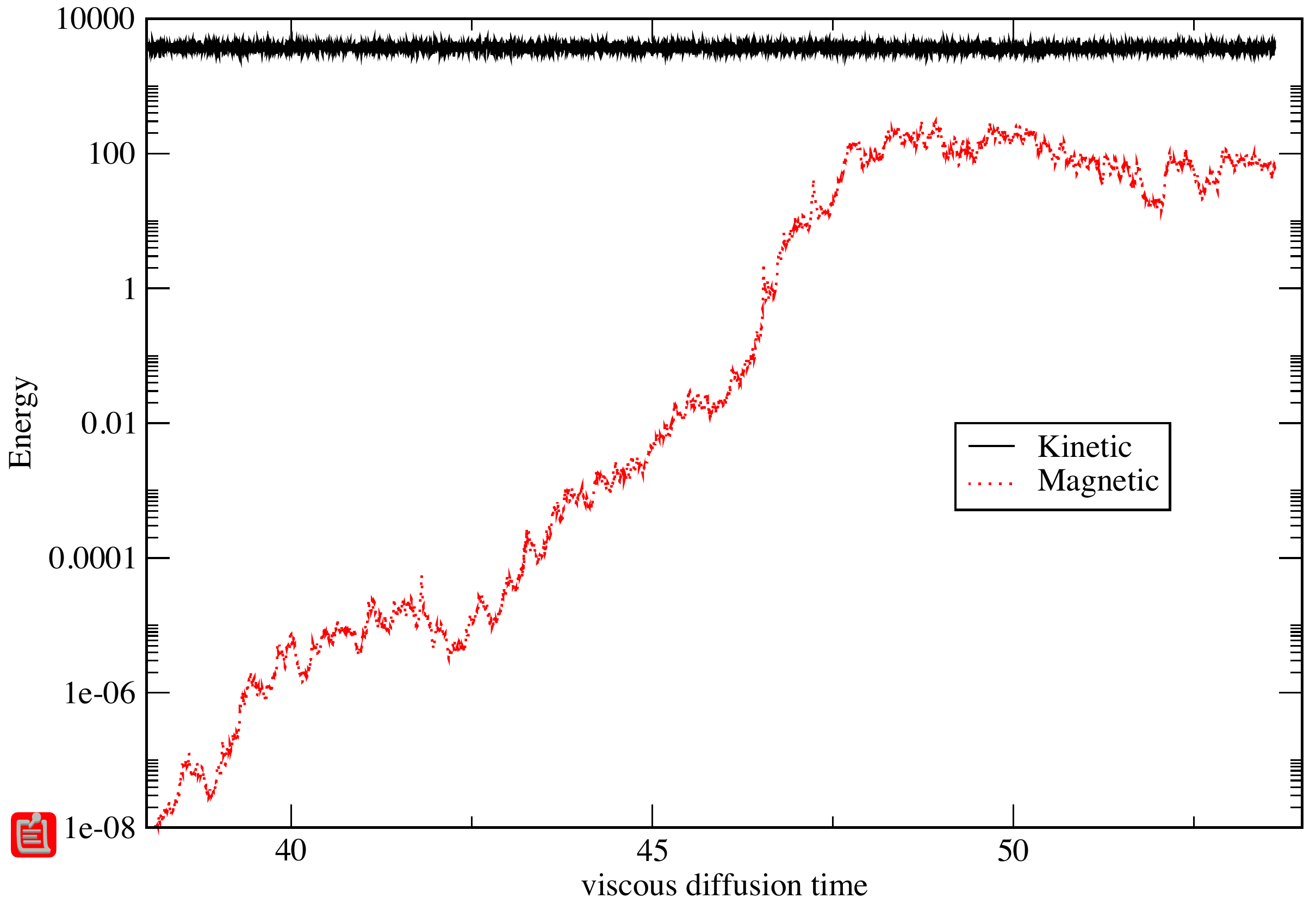}} &
\subfigure[]{\includegraphics[width=6.5cm]{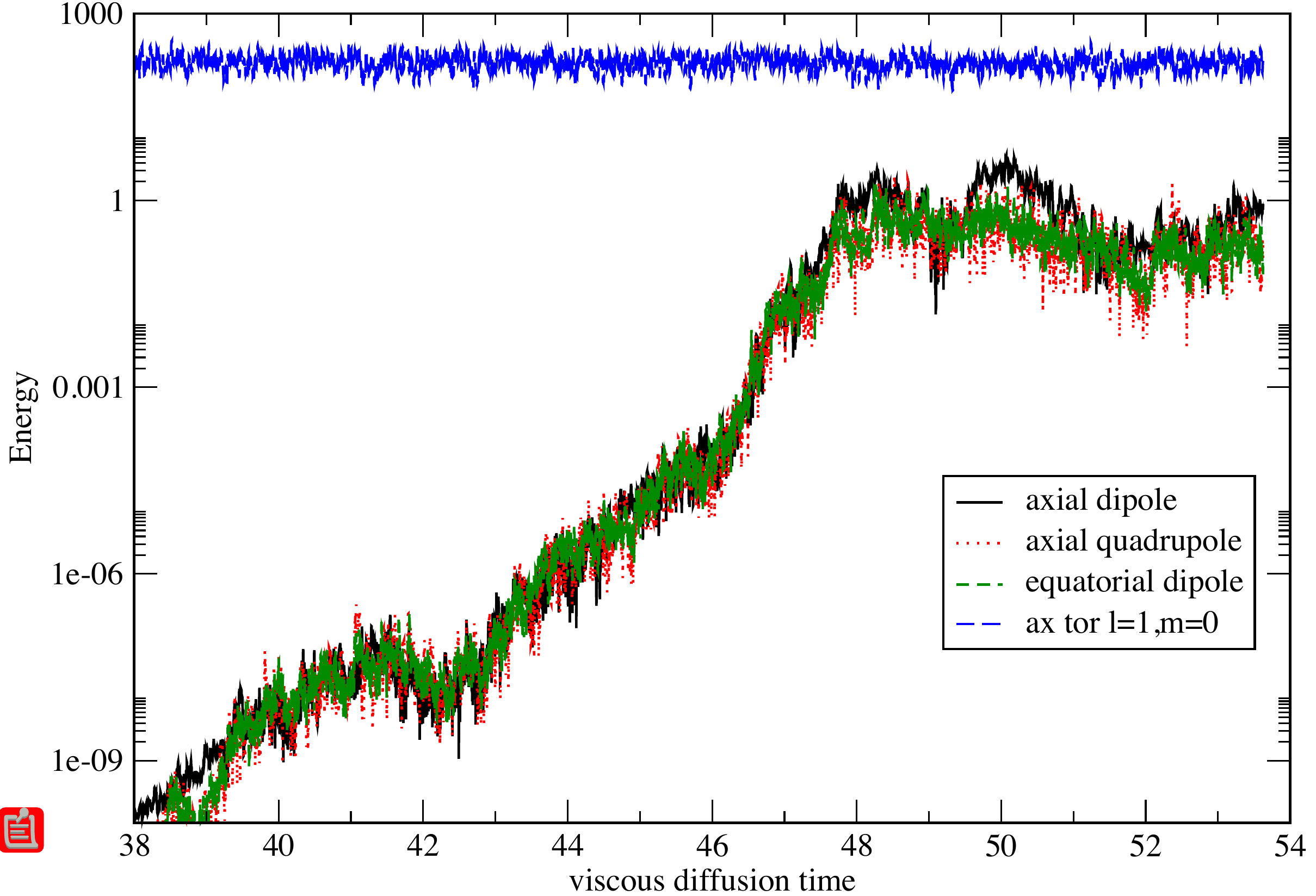}}
\end{tabular}
\caption{Typical behaviour for a multipolar dynamo obtained with a weak initial field. The legend is similar to that of Fig~\ref{run3e4Ra450Pm6}.  The parameters are $E=3\cdot 10^{-4}$, $\Pm=3$ and $Ra=650=10.69Ra_c$. The magnitude of the initial weak magnetic field is first amplified in the kinematic phase. Then, a non-dipole dominated solution is obtained even if $Ro_l=0.091$ and no-slip boundaries are used. }
\label{run3e4Ra650Pm3w}
\end{center}
\end{figure}

\begin{figure}
\begin{center}
\begin{tabular}{cc}
\subfigure[]{\includegraphics[width=7.cm]{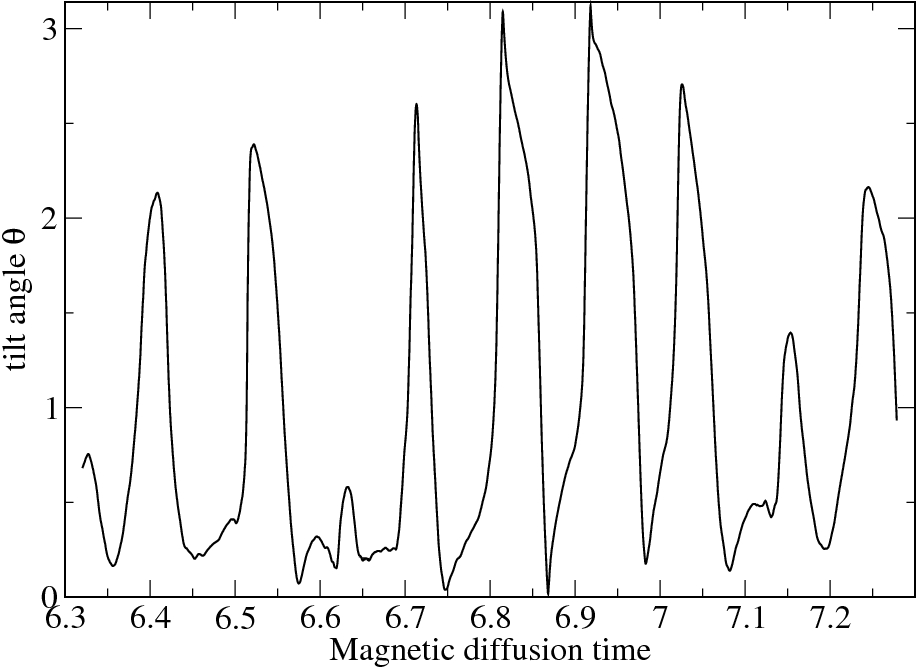}} &
\subfigure[]{\includegraphics[width=7.cm]{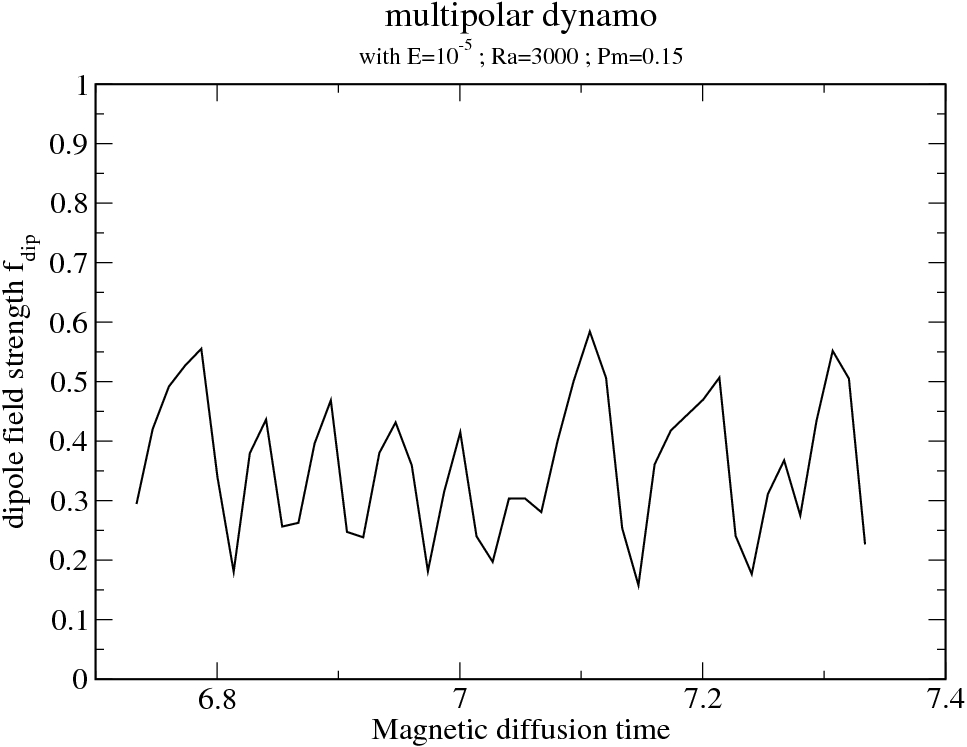}}
\end{tabular}
\caption{Tilt angle (panel a) and dipole field strength $f_{dip}$ (panel b) as a function of time for a multipolar dynamo model close to the dynamo threshold with the parameters $E=10^{-5}$, $Ra=3000=28.4Ra_c$ and $\Pm=0.15$. Dynamo action is lost when $\Pm$ is decreased to $\Pm=0.10$.}
\label{periodrevers}
\end{center}
\end{figure}

   In RIII, by considering a weak initial field we observe the existence of multipolar dynamo solutions which have been integrated for a period of time longer than one magnetic diffusion time (see Fig~\ref{run3e4Ra650Pm3w})  even if the local Rossby number is lower than 0.1 i.e. in the usual dipolar regime (see~\citealt{christensen06}). Such dynamos are typical for simulations performed in RIII with a weak initial field and $Rm$ slightly above $Rm_c$. The relative importance of zonal flows increases as the Ekman number decreases (see the hydrodynamical study). In these models, we observe periodic reversals of the axial dipole component. An example of such models is shown in Fig~\ref{periodrevers}. The importance of the dipolar mode as measured by the dipole field strength can be dominant (when $f_{dip}$ exceeds or approaches 0.5).

\begin{figure}
\begin{center}
\begin{tabular}{c}
\includegraphics[width=8.cm]{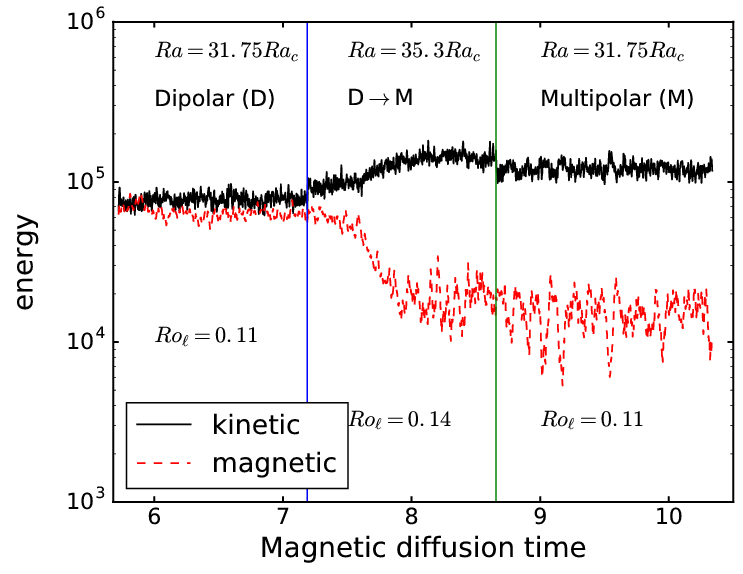}
\end{tabular}
\caption{Illusation of the hysteretic behavior observed when the buoyant forcing $Ra$ is varied such that the the local Rossby number $Ro_\ell$ shifts from one regime (dipolar $Ro_\ell<0.12$ or multipolar $Ro_\ell>0.12$) to another. The other parameters for this models are $E=3\cdot 10^{-5}$ and $\Pm=0.25$. The vertical lines mark a variation of the buoyant forcing $Ra$.}
\label{fighyst}
\end{center}
\end{figure}

  In Fig~\ref{fighyst}, magnetic and kinetic energies as a function of time are given for a numerical experiment which shows the existence of a hysteretic behavior when the buoyant forcing is varied. In the dipolar regime ($Ro_\ell<0.12$), a strong initial dipolar field is maintained in time. The axial dipole collapses when $Ra$ is increased as $Ro_\ell$ crosses the transitional value and the magnetic energy (or $\Lambda$) decreases significantly. For these parameters, the multipolar solution appears to be the only stable solution. Important zonal flows develop close to the dynamo threshold when the Ekman number is sufficently low. If the Rayleigh number is decreased from the multipolar solution $Ra=35.3Ra_c$ to its initial value ($Ra=31.75Ra_c$), the dipolar configuration is not obtained. We cannot exclude that this multipolar solution with $Ro_\ell<0.12$ is a transient state. However, this multipolar solution is maintained longer than one magnetic diffusion time. Most of our models were integrated over one magnetic diffusion time. For some cases, longer time integration periods were considered.

\end{document}